\documentclass[11pt]{article}
\usepackage{amsmath}
\usepackage{amssymb}
\usepackage{amsthm}
\usepackage{graphicx,psfrag,epsf}
\usepackage{psfrag,epsf}
\usepackage{enumitem}
\usepackage{natbib}
\usepackage{url} 
\usepackage[titletoc,title]{appendix}
\usepackage{array}
\usepackage{bm}
\usepackage[dvipsnames]{xcolor}
\usepackage{pifont}
\usepackage{color}
\usepackage{graphicx}
\usepackage{epstopdf}
\usepackage[letterpaper, portrait, margin=1in]{geometry}
\usepackage{lipsum}
\usepackage{listings}
\usepackage{hyperref}
\usepackage{float}
\usepackage[font=footnotesize,labelfont=bf]{caption}
\usepackage{color}
\usepackage{titlesec}
\usepackage{authblk}
 \usepackage{relsize}
 \usepackage{hhline}
\usepackage{multirow}

\usepackage{subcaption}

\usepackage{color}
\usepackage{verbatim}

\usepackage[dvipsnames]{xcolor}
\usepackage{hyperref}
\usepackage{xr}
\newcommand{\E}{\text{E}}
\newcommand{\I}{\mathbb{I}}
\newcommand{\prob}{\text{pr}}
\newcommand{\Var}{\text{var}}
\newcommand{\Cov}{\text{cov}}
\newcommand{\SE}{\text{SE}}
\newcommand{\tr}{^{\mkern-1.5mu\mathsf{T}}}

\newcommand{\given}{\;\middle|\;}
  
\newcommand{\tset}{\mathcal{W}}
\newcommand{\tsettilde}{\tset_{\tilde{z}}}
\newcommand{\tsetone}{\omega}
\newcommand{\z}{\mathbf{z}}
\newcommand{\tauestbibd}{\widehat{\tau}^{\text{BIBD}}}
\newcommand{\tauestfe}{\widehat{\tau}^{\text{BIBD-adj}}}
\newcommand{\tauestk}{\widehat{\tau}_k}
\newcommand{\tauestcbd}{\widehat{\tau}^\text{CBD}}

\newcommand{\tauestclus}{\widehat{\tau}^\text{ClusRD}}

\newcommand{\yestk}{\widehat{{Y}}_k}
\newcommand{\yest}{\widehat{{Y}}}
\newcommand{\yeststar}{\bm{\yest}_{*}}
\newcommand{\HT}{\text{HT}}
\newcommand{\yht}{\yest_{\text{HT}}}
\newcommand{\haj}{\text{Haj}}
\newcommand{\yhaj}{\yest_{\text{Haj}}}
\newcommand{\tauest}{\widehat{\tau}}
\newcommand{\tauhtg}{\tauest_{\HT}(\bm{g})}

\newcommand{\tauestg}{\tauest(\bm{g})}
\newcommand{\tauwg}{\tau_{\bm{w}}(\bm{g})}

\newcommand{\onehat}{\widehat{1}_{\text{HT}}}
\newcommand{\onehatz}{\widehat{1}_{\text{HT}}(z)}
\newcommand{\ybarw}{\bm{\bar{Y}}_{\bm{w}}}
\newcommand{\ybarzw}{\bar{Y}(z;\bm{w})}
\newcommand{\yestkadj}{\widehat{{Y}}_k^{\text{adj}}}
\newcommand{\yestadj}{\widehat{{Y}}^{\text{adj}}}

\newcommand{\ybark}{{\bar{Y}}_{k}}
\newcommand{\ybarj}{{\bar{Y}}_{j}}
\newcommand{\ybar}{{\bar{Y}}}

\newcommand{\bht}{\bm{B}_{\text{HT}}}
\newcommand{\bmw}{\bm{W}}

\newcommand{\varestwbfe}{\widehat{\sigma}^2_{\text{adj-wb}}(z_1,z_2)}
\newcommand{\varestbbfe}{\widehat{\sigma}^2_{\text{adj-bb}}(z_1,z_2)}

\newcommand{\Rk}{\textbf{R}_k}
\newcommand{\Rj}{\textbf{R}_j}
\newcommand{\R}{\textbf{R}}
\newcommand{\bfR}{\mathcal{R}}
\newcommand{\V}[1]{\bar{V}_{k}\left(#1, \tset_{#1}\right)}
\newcommand{\Vtau}{V_{k}\left(z_1, z_2\right)}
\newcommand{\bmmu}{\bm{\mu}}
\newcommand{\bmdelta}{\bm{\delta}}
\newcommand{\bmeta}{\bm{\eta}}

\newcommand{\Shatbb}{\bm{\widehat{S}}^{bb}}
\newcommand{\Shatbbstar}{\Shatbb_{*}}
\newcommand{\Shatbbht}{\Shatbb_{\HT}}
\newcommand{\Shatbbhaj}{\Shatbb_{\haj}}
\newcommand{\Shatwb}{\bm{\widehat{S}}^{wb}}
\newcommand{\Shatwbstar}{\Shatwb_{*}}
\newcommand{\Shatwbht}{\Shatwb_{\HT}}
\newcommand{\Shatwbhaj}{\Shatwb_{\haj}}
\newcommand{\SHT}{S_{\HT}^2}
\newcommand{\sHT}{s_{\HT}^2}
\newcommand{\Shaj}{S_{\haj}^2}
\newcommand{\shaj}{s_{\haj}^2}
\newcommand{\bhaj}{\bm{B}_{\haj}}

\newtheorem{Lemma}{Lemma}[section]
\newtheorem{theorem}{Theorem}[section]

\newtheorem{corollary}{Corollary}[section]
\newtheorem{assumption}{Assumption}
\theoremstyle{definition}

\theoremstyle{remark}

\title{Design-based Causal Inference for Incomplete Block Designs}
\author[$1$]{Taehyeon Koo\thanks{\emph{Email}:\;\texttt{tk3077@cumc.columbia.edu}. This work was supported by the National Science Foundation under Grant No. SES 23169083. Any opinion, findings, and conclusions or recommendations expressed in this material are those of the authors and do not necessarily reflect the views of the National Science Foundation.}}
\author[$2$]{Nicole E. Pashley}

\affil[$1$]{Department of Epidemiology, Columbia University Mailman School of Public Health}
\affil[$2$]{Department of Statistics, Rutgers University}

\begin{document}

\maketitle

\begin{abstract}
Researchers often turn to block randomization to increase the precision of their inference or due to practical considerations, such as in multisite trials.
However, if the number of treatments under consideration is large it might not be feasible or practical to assign all treatments within each block.
We develop novel inference results under the finite-population design-based framework for natural alternatives to the complete block design that do not require reducing the number of treatment arms, the incomplete block design (IBD) and the balanced incomplete block design.
This includes deriving the properties of two design-based estimators, developing a finite-population central limit theorem, and proposing conservative variance estimators.
Comparisons of the design-based estimators are made to linear model-based estimators.
Simulations and a data illustration further demonstrate performance of IBD estimators.
This work highlights IBDs as practical and currently underutilized designs.
\end{abstract}

\section{Introduction}

Increasingly social science researchers are interested in comparing multiple treatments.
These researchers often also use block designs, i.e., blocking, either to increase precision of estimators or due to practicalities of implementation.
A common example of blocking is in multisite experiments, where random assignment of treatments is done separately within each site, which could be, e.g., schools, villages, or hospitals.
Another common form of blocking occurs in crossover designs, where multiple measurements are taken upon the same unit under different treatment conditions and so the unit can be viewed as a block.
Blocks can also be formed by grouping units by covariate values.
See \cite{pashley2020insights} for more discussion of types of blocks.

Complete block designs (CBDs) randomly assign all treatments to some units within each block.
However, assigning all treatments within all blocks may not be practical or feasible.
Consider an educational study where interventions are applied at the classroom level and schools are treated as blocks.
If there are many interventions (e.g., if the interventions are combinations of four 2-level factors, there would be $2^4 = 16$ treatments), there may not be enough eligible classrooms to assign all treatments within each school.
Further, schools may object to having to implement a large number of different treatments and may be more willing to participate if there are fewer treatments to implement.
\cite{graham2001consideration} give another example in the context of policy-capturing research, where individuals in the study are used as blocks and their responses to different scenarios shown in random order are recorded, similar to a conjoint experiment. 
Those authors argue that designs with many treatments that attempt to have individuals view all treatment scenarios can lead to stress and fatigue due to their length.

When a CBD is not desirable, an alternative that still allows researchers to estimate the effects of interest is an incomplete block design (IBD) \citep{yates1936incomplete}.
In an IBD, only a subset of treatments is assigned within each block.
\cite{graham2001consideration} suggest using IBDs and note the advantages in terms of implementation for participants and researchers, such as minimizing time and effort required of participants.
Despite how useful these designs are, they have not been explored within the design-based causal inference literature and appear underutilized in social science experiments.
This paper derives design-based finite-population inference procedures to use with these designs.
After establishing results for the IBD, we focus on a special class of IBDs, called balanced incomplete block designs (BIBDs), which are described in Section~\ref{subsec:bibd_design}, to further illustrate the results and compare to linear-based estimators.

IBDs were initially developed and explored in the 1940s and 1950s as a way to get around the rigid size constraints of a classical CBD \citep[see, e.g.,][]{bosenair_1939, fisher1953statistical, kempthorne_1956, cochran1957experimental}.
These works include detailing designs, exploring their analysis, and deriving efficiency results from the model-based perspective, which relies on linear models and assumptions such as homoskedasticity of errors.
Model-based work on these designs has continued and they can easily be found in standard and modern experimental design textbooks \citep[e.g.,][]{dean2017design, oehlert2010first, wu2021experiments}.

Despite the long history of these designs, to our knowledge, there is no prior work on IBDs within the design-based causal inference literature, but there is a large literature of related work on the analysis of experimental designs within this framework.
\cite{fogarty2018mitigating}, \cite{imai2008variance}, \cite{imai2008misunderstandings},  \cite{liu2020regression}, \cite{pashley2020insights}, \cite{pashley2020block}, and \cite{wu2021design} all discuss the analysis of complete block and matched-pairs designs.
\cite{blackwell2020noncompliance}, \cite{branson2016improving},  \cite{dasgupta2015causal}, \cite{li2020rerandomization}, \cite{lu2016covariate}, \cite{pashley2023causal},  and \cite{zhao2022regression} discuss the analysis of experiments with factorial structure.
Relatedly, the following papers explore conjoint designs: \cite{abramson2022we}, \cite{bansak2018number}, \cite{egami2018causal}, \cite{goplerud2025estimating}, and \cite{hainmueller2014causal}.
Split-plot designs are explored by \cite{zhao2022reconciling} and \cite{zhao2018randomization}.
Latin square designs are discussed in \cite{ding_latin}.
These references are not exhaustive, but demonstrate the large related literature on experimental designs within the potential outcomes framework and the striking lack of work on IBDs.
One other related area is network meta-analysis where experimental data from different sources is combined, and not all experiments contain all treatments.
See \cite{schnitzer2016causal} for a causal treatment of network meta-analysis.

The advantage of design-based inference over standard ANOVA and linear regression-based methods is that there is no model assumed on the outcomes, including no need for assumptions of, e.g., Gaussian errors, homoskedasticity, and linearity.
Underlying standard models is typically an assumption of effect homogeneity, possibly after taking into account covariates.
Although this homogeneity may be appropriate in the industrial and agricultural settings in which IBDs were originally developed, it is generally less appropriate in experiments in fields such as social sciences, with human subjects and their inherent complexity and heterogeneity, as argued by \cite{dasgupta2015causal}.
Additionally, the use of potential outcomes makes the definition of estimands precise in terms of what exactly is being compared and on what population, which can be more difficult to untangle when using model-based parameters involving many interactions.
Further, by focusing on the finite population of the experimental units as our target of inference, there are no assumptions on the random sampling of units, which might be violated in social science experiments that often use convenience samples.
As a final advantage of the potential outcome design-based framework, it builds a base to develop methodology to handle complex issues that arise in experiments with humans, such as noncompliance, in a natural way \citep{dasgupta2015causal}.
Sections~\ref{subsec:fixed-effect} and \ref{subsec:fixed and unadj} discuss a linear model-based approach as a comparison.

\section{Setup and design}\label{sec:setup}

\subsection{Potential outcomes and estimands}
We have $N$ units and each unit belongs (non-stochastically) to one of $K$ blocks.
Let $b_i = k$ indicate that unit $i \in \{1,\dots,N\}$ belongs to block $k \in \{1,\dots, K\}$ with $n_k$ units in block $k$.
Suppose we have $T \geq 3$ treatments which we label $1,\dots,T$.
Assuming the Stable Unit Treatment Value Assumption \citep[SUTVA,][]{rubin_1980}, each unit $i$ has $T$ potential outcomes, corresponding to the $T$ treatments they may receive, which we can collect into a vector: $\bm{Y}_i = (Y_i(1),\dots, Y_i(T))\tr$.
We take the experimental population (and thus the potential outcomes) as fixed and consider the random assignment of treatments, described in the next section, as the only source of randomness.

The average potential outcome in block $k$ under treatment $z$ is $\ybark(z) = {n_k}^{-1}\sum_{i:b_i=k}Y_i(z)$ and the weighted average of potential outcomes under treatment $z$ across the $K$ blocks, using weight vector $\bm{w} = (w_1, \dots, w_K)\tr$ with $\sum_{k=1}^Kw_k=1$, is $\ybarzw = \sum_{k=1}^K w_k\ybark(z)$.
For example, setting $w_k = K^{-1}$ for all $k \in \{1, \dots, K\}$ gives the block-level average, while $w_k = n_k/N$ results in a unit-level average.
We define the vectors $\bm{\ybar}_k = (\ybark(1),\ldots,\ybark(T))\tr$ and $\ybarw = (\ybar(1;\bm{w}),...,\ybar(T;\bm{w}))\tr$.

We focus on estimands that are contrasts of the weighted means.
Let $\bm{g} = (g_1, \dots, g_T)\tr \in \mathbb{R}^T$ with $\sum_{z=1}^T g_z = 0$ be the contrast vector, then the estimand is $\tau_w(\bm{g}) =\sum_{z=1}^T g_z \ybarzw = \bm{g}\tr\ybarw$.

\subsection{Incomplete and balanced incomplete block designs}\label{subsec:bibd_design}
Under an IBD, each block only assigns a subset of $t$ out of the total $T$ treatments with $T > t \geq 2$.
Randomization proceeds in two stages, which distinguishes an IBD from a CBD or a completely randomized design (CRD), and increases the challenge of inference.

In the first stage, we randomly assign which blocks will receive which subset of $t$ treatments.
Let $\Rk=\{R_{k,1},\dots,R_{k,t}\}$, with $R_{k,j} \in \{1,\dots, T\}$, represent the subset of treatments that block $k$ receives.
Not all possible subsets of $t$ treatments will necessarily be used in the IBD.
We denote the set of treatment subsets used in the design as $\tset$ so that $\Rk \in \tset$.
$\tset$ is chosen a priori and taken as fixed.
A CRD for assigning subsets of treatments to blocks is used.
We begin by describing a general version of the IBD in which treatment subset $\tsetone \in \tset$ is assigned to $r_{\tsetone}$ blocks, with $r_{\tsetone}$ fixed before randomization.
The assignment mechanism of this first stage is
\[\prob(\R_1 = \tsetone_1, \dots,\R_K = \tsetone_K) = \begin{cases}
\frac{\prod_{\tsetone \in \tset}(r_{\tsetone}!)}{K!} & \text{if } \sum_{k=1}^K\mathbb{I}[\tsetone_k=\tsetone] = r_{\tsetone} \;\text{ for all }\;\tsetone \in \tset,\\
0 & \text{otherwise,}
\end{cases}
\]
where $\mathbb{I}[\cdot]$ is an indicator function which is 1 if the argument input is true and 0 otherwise.

Each treatment $z$ occurs in $L_{z} > 0$ blocks and each pair of treatments $z$ and $z'$ appears together in $l_{z, z'}$ blocks.
These quantities are determined by $\tset$ and $\{r_{\tsetone}\}$.
In a standard IBD, the number of units in each block is $n_k = t$, the number of treatments assigned to the block. 
We consider a generalization and just take $n_k$ to be divisible by $t$ for $k \in \{1,\dots, K\}$.

In the second stage of randomization, units within the block are randomly assigned to receive one of the $t$ treatments assigned to the block in the first stage.  
Let $n_k/t$ units in block $k$ receive each of the $t$ treatments.
Like a CBD, this randomization is done independently for each block and is a CRD within block.
Let $Z_i$ be the treatment assignment for unit $i$.
If unit $i$ belongs to block $k$, $Z_i \in \Rk$.
Letting $\bfR = (\R_1, \dots, \Rk)$, the assignment mechanism of this second stage is
\[\prob(Z_1=z_1, \dots, Z_N = z_N\mid\bfR) = \begin{cases}
\prod_{k=1}^K\frac{[(n_k/t)!]^t}{n_k!}  & \text{if }  \sum_{i:b_i=k}\mathbb{I}[Z_i=z] = {n_k}/{t} \text{ for all }z \in \Rk,\\
0 & \text{otherwise.}
\end{cases}
\]
We can now connect the observed outcomes in an experiment to the potential outcomes defined previously as follows: For unit $i$, their observed outcome is $Y_i^{\text{obs}} = Y_i(Z_i) = \sum_{z=1}^T\mathbb{I}[Z_i = z]Y_i(z)$.

There is often an additional third stage of randomization in which the treatments are randomly assigned to the labels $1,\dots T$.
For example, in an experiment with a control and two active treatments, it would be randomized whether control is labeled 1, 2, or 3.
This implies that the set of treatments used in $\tset$ is random.
Throughout we will condition on this randomization, which is the same as taking the treatment labels as fixed (i.e., skipping this step).
Because we show our estimators are unbiased under this conditioning, they are also unbiased unconditionally with the additional randomization.
Further, we appeal to the principles of relevance and precision in conditioning on the ancillary treatment label randomization \citep[see, e.g.,][]{pashley2021conditional}.

A BIBD is a special case of an IBD in which each subset of treatments is assigned to the same number of blocks (i.e., $r_{\tsetone} = r$ for all $\tsetone \in \tset$), with $\tset$ chosen such that $L_z = L$ for all $z$ and $l_{z, z'} =l$ for all pairs $z, z'$ \citep[see, e.g.,][]{wu2021experiments}.
A key feature of a BIBD is that $l$ is the same for all pairs of treatments.
A BIBD implies \citep[see, e.g.,][]{wu2021experiments}
$TL=tK$ and $ l=L(t-1)/(T-1).$ 
These conditions are necessary for a BIBD to exist; in particular it is necessary that $l$ is a whole number.
However, there is not in general a single, unique BIBD for a given setting, so we first choose an appropriate design given parameter vector $\bm{\lambda} = (K,T,t,L,l)$, which amounts to picking $\tset$. 

For example, Table~\ref{tab:(5,3,6,3) BIBD} gives as columns the $\tset$ for a BIBD for $T = 5$ and $t = 3$.
\begin{table}[t] 
\centering
\caption{Columns give $\tset$ for a BIBD with $(T,t,L,l) = (5,3,6,3)$ \citep[C.2,][]{oehlert2010first}.}
\resizebox{0.5\textwidth}{!}{%
\begin{tabular}{|c|c|c|c|c|c|c|c|c|c|}
1 & 1 & 3 & 2 & 1 & 2 & 1 & 1 & 2 & 1  \\
2 & 3 & 4 & 3 & 2 & 3 & 3 & 2 & 4 & 4 \\
3 & 4 & 5 & 5 & 4 & 4 & 5 & 5 & 5 & 5 
\end{tabular}%
}
\label{tab:(5,3,6,3) BIBD}
\end{table}
If $K = 10r$ and $r$ blocks are assigned to each of the prior treatment subsets, we can verify that $L= 6r$ and $l = 3r$, satisfying the BIBD criterion.
In the case that $T=3$ and $t=2$, $\tset = \{\{1,2\}, \{1,3\}, \{2,3\}\}$ is the only valid $\tset$ for a BIBD.
On the other hand, a more general IBD could use $\tset = \{\{1,2\}, \{1,3\}\}$, with an unequal number of blocks assigned to each subset of treatments.
A list of BIBDs can be found in \cite{cochran1957experimental} and \cite{fisher1953statistical}.

\section{Estimation in IBD}\label{sec:est_inf}
\subsection{The Horvitz-Thompson and H\'ajek Estimators in IBD}\label{subsec: ht_haj}
We first propose a design-based Horvitz-Thompson (HT) type estimator \citep{horvitz1952generalization} for the general contrast estimand under an IBD.
We calculate the observed average outcome for the $j$th treatment assigned in block $k$ as
$\yestk(R_{k,j}) = (n_k/t)^{-1}\sum_{i: b_i = k}\mathbb{I}[Z_i = R_{k,j}]Y_i(R_{k,j})$.
The HT estimator of $\ybarzw$ is the weighted average of these means across blocks of all units assigned to treatment $z$,
$\yht(z) =  \sum_{k=1}^K w_k\yestk(z)\I[z\in\Rk]/\prob(z\in\Rk)$ where $\prob(z\in\Rk) = L_z/K$, and $L_z$ denotes the number of blocks receiving treatment $z$ as defined earlier. 

Key criticisms of the HT estimator are that it is not invariant to location shifts in general \citep{middleton2015unbiased,su2021model} and it is unstable when certain treatments have low inclusion probabilities.
In contrast, the H\'ajek estimator \citep{Hajek1971comment}, defined as
\begin{align*}
\yhaj(z) =  \sum_{k=1}^K\frac{w_k\I[z\in\Rk]\yestk(z)}{\sum_{k=1}^Kw_k\I[z\in\Rk]} = \frac{\yht(z)}{\onehatz},\quad \text{with}\quad\onehatz = \sum_{k=1}^K\frac{w_k\I[z\in\Rk]}{\sum_{k=1}^K\prob(z\in\Rk)},
\end{align*}
normalizes the HT estimator by the sum of the weights, ensuring location invariance. 
Because $\onehatz$ is the HT estimator of the constant 1, the H\'ajek estimator is a ratio estimator for $\ybarzw = \ybarzw/1$, with the numerator and denominator estimated by $\yht(z)$ and $\onehatz$, respectively.

We write $\yeststar = (\yest_{*}(1), \dots, \yest_{*}(T))\tr$ and define estimators for $\tau_w(\bm{g})$ as $\widehat{\tau}_{*}(\bm{g}) = \bm{g}\tr\yeststar= \sum_{z=1}^Tg_z\yest_{*}(z)$ for $* = \HT$ or Haj.

\subsection{The finite-sample properties of the Horvitz-Thompson Estimator}\label{subsec: finite_ht}
The covariance matrix of $\bm{\yht}$ requires some additional notation.
We have within-block (subscripted $k$) and appropriately weighted between-block (subscripted $\HT$) variance expressions
\begin{align*}
&S^2_k(z) = \frac{1}{n_k-1}\sum_{i: b_i = k}(Y_i(z) - \ybark(z))^2,\quad S^2_{\HT}(z) = \frac{1}{K-1}\sum_{k=1}^K\left(Kw_k\ybark(z) -  \ybarzw\right)^2,\\
&S^2_k(\tau(z, z')) = \frac{1}{n_k-1}\sum_{i: b_i = k}\left[Y_i(z) - Y_i(z') -\left( \ybark(z) - \ybark(z')\right)\right]^2,\\
&S^2_{\HT}(\tau(z,z')) = \frac{1}{K-1}\sum_{k=1}^K\left[Kw_k\left(\ybark(z) - \ybark(z')\right) -\left(  \ybarzw - \ybar(z';\bm{w})\right)\right]^2.
\end{align*}
Also define $p_z = \prob(z\in\Rk)=L_z/K$ and $q_{z,z'} = \prob(z,z'\in\Rk)=l_{z,z'}/K$. 
We then define a between-block covariance matrix and a within-block covariance matrix:
$\bht = \left(\bht(z,z')\right)_{1\leq z,z' \leq T}$ is a matrix of between-block variability with
\begin{align*}
&\bht(z,z')=\frac{1}{2}\left[\left(\frac{q_{z,z'}}{p_z p_{z'}}-1\right)\left(S^2_{\HT}(z)+S^2_{\HT}(z')-S^2_{\HT}(\tau(z,z'))\right)\right],
\end{align*}
and $\bmw = \left(\bmw(z,z')\right)_{1\leq z,z' \leq T}$ is a matrix of within-block variability with
\begin{align*}
&\bmw(z,z) = \frac{1}{p_z}\left(\frac{1}{K}\sum_{k=1}^K\frac{(Kw_k)^2S_k^2(z)}{n_k/(t-1)}\right),\\
 &\bmw(z,z')=-\frac{q_{z,z'}}{2p_z p_{z'}}\left[\frac{1}{K}\sum_{k=1}^K(Kw_k)^2\left(\frac{S_k^2(z)}{n_k}+\frac{S_k^2(z')}{n_k}-\frac{S_k^2(\tau(z,z'))}{n_k}\right)\right].
\end{align*}

\begin{theorem}\label{theorem:var_ht}
The expectation and the covariance matrix of $\bm{\yht}$ are $\E[\bm{\yht}] = \ybarw$ and $ \bm{S}_{\HT} = \Cov(\bm{\yht}) = K^{-1}\left(\bht+\bmw\right).$ Consequently, $\E[\tauhtg]=\tauwg$ and $\Var(\tauhtg) = K^{-1}\bm{g}\tr\left(\bht+\bmw\right)\bm{g}$.
\end{theorem}
The proof can be found in Supplementary Material~\ref{append:proof of var_ht}.

This result is similar to the breakdown of covariance in a split-plot design into split-plot and whole-plot components \citep{zhao2022reconciling}.
The variance of the HT estimator increases as both within-block variability and between-block variability increase, with the relative importance of these two components determined by design components, such as the number of units assigned to each treatment.
This result should align with our intuition, given the following standard variance decomposition conditioning on $\bfR$: 
\begin{align*}
   \Var(\tauhtg) = \Var[\E(\tauhtg\mid\bfR)] + \E[\Var(\tauhtg\mid\bfR)].
\end{align*}
In this decomposition, the first term will correspond to variability between-block averages and the second term will represent average variability within blocks.
The tradeoff of within- and between-block variability is exemplified through discussion of the special case of a pairwise treatment comparison under a BIBD in Section~\ref{sec:bibds}.

Because of the difficulty of the random denominator in the H\'ajek estimator, we will rely on asymptotic results in the following section to understand its properties.

\section{Inference in IBD}\label{sec:inf_ibd}
\subsection{Asymptotic normality of the Horvitz-Thompson and H\'ajek estimators}\label{subsec:asymp normal}
We derive the asymptotic behavior of the HT and H\'ajek estimators.
Proofs can be found in Supplementary Material~\ref{append:proofs}.
Commonly, notation for asymptotic arguments uses indexing for the population size, which here would be based on $K$, the number of blocks.
We suppressed this for simplicity, but all estimators and estimands depend on the sample and thus $K$.
The work presented here builds on results for split-plot designs in \cite{zhao2022reconciling}, which can be shown to be a special case of our general IBD definition.

We define between-block variance expressions for the H\'ajek estimator as
\begin{align*}
&S^2_{\haj}(z) = \frac{1}{K-1}\sum_{k=1}^KK^2w_k^2\left(\ybark(z) -  \ybarzw\right)^2,\\
&S^2_{\haj}(\tau(z,z')) = \frac{1}{K-1}\sum_{k=1}^KK^2w_k^2\left[\ybark(z) - \ybark(z') -\left(  \ybarzw - \ybar(z';\bm{w})\right)\right]^2,
\end{align*}
and define the between-block covariance matrix for the H\'ajek estimator as $\bhaj$ with elements
\begin{align*}
&\bhaj(z,z')=\frac{1}{2}\left[\left(\frac{q_{z,z'}}{p_z p_{z'}}-1\right)\left(S^2_{\haj}(z)+S^2_{\haj}(z')-S^2_{\haj}(\tau(z,z'))\right)\right].
\end{align*}

We introduce the following conditions for the asymptotic normality of the estimators.
\begin{assumption}\label{assump:clt}
For all $z,z'\in \{1,...,T\}$, as $K\to\infty$:
\begin{enumerate}[label=\upshape(\roman*)]
\item \label{itm:T and W fixed}$T$ and $|\tset|$ are fixed,
\item \label{itm:kw conv} $K^{-1}\sum_{k=1}^K(Kw_k)^2$ has a finite limit,
\item \label{itm:p conv}  $p_z\rightarrow p_z^{\infty}\in (0,1)$ and $q_{z,z'}\rightarrow q_{z,z'}^{\infty}\in [0,1)$,
\item \label{itm:bb conv}$\ybarw$ has a finite limit; $\SHT(z)$,  $\SHT(\tau(z,z'))$, $\Shaj(z)$, and $\Shaj(\tau(z,z'))$ have finite limits,
\item \label{itm:wb conv}$K^{-1}\sum_{k=1}^Kn_k^{-1}(Kw_k)^2S_k^2(z)$ and $K^{-1}\sum_{k=1}^Kn_k^{-1}(Kw_k)^2S_k^2(\tau(z,z'))$ have finite limits,
\item \label{itm:max}$\max_{1\leq k\leq K}K^{-1}(Kw_k\ybark(z)-\ybarzw)^2\rightarrow 0$,
\item \label{itm:fourth moment}$K^2\sum_{k=1}^Kw_k^4n_k^{-1}\sum_{i:b_i=k}Y_i(z)^4 \rightarrow 0$.\end{enumerate}
\end{assumption}

These assumptions are effectively generalizations of those given in \cite{zhao2022reconciling} for split-plots.
Assumption~\ref{assump:clt}\ref{itm:T and W fixed} specifies that although the number of blocks increases, the number of treatments and the collection of treatment subsets used in the experiment remain fixed.
Assumption~\ref{assump:clt}\ref{itm:kw conv} safeguards against situations where, in large samples, a few blocks could receive excessively large weights.
Assumption~\ref{assump:clt}\ref{itm:p conv} requires that the probability of each block being assigned a particular treatment or pair of treatments converges to a value strictly less than one, with the former also being required to be strictly greater than zero.
Taken together with Assumption~\ref{assump:clt}\ref{itm:p conv}, Assumptions~\ref{assump:clt}\ref{itm:bb conv} and \ref{itm:wb conv} guarantee that the covariance terms of the HT and H\'ajek estimators ($\bht$, $\bhaj$, and $\bm{W}$) converge to finite limits as $K \to \infty$, ensuring their asymptotic covariances are well-defined.
Assumption~\ref{assump:clt}\ref{itm:max} serves as a regularity condition that controls the maximum deviation of $K w_k \ybark(z)$ from its average uniformly over $k$, analogous to the regularity condition in Theorem 5 of \cite{li2017general}.
Finally, Assumption~\ref{assump:clt}\ref{itm:fourth moment} imposes a bounded fourth-moment condition to support the asymptotic properties of the HT and H\'ajek estimators.

The conditions in Assumption~\ref{assump:clt} are formulated under an asymptotic regime where the number of blocks $K$ tends to infinity. 
This framework allows flexibility in the behavior of the block sizes $\{n_k\}_{k=1}^K$, which may either remain bounded or diverge, provided that all conditions in Assumption~\ref{assump:clt} are satisfied.
 However, this regime does not accommodate settings in which $K$ is fixed and the block sizes increase \citep[e.g.,][]{liu2014large}, due to the loss of asymptotic normality, as in \cite{zhao2022reconciling}.

Now we present the main theorem for the inference based on asymptotic normality.
\begin{theorem}\label{thm:clt}
Let $\Sigma_{*}^{\infty} = \lim_{K\rightarrow\infty}\Sigma_{*}$ where $\Sigma_{*} = \bm{B}_{*}+\bm{W}$ for $* =  \HT$ or $\haj$. Under Assumption~\ref{assump:clt},
$\sqrt{K}(\bm{\yeststar} - \ybarw)\overset{d}{\rightarrow}\mathcal{N}(0,\bm{\Sigma}_{*}^{\infty})$
and $\sqrt{K}(\widehat{\tau}_{*}(\bm{g})-\tauwg)\overset{d}{\rightarrow}\mathcal{N}(0,\bm{g}\tr\Sigma_{*}^{\infty}\bm{g})$ as $K\to\infty$.
\end{theorem}

Theorem~\ref{thm:clt} establishes the consistency of the HT and H\'ajek estimators and provides a finite-population central limit theorem \citep{li2017general,zhao2022reconciling} for both estimators. Furthermore, it ensures that $\bm{\Sigma}_{\haj}$ serves as a finite-sample approximation of the asymptotic covariance of $\sqrt{K}(\bm{\yhaj}- \ybarw)$. 

With expressions for the asymptotic variances of the HT and H\'ajek estimators established, a natural question arises: which estimator exhibits better asymptotic performance, which we take as a proxy for large-sample performance? When estimating the block-level effect, where $w_k=1/K$, the asymptotic precisions of the HT and H\'ajek estimators are identical. However, for the unit-level effect, their precisions may differ.

For intuition and clarity, we focus on the diagonal entries of $\bm{\Sigma}_{\HT}$ and $\bm{\Sigma}_{\haj}$, which correspond to the asymptotic variances of $\sqrt{K}\yht(z)$ and $\sqrt{K}\yhaj(z)$ for $z=1,...,T$, respectively.
We find analogous results to Corollary 1 of \cite{zhao2022reconciling}, which we reproduce below in our notation and setting.
\begin{corollary}\label{cor:var comp} Under Assumption~\ref{assump:clt}, we have
\begin{align*}
&\bm{\Sigma}_{\HT}^{\infty}(z,z)-\bm{\Sigma}_{\haj}^{\infty}(z,z)=\left((p_z^{\infty})^{-1}-1\right)\lim_{K\rightarrow\infty}\left[S^2_{\HT}(z)-S^2_{\haj}(z)\right].
\end{align*}
In particular,
\begin{enumerate}[label=\upshape(\roman*)]
\item $S^2_{\HT}(z)-S^2_{\haj}(z)=0$ if $w_k = 1/K$ for all $k$,
\item $S^2_{\HT}(z)-S^2_{\haj}(z)\geq0$ if $\ybark(z)$ is constant over all $k$,
\item $S^2_{\HT}(z)-S^2_{\haj}(z)\leq0$ if $Kw_k\ybark(z)$ is constant over all $k$.
\end{enumerate}
\end{corollary}
When $w_k=1/K$ for all $k$, the HT and H\'ajek estimators are identical, so (i) follows immediately. 
On the other hand, $\yhaj(z)$ is asymptotically more efficient than $\yht(z)$ when the blocks are similar in their average potential outcomes. 
In contrast, when the weights in the estimand are inversely related to the block-specific averages, so that the summands forming $\ybarzw$ are roughly constant across blocks, $\yht(z)$ can be more efficient asymptotically. 
Although the results of Corollary~\ref{cor:var comp} are only for the variances of $\yht(z)$ and $\yhaj(z)$, our simulation results in Section~\ref{subsec: infer ht haj} demonstrate that the relative efficiency of the HT and H\'ajek estimators for $\tauwg$ exhibits similar patterns under conditions analogous to those in Corollary~\ref{cor:var comp}.

\subsection{Estimation of the sampling covariances}\label{subsec:est_cov}
In this section, we provide estimators of the variances of the HT and H\'ajek estimators.
As usual, the treatment effect heterogeneity terms $S_k^2(\tau(z,z'))$, $S_{\haj}^2(\tau(z,z'))$ and $S_{\HT}^2(\tau(z,z'))$ are not identifiable.
Thus, we will find conservative variance estimators.
We start by defining an estimator of within-block variance for each block $k$ assigned treatment $z$.
For $k$ such that $z \in \Rk$, we define a sample version of the within-block variability term as follows: For $z \in \{1,\dots,T\}$, $s_k^2(z) = (n_k/t-1)^{-1}\sum_{i:b_i=k}\I[Z_i=z](Y_i(z)-\yestk(z))^2$.
To make $s_k^2(z)$ always well defined, we can take $s_k^2(z) = 0$ if $z \notin \Rk$ and require $n_k/t\geq2$.

We next define sample versions of the between-block variability terms in the variance of the HT and H\'ajek estimators.
We define the sample estimators of $\SHT(z)$ and $\SHT(\tau(z,z'))$ for $z,z'\in\{1,\dots,T\}$ as $\sHT(z) = (L_z-1)^{-1}\sum_{k=1}^K\I[z\in\Rk](Kw_k\yestk(z)-\yht(z))^2$ and $\sHT(\tau(z,z')) = (l_{z,z'}-1)^{-1}\sum_{k=1}^K\I[z,z'\in\Rk][Kw_k(\yestk(z)-\yestk(z'))-(\yest_A(z)-\yest_A(z'))]^2$, where $ \yest_A(z) = l_{z,z'}^{-1}\sum_{k=1}^KKw_k\I[z,z'\in\Rk]\yestk(z)$.
We define the corresponding quantities for the H\'ajek estimator: For $z,z'\in\{1,\dots,T\}$, $\shaj(z) = (L_z-1)^{-1}\sum_{k=1}^K\I[z\in\Rk](Kw_k)^2(\yestk(z)-\yhaj(z))^2$, and $\shaj(\tau(z,z')) = (l_{z,z'}-1)^{-1}\sum_{k=1}^K\I[z,z'\in\Rk](Kw_k)^2[\yestk(z)-\yestk(z')-(\yest_{A'}(z)-\yest_{A'}(z'))]^2$, where $\yest_{A'}(z) = \sum_{k=1}^Kw_k\I[z,z'\in\Rk]\yestk(z)/\sum_{k=1}^Kw_k\I[z,z'\in\Rk]$. 
We remark that $s_{*}^2(z)$ and $s_{*}^2(\tau(z,z'))$ are well-defined for $L_z\geq 2$ and $l_{z,z'}\geq 2$, respectively, for $* = \HT$ or $\haj$.

With the above quantities, we define two estimators of the covariance matrices of the HT and H\'ajek estimators: For $z,z'\in\{1,...,T\}$,
\begin{align*}
&\Shatbbstar(z,z') = \frac{l_{z,z'}\I[l_{z,z'}\geq 2]}{2L_zL_{z'}}(s_{*}^2(z)+s_{*}^2(z')-s_{*}^2(\tau(z,z'))),\\
&\Shatwbstar(z,z') = \I(l_{z,z'}\geq2)\left(1-\frac{L_zL_{z'}}{Kl_{z,z'}}\right)\Shatbbstar(z,z') +\frac{\I[z=z']}{p_z}\left(\frac{1}{K^2}\sum_{k=1}^K\frac{\I[z\in\Rk](Kw_k)^2s_k^2(z)}{n_k/t}\right),
\end{align*}
and let $\Shatbbstar = (\Shatbbstar(z,z'))_{1\leq z,z'\leq T}$, and $\Shatwbstar= (\Shatwbstar(z,z'))_{1\leq z,z'\leq T}$, for $*=\HT$ or $\haj$. 

$\Shatbbstar$ serves as a scaled sample analog of $\bm{B}_*$. 
An analogous (co)variance estimator that uses only the between-block component is employed in split-plot designs \citep[e.g.,][]{zhao2022reconciling}.
On the other hand, $\Shatwbstar$ consists of the weighted sum of the sample analog of $\bm{B}_*$ and a diagonal matrix capturing within-block variability, which corresponds to the sample analog of $\bmw$. 
A limitation of $\Shatwbstar$ in small block scenarios is that to define $s_k^2(z)$ in $\Shatwbstar(z,z)$, the number of units assigned to each treatment within each block must be greater than $1$. 
$\Shatbbstar$ avoids this limitation by using only between-block variability estimates.

Now we present probability limits of the variance estimators.
\begin{theorem}\label{thm:var est}
Under Assumption~\ref{assump:clt}, and assuming $L_z\geq 2$ for $z$ with nonzero $g_{z}$ and for all $K$, and $q_{z,z'}\to 0$ as $K\to\infty$ whenever $l_{z,z'} \leq 1$ for all $K$, we have 
$
\bm{g}\tr(K\Shatbbstar-\bm{\Sigma}_{*})\bm{g}-S_{*}^2(\tauwg) \overset{p}{\rightarrow} 0
$ 
for $*=\HT$ or $\haj$ as $K \to\infty$ where $S^2_{\HT}(\tauwg) = (K-1)^{-1}\sum_{k=1}^K[(Kw_k)\bm{g}\tr\bm{\ybar}_k -\bm{g}\tr\ybarw]^2$ and $S^2_{\haj}(\tauwg) = (K-1)^{-1}\sum_{k=1}^K[Kw_k(\bm{g}\tr\bm{\ybar}_k -\bm{g}\tr\ybarw)]^2$.
If we further assume that $l_{z,z'}\geq 2$ for all pairs $z, z'$ with nonzero $g_z$ and $g_{z'}$, and that $n_k/t \geq 2$ for all $k = 1, \dots, K$, we have
$
\bm{g}\tr(K\Shatwbstar-\bm{\Sigma}_{*})\bm{g}-K^{-1}\sum_{k=1}^Kn_k^{-1}S_k^2(\tauwg) \overset{p}{\rightarrow} 0
$
for $*=\HT$ or $\haj$ as $K\to\infty$ where $S^2_{k}(\tauwg) = (n_k-1)^{-1}\sum_{i:b_i=k}[Kw_k(\bm{g}\tr\bm{Y}_i -\bm{g}\tr\bm{\ybark})]^2$.\end{theorem}

The estimation of variance puts additional constraints on the design not required for point estimation or the central limit theorem results: each treatment with nonzero weight in the contrast must be assigned to at least two blocks. 
Additionally, for the estimator using $\Shatwbstar$, at least two units within each block must be assigned to each treatment assigned to the block and each pair of treatments that both have nonzero weight in the contrast must be assigned to at two of the same blocks.
While $\Shatbbstar$ does not require $l_{z,z'} \geq 2$ for pairs $(z, z')$ with nonzero $g_z$ and $g_{z'}$, its validity rests on an assumption concerning the limiting behavior of $q_{z,z'}$: 
we assume that $q_{z,z'}$ converges to zero whenever $l_{z,z'} \leq 1$.
Under this condition, the covariance term $\bm{\Sigma}(z, z')$ is asymptotically negligible, justifying its estimation as zero.
See Supplementary Material~\ref{supp_mat:finite_bias_ht} for finite-sample biases of the HT variance estimators without assuming $q_{z,z'}\to 0$ when $l_{z,z'}\leq1$.

We remark that $S^2_{\HT}(\tauwg)$, $S^2_{\haj}(\tauwg)$, and $S^2_{k}(\tauwg)$ are nonnegative, ensuring that the variance estimators $K\Shatbbstar$ and $K\Shatwbstar$ are both asymptotically conservative for $\bm{\Sigma}_{*}$, for $* = \HT$ and $\haj$.
For $* = \HT$ or $\haj$, $\bm{\widehat{S}}_{*}^{wb}$ exhibits the same bias for $\Sigma_*$, meaning that the difference in length between their associated confidence intervals is determined by $\Sigma_*$. 
In contrast, $\bm{\widehat{S}}_{*}^{bb}$ has a bias that varies depending on the choice of $*$. 
This difference in bias will be influenced by the discrepancy between $\SHT$ and $\Shaj$, which can be assessed with the help of Corollary~\ref{cor:var comp}.
However, since the exact bias is unknown, we suggest selecting variance estimators based on beliefs about the relative magnitudes of variability: if one expects the between-block variability to be larger, an estimator whose bias depends on the within-block variability is likely to yield better precision, and vice versa.
Most often, blocks are purposely formed to make units more similar within than between, which should lead the estimator with $\Shatwbstar$ to have smaller bias; however, with structural blocks, such as in multisite trials, the blocks come pre-formed and so this should be considered carefully with subject-matter knowledge \citep[see][for related discussion]{pashley2020block}.

Combining Theorem~\ref{thm:clt} and \ref{thm:var est}, we can construct two level $1-\alpha$ confidence intervals (CIs) for $\tauwg$ with $* = \HT$ or $\haj$ as
\begin{align}
\text{CI}^{bb}_{*}:&\left[\bm{g}\tr\bm{\yeststar} - z_{\alpha/2}\sqrt{\bm{g}\tr\Shatbbstar\bm{g}},\ \bm{g}\tr\bm{\yeststar} + z_{\alpha/2}\sqrt{\bm{g}\tr\Shatbbstar\bm{g}}\right],\label{eq:CI star bb}\\
\text{CI}^{wb}_{*}:&\left[\bm{g}\tr\bm{\yeststar} - z_{\alpha/2}\sqrt{\bm{g}\tr\bm{\widehat{S}}_{*}^{wb}\bm{g}},\ \bm{g}\tr\bm{\yeststar} + z_{\alpha/2}\sqrt{\bm{g}\tr\bm{\widehat{S}}_{*}^{wb}\bm{g}}\right],\label{eq:CI star wb}
\end{align}
which are asymptotically conservative (i.e., will have asymptotic coverages at least $1-\alpha$).

\section{BIBD: Illustration and comparisons}\label{sec:bibds}

\subsection{Design-based estimator} \label{subsec:unadj_est}
In this section, we focus on the special case of the BIBD and the pairwise difference estimand with equal block weights ($w_k = 1/K$).
This special case helps illustrate the prior results to build understanding and intuition.
This case also simplifies comparison to a linear model-based estimator, as discussed in the next section. 
The pairwise difference estimand is
$\tau(z_1, z_2) = \ybar(z_1) - \ybar(z_2),$
with $z_1, z_2 \in \{1,\dots,T\}$ and $\ybar(z)=K^{-1}\sum_{k=1}^K\ybark(z).$
When discussing this pairwise comparison, we will use the shorthand $\z = \{z_1, z_2\}$ to refer to the two treatments under comparison, with $\z^c = \{1,\dots, T\} \setminus \z$ representing the other treatments not under consideration in this estimand.
Because the HT and H\'ajek estimators are equivalent when $w_k = 1/K$ for all $k$, we omit the subscripts $\HT$ and $\haj$ from the point estimator and denote between-block variance expressions and their sample estimators using the subscript $bb$ (e.g., $s_{bb}^2(z)$ instead of $s_{*}^2(z)$).

We first elucidate simplifications for the design-based estimator presented in Section~\ref{sec:est_inf} in this simpler setting.
It is immediate that for estimating a pairwise difference $\tau(z_1, z_2)$, the estimator $\tauestbibd(z_1, z_2) = \yest(z_1) - \yest(z_2)$ is unbiased.
The other results extend directly as well.

\begin{corollary}\label{theorem:var_bibd}
Through a direct simplification of Theorem~\ref{theorem:var_ht}, the variance of $\tauestbibd(z_1, z_2)$ is
\begin{align*}
\Var(\tauestbibd(z_1,z_2))&=\frac{T-t}{t(T-1)}\left(\frac{S_{bb}^2(z_1)}{K/T}+\frac{S_{bb}^2(z_2)}{K/T}-\frac{S_{bb}^2(\tau(z_1,z_2))}{K}\right)\\
&+\frac{T(t-1)}{t(T-1)}\left[\frac{1}{K^2}\sum_{k=1}^K\left(\frac{S_k^2(z_1)}{n_k/T}+\frac{S_k^2(z_2)}{n_k/T}-\frac{S_k^2(\tau(z_1,z_2))}{n_k}\right)\right].
\end{align*}
\end{corollary}

The variance is a weighted average of variation between blocks (first term) and variation within blocks (second term), with the relative importance of these two components determined by $T-t$ and $T(t-1)$.
For $t \geq 2$, $T-t <T(t-1)$, implying that within-block variability plays a relatively larger role in general.
The between-block and within-block variances have analogous components depending on the variability of potential outcomes and treatment effect heterogeneity, either between blocks or within blocks.
To build intuition, consider the case in which the blocks are formed by randomly assigning units into groups.
Then, in expectation, the between-block component (first term in parentheses) will equal the within-block component (second term in brackets).
In practice, we often expect units to be more similar within blocks than between, implying the within-block variability component would be smaller.
However, there may be cases where there is quite a bit of heterogeneity within blocks but the blocks all have similar distributions of potential outcomes, implying the between-block variability component would be smaller.

\subsection{Adjusted estimator}\label{subsec:fixed-effect}
Our focus has been on design-based estimators.
However, a standard method for analyzing BIBDs is to use an ANOVA linear model with an additive treatment effect and block term, i.e., no interaction between treatments and blocks \citep[see, e.g.,][]{wu2021experiments}.
Although we do not assume such a model holds, we can explore the resulting estimator under the design-based framework.
Standard model-based analyses such as these typically rely on orthogonality, balanced designs, and homoskedasticity.
And, in fact, deriving the form of the linear estimator for a general IBD so that we may proceed with interpreting its properties from the design-based perspective is challenging \citep[][p. 357]{dean2017design}.
Thus, it is most sensible to focus on this estimator only in the BIBD setting where it has a known form, to illustrate the idea of the comparison.
Happily, BIBDs are a common form of IBDs, making this comparison of relevance.
For simplicity, we again focus the estimand ${\tau}(z_1,z_2)$.
We define the linear model adjusted estimator as $
\tauestfe(z_1,z_2)= t(\yestadj(z_1)-\yestadj(z_2))/(lT)$ where $\yestadj(z) = \sum_{k=1}^K\yestkadj(z)$ and $\yestkadj(z) = \I[z\in \Rk][\yestk(z)-t^{-1}\sum_{m=1}^t\yestk(R_{k,m})]$ for $z = z_1$ or $z_2$.
The $\yestkadj(z)$ are adjusted observed average outcomes,  
with an adjustment for the block effect by subtracting the average of the observed outcomes within each block. 
This estimator is equivalent to a weighted least squares estimator with weights proportional to the inverse of the block sizes. 
This weighting makes it unbiased for the block-level estimand and thus directly comparable to the design-based estimator.
If all blocks are the same size ($n_k = n$ for all $k$), then the estimator is the same as the classical least squares estimator from the standard additive linear model. 
Proof of this equivalence is provided in Supplementary Material~\ref{append:equiv_adj}.

\begin{theorem}\label{theorem:bias_fe}
Under a BIBD, $\tauestfe(z_1,z_2)$ is unbiased: $\E[\tauestfe(z_1,z_2)]={\tau}(z_1,z_2).$
\end{theorem}
For this particular form of the estimator, the balance criterion of a BIBD plays a role in unbiasedness that was not needed with our design-based estimator, ensuring that the adjustments cancel out in expectation.

To understand the variance of the adjusted estimator, we recall that the variance can be broken down through conditioning into a term for variability between block averages and a term for the average of within-block variances.
Unlike the design-based setting, variability between blocks accounts for the adjusted average outcomes, while variability within blocks depends on the assignment of all other treatments in the block.

To account for these adjustments, we need to define a few extra quantities.
Recall that $\z = \{z_1, z_2\}$.
Let $\tsettilde \in \tset$ be the subsets of the treatments sets in the design that include $\tilde{z} \in \z$ but not $\z \setminus\{\tilde{z}\}$ and $W_{\tilde{z}} = |\tsettilde|$ be the number of such treatment sets.
For example, $W_{\tilde{z}}=L-l$ if each treatment subset is only assigned to one block.
Consider one particular element of this set of subsets, $\tsetone \in \tsettilde$.
For example, if $z_1 = 1$ and $z_2 =2$, the design in Table~\ref{tab:(5,3,6,3) BIBD} has $W_{z_1}= 3$ and one element $\tsetone \in  \tset_{z_1}$ is $\tsetone = \{1,3,4\}$ because this treatment set contains $z_1 = 1$ but not $z_2 = 2$.

To understand the impact of adjustment, condition on $\Rk = \tsetone \in \tsettilde$ and estimate the adjusted mean $\tauestk(\tilde{z}, \tsetone) =t \yestadj(\tilde{z})/(t-1)$.
The conditional variance of this estimator is
\begin{align*}
&\Var(\tauestk(\tilde{z}, \tsetone)\mid\Rk = \tsetone) = \frac{S^2_k(\tilde{z})}{n_k/t} + \frac{1}{(t-1)^2}\sum_{z \in \tsetone\setminus\{\tilde{z}\}} \frac{S^2_k(z)}{n_k/t} - \frac{S^2_k(\tau(\tilde{z}, \tsetone)) }{n_k},\quad
\text{where}\\
&S^2_k(\tau(\tilde{z}, \tsetone)) =\frac{1}{n_k-1}\sum_{i:b_i=k}\left[Y_i(\tilde{z})-\frac{1}{t-1}\sum_{z \in \tsetone\setminus\{\tilde{z}\}}Y_i(z)-\left(\ybar(\tilde{z})-\frac{1}{t-1}\sum_{z \in \tsetone\setminus\{\tilde{z}\} }\ybar(z)\right)\right]^2.
\end{align*}
If we average over all treatment subsets in $\tsettilde$ we get
\begin{align*}
&\V{\tilde{z}} = \frac{1}{W_{\tilde{z}}} \sum_{\tsetone \in \tsettilde} \Var(\tauestk(\tilde{z}, \tsetone)\mid\Rk = \tsetone)=\frac{S^2_k(\tilde{z})}{n_k/t} + \frac{1}{t-1}\frac{\bar{S}^2_k(\tsettilde)}{n_k/t} -\frac{\bar{S}^2_k(\tau(\tilde{z}, \tsettilde))}{n_k},\\
&\text{where}\quad\bar{S}^2_k(\tsettilde) = \frac{1}{W_{\tilde{z}} }\sum_{\mathcal{C} \in \tset_{\tilde{z}}}\sum_{z\in \mathcal{C}\setminus\{\tilde{z}\}}\frac{1}{t-1} S_k^2(z)\quad\text{and}\quad \bar{S}^2_k(\tau(\tilde{z}, \tsettilde)) = \frac{1}{W_{\tilde{z}} }\sum_{\tsetone \in \tsettilde}S^2_k(\tau(\tilde{z}, \tsetone)) .
\end{align*}
Define a between-block variability piece corresponding to these adjusted means as
\[S_{bb}^2(\tau(\tilde{z},\tsetone)) = \frac{1}{K-1}\sum_{k=1}^K\left[\ybar(\tilde{z}) - \frac{1}{t-1}\sum_{z \in \tsetone\setminus\{\tilde{z}\}}\ybar(z) - \left(\ybar(\tilde{z}) - \frac{1}{t-1}\sum_{z \in \tsetone\setminus\{\tilde{z}\}}\ybar(z)\right)\right]^2,\]
which we can similarly average as $\bar{S}^2_{bb}(\tau(\tilde{z}, \tsettilde)) = W_{\tilde{z}}^{-1} \sum_{\tsetone \in \tset_{\tilde{z}}}S_{bb}^2(\tau(\tilde{z},\tsetone))$.
Define $A_k$ as an indicator of whether block $k$ is assigned both $z_1$ and $z_2$, with $A_k = 1$ in that case, and let $\tauestk(z_1, z_2) = \yestkadj(z_1)-\yestkadj(z_2) = \yestk(z_1) - \yestk(z_2)$.
Also define
\begin{align*}
\Vtau &= \Var(\tauestk(z_1, z_2) \mid A_k = 1) = \frac{S_k^2(z_1)}{n_k/t}+\frac{S_k^2(z_2)}{n_k/t}-\frac{S_k^2(\tau(z_1,z_2))}{n_k}.
\end{align*}

\begin{theorem}\label{thrm:adj_var}
The variance of the adjusted estimator is
    \begin{align*}
&\Var(\tauestfe(z_1,z_2))\\
&=\frac{T-t}{T(t-1)}\left[\frac{1}{K}\left\{S_{bb}^2(\tau(z_1,z_2))+(T-1)\left(\frac{\bar{S}_{bb}^2(\tau(z_1,\tset_{z_1}))}{t/(t-1)}+\frac{\bar{S}_{bb}^2(\tau(z_2,\tset_{z_2}))}{t/(t-1)}\right)\right\}\right]\\
&+\frac{T-1}{T(t-1)}\left[\frac{1}{K^2}\sum_{k=1}^K\left\{t\Vtau+(T-t)\left(\frac{\V{z_1}}{t/(t-1)}+\frac{\V{z_2}}{t/(t-1)}\right)\right\}\right].
    \end{align*}
  \end{theorem}
The variance in Theorem~\ref{thrm:adj_var} has a term related to between-block variability and a term related to within-block variability, similar to the variance of the design-based estimator.
The first line of the variance corresponds to between-block variability.
Here, unlike with the design-based estimator, variability only depends on the variability of average treatment effects between blocks, not average outcomes.
Thus, variability will be reduced in a BIBD using the adjusted estimator when blocks have similar average treatment effects, although the average outcomes may vary across the blocks.

The second line of the variance in Theorem~\ref{thrm:adj_var} corresponds to within-block variability.
The first set of terms on the second line corresponds to the variance we would get if running an experiment with design-based means for $z_1$ and $z_2$ within each block, similar to the within-block variance terms for the design-based estimator.
But now we have an additional set of terms corresponding to the average variability within each block for estimating adjusted means, $\yestkadj(\tilde{z})$, averaged over all possible assignments to the block that include $\tilde{z}$ but not $\z \setminus\{\tilde{z}\}$.
The term corresponding to the design-based means has a coefficient $t$ whereas the terms corresponding to adjusted means have a coefficient of $T-t$.
This implies that with a larger total number of treatments compared to the number of treatments assigned within each block, the variability of the adjusted means becomes more important, because there are relatively fewer blocks with both $z_1$ and $z_2$.

We construct conservative variance estimators for the adjusted estimator, following the approach in Section~\ref{subsec:est_cov}.
These estimators exhibit the same type of finite-sample bias as those in  Section~\ref{subsec:est_cov}, with one capturing between-block variability and the other capturing within-block variability; see Supplementary Material~\ref{supp_mat:var_est_fe} for details.

\subsection{Design-based vs adjusted estimator}\label{subsec:fixed and unadj}
Consider comparing the variances of the design-based and adjusted linear regression estimators.
The adjusted estimator's between-block component depends only on treatment effect heterogeneity between blocks, whereas the design-based estimator's between-block component also relies on mean outcome heterogeneity between blocks.
In particular, the difference in the between-block component between the adjusted estimator and the design-based one is
\begin{align*}
&\frac{T-t}{Kt}\Bigg[ \frac{2Tt-T-t}{T(T-1)(t-1)}S_{bb}^2(\tau(z_1,z_2))+\sum_{z\in\{z_1,z_2\}}\left\{\frac{T-1}{T}\bar{S}_{bb}^2(\tau(z,\tset_{z}))- \frac{T}{T-1}S_{bb}^2(z)\right\} \Bigg].
\end{align*}
The between-block component for the adjusted estimator will be smaller if there is no treatment effect heterogeneity between blocks but there is between-block heterogeneity of mean potential outcomes.
When this is not true, the comparison depends on the relative magnitudes of the different heterogeneities across blocks and design components $T$ and $t$.

The difference in the within-block components is
\begin{align*}
   &\frac{T-t}{K^2t}\sum_{k=1}^K\left[\frac{2Tt-T-t}{T(T-1)(t-1)}V_k(z_1,z_2)+\sum_{z\in\{z_1,z_2\}}\left\{\frac{T-1}{T}\V{z}-\frac{T(t-1)}{T-1}\left(\frac{S_k^2(z)}{n_k}\right)\right\}\right].
\end{align*}
While its interpretation is more difficult than the between-block component, largely due to the different weighting of observed outcomes in the two estimators, we see that the within-block component for the design-based estimator will be smaller if there is no within-block treatment effect heterogeneity, with $Y_i(z)-Y_i(z')$ constant for all $z,z'$; see Supplementary Material~\ref{append:unadj_fe_comp} for further details.

To build intuition, we compare the variances under finite-population versions of common models for the outcomes.
We first consider the case where potential outcomes follow an additive linear fixed-effects relationship, which is the model the adjusted estimator is based on, such that for unit $i$ with $b_i=k$
\begin{align}\label{eq: fixed-effects model}
    Y_i(z) = \alpha_z +\beta_k +\epsilon_{zi}.
\end{align}
We take $\sum_{i:b_i=k}\epsilon_{zi}=0$, and this is still a finite-population model; there is no distributional assumption on the $\epsilon$ terms.
In words, this model implies that all blocks have the same average treatment effects but average outcomes can vary across blocks and individual outcomes can vary as well.
Then, $\ybark(z) = \alpha_z +\beta_k$, and $\ybar(z) = \alpha_z +\bar{\beta}$ with $\bar{\beta} = K^{-1}\sum_{k=1}^K\beta_k$.
Thus, for $z\neq z'$,
\begin{align*}
    &S_k^2(z) = \frac{1}{n_k-1}\sum_{i:b_i=k}\epsilon_{zi}^2,\quad S_k^2(\tau(z,z')) = \frac{1}{n_k-1}\sum_{i:b_i=k}(\epsilon_{zi}-\epsilon_{z'i})^2,\\
    &S_{bb}^2(z) = \frac{1}{K-1}\sum_{k=1}^K(\beta_k-\bar{\beta})^2\equiv S_{bb}^2,\quad S_{bb}^2(\tau(z,z')) = 0.
\end{align*}

We will consider two additional assumptions.
First, further assume additive treatments, that is, $Y_i(z)-Y_i(z')$ is constant for all $z,z'$, which occurs when $\epsilon_{zi} = \epsilon_i$ for all $z$, i.e., the potential outcomes are perfectly correlated.
Then $S_k^2(z)\equiv S_k^2$ for all $z$ and 
  $S_k^2(\tau(z,z')) =0$ for all $z\neq z'$. In this case, $\Var(\tauestbibd(z_1,z_2)) >\Var(\tauestfe(z_1,z_2))$ holds when $S_{bb}^2> c(K^{-1}\sum_{k=1}^KS_k^2/n_{k})$ with $c = 1 + {t(T-1)}/{(T(t-1))}$, which is the coefficient of $\Vtau$ in $\Var(\tauestfe(z_1,z_2))$.
For the adjusted estimator to have higher precision than the design-based estimator, the between-block variability must be greater than the within-block variability by a multiplicative factor $c$. 

To understand the relative magnitudes of the between-block and within-block variability components, again consider if we grouped units into $K$ blocks at random, fixing the block sizes (i.e., fixing $n_k$'s).
Recall that if blocking were done randomly, $S^2_{bb} \approx K^{-1}\sum_{k=1}^KS_k^2$.
This implies that the adjusted estimator has lower variance when blocking separates units into homogenous blocks than random blocking by a factor of $c$.

Second, instead of additive effects, assume that potential outcomes are independent in the sense that $\sum_{i}\epsilon_{zi}\epsilon_{z'i} = 0$, so that $S_k^2(\tau(z,z')) = 2S_k^2$ for all $z\neq z'$.
Then $Var(\tauestbibd(z_1,z_2)) >\Var(\tauestfe(z_1,z_2))$ holds when $S_{bb}^2> [(T-1)/T](K^{-1}\sum_{k=1}^KS_k^2/n_{k})$.
In contrast to the previous case, the multiplicative factor $(T-1)/T$ is less than 1, implying blocking need only be correlated with outcomes for the adjusted estimator to have better precision under this model.

 We can also consider model misspecification of the adjusted estimator.
 For example, we can assume the following interactive fixed-effects model for each unit $i$ with $b_i=k$:
\begin{align}\label{eq: interactive model}
    Y_i(z) = \alpha_z +\beta_k + \gamma_z\delta_k+ \epsilon_{zi}.
\end{align}
Then $S_k^2(z)$ and $S_k^2(\tau(z,z'))$ are the same as in the additive model without extra assumptions, but
\begin{align*}
    &S_{bb}^2(z) = \frac{1}{K-1}\sum_{k=1}^K(\beta_k-\bar{\beta}+\gamma_z(\delta_k-\bar{\delta}))^2,\quad\text{and}\quad S_{bb}^2(\tau(z,z')) =\frac{ (\gamma_z-\gamma_{z'})^2}{K-1}\sum_{k=1}^K(\delta_k-\bar{\delta})^2.
\end{align*}
Because the $S_{bb}^2(\tau(z,z'))$ terms are no longer zero under this model, the between-block component of $\Var(\tauestfe(z_1,z_2))$ increases but still does not depend on $\beta_k$ .

To further understand the tradeoffs of the estimators under this interactive model, we conduct simulations in Section~\ref{sec: simulation}.
We also provide additional mathematical results for these comparisons, including the case of imperfectly correlated potential outcomes, in Supplementary Material~\ref{append:unadj_fe_comp}.

\section{Simulations}\label{sec: simulation}
\subsection{Simulation design}
We next explore the performance of the adjusted and design-based estimators and the variance estimators through simulations.
Throughout we define the estimand $\tauwg$ by taking $w_k=n_k/N$ for each $k$ and $\bm{g} = (1,-1,0,...,0)\tr$.
To generate data, we consider a BIBD corresponding to the design in Table~\ref{tab:(5,3,6,3) BIBD}, under three different settings with varying block sizes:
\begin{align*}
&\textbf{S1}: n_k = n = 15,\quad \textbf{S2}: n_k = 3\max\{2,X_k\}, \quad \textbf{S3}:n_k = 3\max\{2,X_{(k)}\}, \quad X_k\overset{i.i.d}{\sim} {\rm Poi}(5), 
\end{align*}
where $X_{(k)}$ denotes $k$-th order statistic of $\{X_k\}_{k=1}^K$.
In \textbf{S1}, every block has the same size. 
In \textbf{S2}, the block sizes are $i.i.d.$ draws from a Poisson distribution, subsequently adjusted to be divisible by 3 and greater than 6.
\textbf{S3} uses a similar random generation mechanism as in S2, but then orders the block sizes so that they increase with $k$, the block index.
Throughout the simulation studies, we vary $K$ over $\{10,20,50,100\}$.

For each set, we generate potential outcomes using model $Y_i(z) = \beta_k + \gamma_z\delta_k+ \epsilon_{zi}$, with $\beta_k = \beta \chi^2_{10}(1-{k}/{(K+1)})$,  $\gamma_z = \gamma z$, $\delta_k = \chi^2_{10}(1-{k}/{(K+1)})$, for some constants $\beta$ and $\gamma$ where $\chi^2_{d}(\alpha_0)$ denotes the $\alpha_0$-quantile of chi-square distribution with $d$ degree of freedom.
When $\gamma = 0$ this is an additive model as in Equation~\eqref{eq: fixed-effects model} and otherwise it is an interactive model as in Equation~\eqref{eq: interactive model}.
Additionally, the block average treatment effects are spread further apart as $\gamma$ increases.
Referring to those Equations \eqref{eq: fixed-effects model} and \eqref{eq: interactive model}, the variances of the estimators do not depend on $\alpha_z$, the average outcome under $z$, so we set $\alpha_z=0$.
Further, our model implies that the size of treatment effect shrinks with larger $k$ because both $\beta_k$ and $\gamma_k$ decrease as $k$ increases. 
The block averages are spread further apart as $\beta$ increases and are exactly the same if $\beta = 0$.

We generate $\epsilon_{zi}$ for all $z=1,\dots,T$ and $i$ with $b_i=k$ as
$(\epsilon_{1i}, \dots,\epsilon_{Ti})\tr
		\sim \mathcal{N}(\mathbf{0}_T,S_k^2\Sigma_{\rho})$,
where $\mathcal{N}$ denotes the multivariate normal distribution, $\mathbf{0}_T = (0, \dots, 0)\tr\in \mathbb{R}^{T}$, and $\Sigma_\rho  \in \mathbb{R}^{T\times T}$ is a matrix with 1's on the diagonal and $\rho$ on the off-diagonal entries.
We fix $S_k^2=100$, and vary $\gamma \in \{0,0.5,1\}$, $\beta \in \{0,0.1,\ldots,0.8\}$ and $\rho \in \{0,0.5,1\}$.

The simulations are for the finite population and thus the potential outcomes are generated once from the distribution above for each combination of $\gamma$, $\beta$, and $\rho$.
The random components $\epsilon_{zi}$ are generated such that the empirical averages and variances for each block match the theoretical values set for that simulation, to avoid draws far from the intended setting.
Once the potential outcomes are generated for each setting, the simulations are implemented over 10,000 random treatment assignments. 
We focus on comparing point estimators and the coverage of CIs.
 Additional comparisons, such as those between variance estimators, the lengths of CIs, and the precision of HT versus H\'ajek estimators, are provided in Supplementary Material~\ref{supp_mat:add sim}.

\subsection{Comparison of inference results from the Horvitz-Thompson and H\'ajek estimators}\label{subsec: infer ht haj}
In this section, we compare the coverage of CIs ${\rm CI}_{*}^{bb}$, defined in \eqref{eq:CI star bb} using $\Shatbbstar$, and ${\rm CI}_{*}^{wb}$, defined in \eqref{eq:CI star wb} using $\Shatwbstar$, with the HT and H\'ajek estimators.
Throughout this section, we set the nominal coverage to 95\%. 
For simplicity, we focus on a subset of settings, with additional results including comparisons of CI lengths, provided in Supplementary Material~\ref{supp_mat:add sim}.

\begin{figure}
\centering
\includegraphics[width=\textwidth]{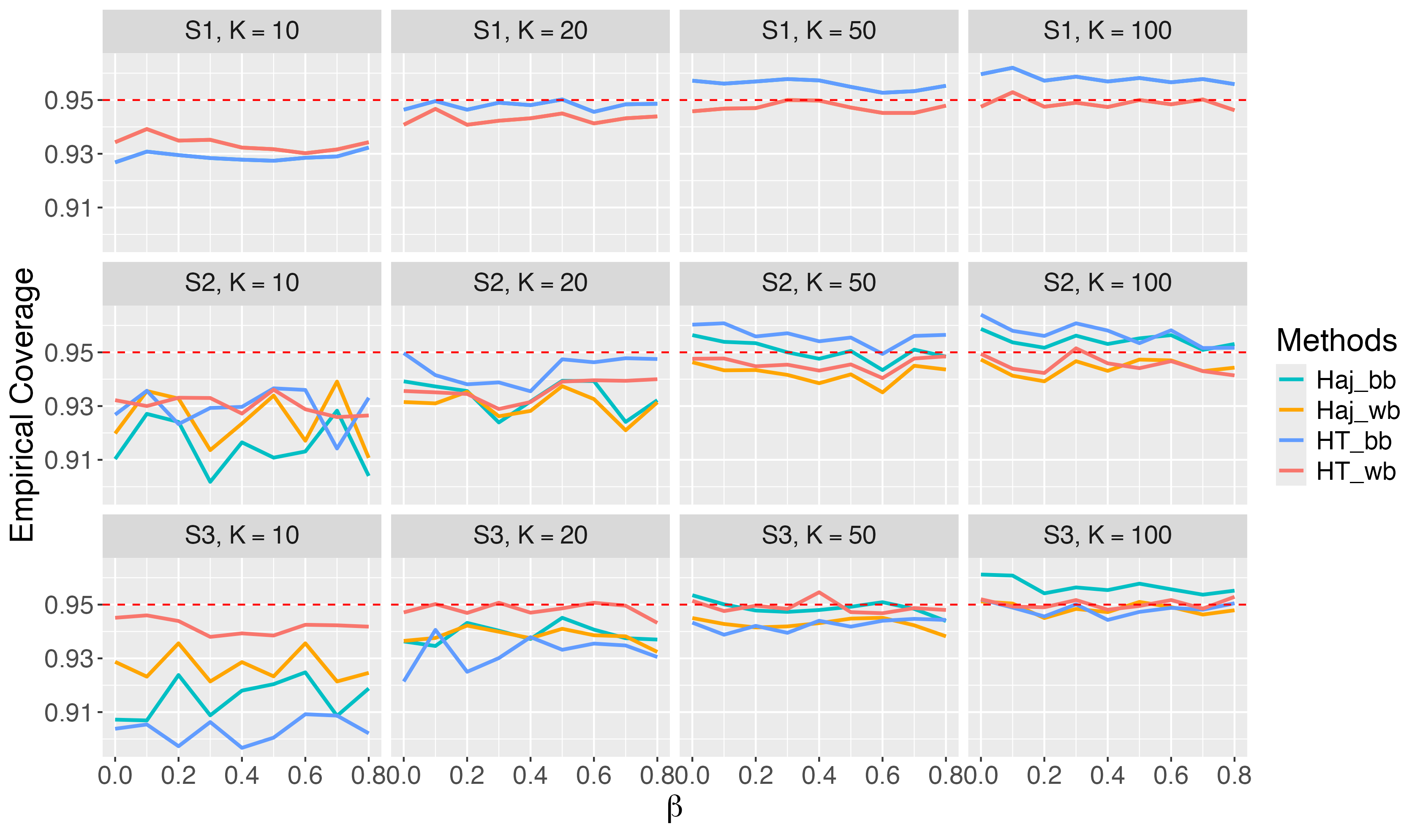}
\caption{Simulations with $\gamma=1$ and $\rho=1$.
Each line corresponds to the empirical coverage of the confidence interval in one simulation.
\texttt{HT} and \texttt{Haj} denote the Horvitz-Thompson and H\'ajek estimators respectively, and \texttt{bb} and \texttt{wb} denote corresponding variance estimators with between-block and within-block biases respectively.
The dashed horizontal red lines are at 0.95.}
\label{fig: cov_comp}
\end{figure}

We start by comparing the empirical coverage of CIs under $\gamma=1$ and $\rho=1$, i.e., larger interactions and no treatment effect heterogeneity within blocks, with results presented in Figure~\ref{fig: cov_comp}. 
Under \textbf{S1}, the HT and H\'ajek estimators are equivalent, and thus their CI coverage is identical.
However, under \textbf{S2} and \textbf{S3}, they exhibit different behavior.
Under \textbf{S2}, the H\'ajek-based CIs, tend to be more anti-conservative than those based on the HT estimator. Under \textbf{S3}, patterns depend on both the point estimator and variance method, but HT-based CIs are generally shorter than H\'ajek-based CIs—reversing the \textbf{S2} pattern; see Supplementary Material~\ref{supp_mat:len_ht_haj}.
When $K=10$, the CIs tend to undercover, but as $K$ increases, the empirical coverage approaches or exceeds 95\%, supporting the validity of our central limit theorem and variance estimation procedures in Theorem~\ref{thm:clt} and Theorem~\ref{thm:var est}.
When $\rho=1$, $S_k^2(\tauwg)=0$, leading $\Shatwbstar$ to be consistent and the empirical coverage of the CIs of ${\rm CI}_{*}^{wb}$ to align closely with the nominal level.
However, when $\gamma \neq 0$, $S_{*}^2(\tauwg) \neq 0$ in general, leading to conservative CIs—i.e., those with empirical coverage exceeding 0.95—when using the CIs ${\rm CI}_{*}^{bb}$. 
We provide additional results to compare the lengths of ${\rm CI}_{*}^{wb}$ and ${\rm CI}_{*}^{bb}$ in Supplementary Material~\ref{supp_mat:var_est_sims}.

\graphicspath{ {./Images/} }

\subsection{Comparison between BIBD estimators}\label{subsec: comp est}
\begin{figure}
\centering
\includegraphics[width=\textwidth]{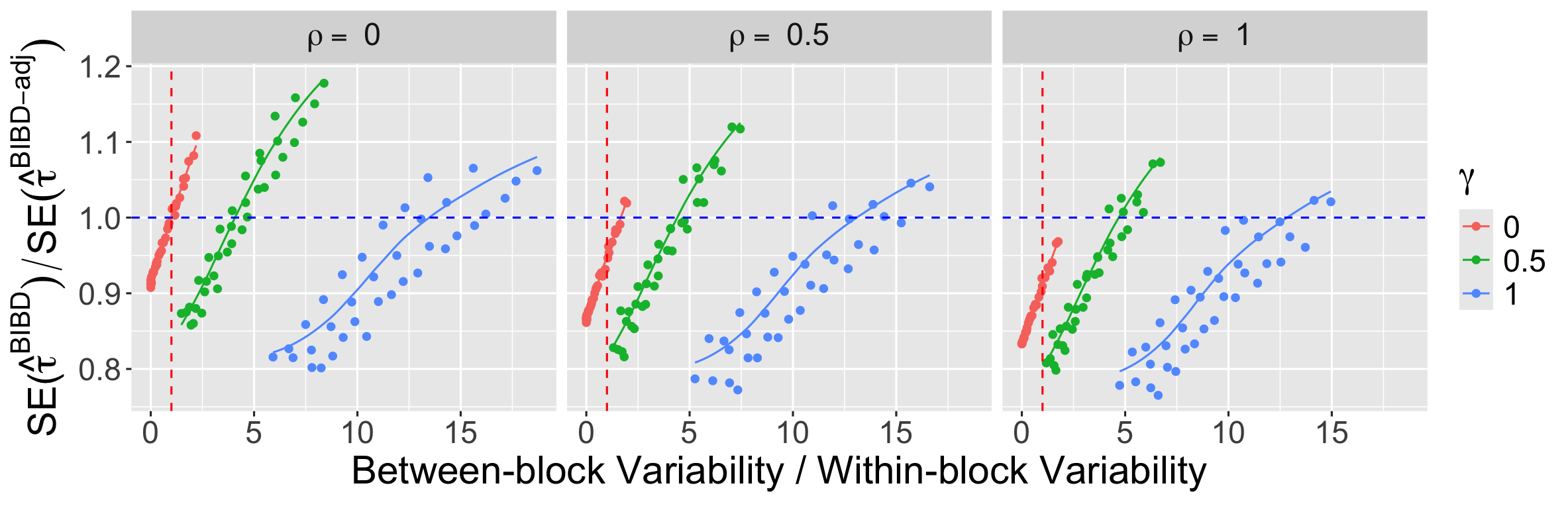}
\caption{Simulations under \textbf{S1}.
Each point corresponds to the empirical ratio of standard errors ($\SE = \sqrt{\Var}$) in one simulation.
Lines are fitted to the theoretical standard error ratios across the simulation settings.
The dashed horizontal blue and vertical red lines are at 1.}
\label{fig: se_comp_est_S1}
\end{figure}

We now compare the standard error of the design-based estimator, $\tauestbibd(z_1,z_2)$, and the adjusted estimator, $\tauestfe(z_1,z_2)$ with $z_1=1$ and $z_2=2$.
Because $\tauestfe(z_1,z_2)$ is defined under BIBD with $w_k=1/K$, we focus on \textbf{S1}, with results presented in Figure~\ref{fig: se_comp_est_S1}.

\graphicspath{ {./Images/} }

First consider the case when $\gamma=0$, recalling that this implies that the potential outcomes are generated following the additive model.
As discussed in Section~\ref{subsec:fixed and unadj}, the precision of $\tauestfe(z_1,z_2)$ depends on the correlation of potential outcomes, denoted by $\rho$.
In Figure~\ref{fig: se_comp_est_S1}, we see that under the values of $\rho$ considered, the adjusted estimator performs relatively better in terms of variability when the between-block variability is low compared to within-block variability, matching the intuition from Section~\ref{subsec:fixed and unadj}.
However, when exactly $\tauestfe(z_1,z_2)$ has an advantage compared to $\tauestbibd(z_1,z_2)$ depends on $\rho$.
With uncorrelated potential outcomes ($\rho = 0$), there simply needs to be more between-block variability relative to within-block variability than expected by random blocking (which is when the between and within variability would be equal, given by the dashed vertical line).
This aligns with the results from Section~\ref{subsec:fixed and unadj}.
However, with additive effects ($\rho=1$) or partial correlation between potential outcomes ($\rho=0.5$), we need higher between-block heterogeneity, relative to within-block heterogeneity, for $\tauestfe(z_1,z_2)$ to be beneficial compared to $\tauestbibd(z_1,z_2)$.

Second, consider when $\gamma$ is not zero, implying that the potential outcomes are generated following the interactive model with $\gamma$ representing the level of interaction between the treatment and block.
In Figure~\ref{fig: se_comp_est_S1}, we can see that $\tauestbibd(z_1,z_2)$ performs better than $\tauestfe(z_1,z_2)$ in terms of precision as the size of the interaction increases.
This aligns with the intuition that the adjusted estimator, based on an additive model, should perform poorly when the outcomes do not follow an additive model.
For example, when $\gamma=1$, even very high between-block heterogeneity does not lead to $\tauestfe(z_1,z_2)$ being advantageous compared to $\tauestbibd(z_1,z_2)$.

Overall, these simulations show the possibility of precision gains using $\tauestfe(z_1,z_2)$ compared to $\tauestbibd(z_1,z_2)$ when the additive model holds and the blocking lowers within-block variability compared to between-block variability.
However, this advantage quickly disappears when the true model is interactive or when there is a higher level of correlation between the potential outcomes.
Because the blocks do not contain all treatments, an assumption of no interaction is difficult to test empirically \cite[p.356]{dean2017design}.
This leads us to recommend the design-based estimator as potentially having relatively better precision across a larger variety of scenarios.
We also note that simply calculating estimates of the effect based on both estimators and picking the one with the smallest standard error estimates can lead to systematically anti-conservative inference.

\section{Data illustration}\label{sec:data_ill}

We revisit Tennessee's Student/Teacher Achievement Ratio (STAR) Project \citep{word1990state,star_data} to illustrate the use of IBDs and BIBDs.
This study examined the impact of three classroom types on elementary school students’ academic achievement: small classes (13-17 students per teacher), regular classes (22-25 students per teacher), and regular classes with a full-time aide.
In 1985-86, a cohort of kindergarten students from 79 schools was randomly assigned to one of these conditions and remained in their assigned type through grade 3, with the exception of some students in regular-sized classes moving to or from classes with an aide between kindergarten and first grade in addition to some student mobility between schools. 
A CBD was used, with schools serving as blocks.
Each school needed at least 57 kindergarten students to ensure sufficient enrollment for at least one small class and two regular classes.
In total, 328 kindergarten classrooms were included in the study.
This design was chosen to control for school-level differences and incentivize full school participation by ensuring each school experienced all three conditions \citep[see][]{word1990state}.

However, a BIBD with only two treatments per school could have reduced the minimum enrollment requirement and allowed more schools to participate.
In fact, 180 schools expressed interest in participating, but only 100 met the minimum number of kindergarteners required \citep{word1990state}.
In our re-analysis, we focus on the impact of first-grade classroom assignment on first-grade Stanford Achievement Test (SAT) scores. We average each student’s scores across four test components (reading, math, listening, and word study skills), and the classroom serves as the unit of analysis. Specifically, we compare small classes ($z_1$) to regular classes without an aide ($z_2$). For simplicity, we excluded students with missing data on test scores, classroom type, or school, leaving us with 75 schools with adequate data. Among these, 71 schools included all three treatment types, while 4 schools had only small and regular classes without an aide.

To create a BIBD, we (i) retained the 4 schools with only two treatments (small and regular classes without an aide), (ii) randomly assigned a pair of treatments to each of the 71 remaining schools, ensuring each of the three treatment pairs was assigned to exactly 25 schools, and (iii) randomly dropped classrooms as necessary to maintain an equal number of classes per treatment in each school. Using this subset of 75 schools, we applied the treatment effect and variance estimators from Section~\ref{sec:bibds}.
This process was repeated 500 times and averaged.
Results are shown at the top of Table~\ref{tab:STAR}.

Because schools may have been incentivized to participate by the support provided for small classes, we also recreated an (unbalanced) IBD in which each school was assigned at least one small class. 
We will refer to this as the IBD in this section.
For the IBD, only two treatment pairs were used: small and regular classes without an aide (randomly assigned to 33 schools to get a total of 37 with the four schools that only had this pairing) and small and regular classes with an aide (randomly assigned to 38 schools).
As before, we adjusted classroom numbers as needed and applied estimators from Section~\ref{sec:est_inf}, repeating this process 500 times for an average estimate.

For schools with only one classroom per treatment, within-block variance estimators, $\bm{g}\tr\bm{\widehat{S}}_{*}^{wb}\bm{g}$, are not applicable.
To illustrate these estimators, for the BIBD we subsetted to 16 big schools with at least two classrooms per treatment, randomly dropping one school to maintain equal treatment pairs.
For the IBD, we similarly assigned each of the two treatment pairs equally across the 16 big schools. 
Results for BIBD and IBD from these 500 replicates are at the bottom of Table~\ref{tab:STAR}. 
As a further comparison, the CBD subsetted to the same set of 75 or 16 schools, but without dropping schools or classrooms, yields point estimates of 11.26 and 18.33, respectively, for the average block-level treatment effects, similar to those from the BIBD and IBD.

\begin{table}[]
\centering
\begin{tabular}{c|c|c|ccc|ccc}
\multirow{2}{*}{Blocks included} &\multirow{2}{*}{Designs} &  \multirow{2}{*}{Estimator}                     & \multicolumn{3}{c|}{Block-level}                     & \multicolumn{3}{c}{Unit-level}                       \\  \cline{4-9}
 &               &  & \multicolumn{1}{c|}{$\widehat{\tau}$} & \multicolumn{1}{c|}{$\widehat{\sigma}_{bb}$} & $\widehat{\sigma}_{wb}$ & \multicolumn{1}{c|}{$\widehat{\tau}$} & \multicolumn{1}{c|}{$\widehat{\sigma}_{bb}$} & $\widehat{\sigma}_{wb}$ \\ \hline
\multirow{5}{*}{All blocks}  & \multirow{3}{*}{BIBD} & HT        & \multicolumn{1}{c|}{11.290}          & \multicolumn{1}{c|}{4.030}& N/A         & \multicolumn{1}{c|}{9.628}           & \multicolumn{1}{c|}{28.719}    & N/A    \\
                    &                   & Haj       & \multicolumn{1}{c|}{11.290}          & \multicolumn{1}{c|}{4.030}    & N/A     & \multicolumn{1}{c|}{13.061}          & \multicolumn{1}{c|}{4.107}    & N/A     \\
                    &                  & Adjusted       & \multicolumn{1}{c|}{11.097}          & \multicolumn{1}{c|}{3.422}     & N/A    & \multicolumn{1}{c|}{N/A}         & \multicolumn{1}{c|}{N/A}      & N/A  \\ \cline{2-9} 
                   &\multirow{2}{*}{IBD}  & HT        & \multicolumn{1}{c|}{12.369}          & \multicolumn{1}{c|}{3.607}     & N/A    & \multicolumn{1}{c|}{13.811}          & \multicolumn{1}{c|}{22.826}     & N/A   \\
                     &                    & Haj       & \multicolumn{1}{c|}{12.369}          & \multicolumn{1}{c|}{3.607}   & N/A      & \multicolumn{1}{c|}{13.749}          & \multicolumn{1}{c|}{3.638}        & N/A             \\ \hline
                     \multirow{5}{*}{Big blocks} & \multirow{3}{*}{BIBD} & HT        & \multicolumn{1}{c|}{18.149}           & \multicolumn{1}{c|}{8.350}                   & 8.721                   & \multicolumn{1}{c|}{18.734}           & \multicolumn{1}{c|}{22.511}                  & 22.650                  \\
                    &                   & Haj       & \multicolumn{1}{c|}{18.149}           & \multicolumn{1}{c|}{8.350}                   & 8.721                   & \multicolumn{1}{c|}{18.358}           & \multicolumn{1}{c|}{8.168}                    & 8.578                    \\ 
                    &                  & Adjusted       & \multicolumn{1}{c|}{18.481}   & \multicolumn{1}{c|}{6.109}       & 5.527         & \multicolumn{1}{c|}{NA}         & \multicolumn{1}{c|}{NA} &    NA    \\  \cline{2-9} 
       &\multirow{2}{*}{IBD}  & HT        & \multicolumn{1}{c|}{17.858}           & \multicolumn{1}{c|}{7.113}                   & 7.495                   & \multicolumn{1}{c|}{18.921}           & \multicolumn{1}{c|}{15.143}                  & 15.388                  \\
                    &                   & Haj       & \multicolumn{1}{c|}{17.858}           & \multicolumn{1}{c|}{7.113}                   & 7.495                   & \multicolumn{1}{c|}{18.095}           & \multicolumn{1}{c|}{6.962}                   & 7.383      
\end{tabular}
\caption{Estimates based on mimicking a BIBD and IBD using the data from the Tennessee STAR Project. $\hat{\sigma} = \sqrt{\hat{\sigma}^2}$ is the estimated standard error, with subscripts $bb$ and $wb$ denoting whether the variance is the within-block or between-block version for each estimator. Data from \texttt{https://doi.org/10.7910/DVN/SIWH9F}.}
\label{tab:STAR}
\end{table}


Focusing on the BIBD, the averaged estimated variability for the adjusted estimator is smaller than that for the design-based estimators, especially for the analysis with only ``big'' blocks.
Although the variance estimators are biased (conservative), this gives some indication that the adjusted estimator has lower variance.
Based on our results, we expect the adjusted estimator to have lower variance when the block means differ across blocks but there is less variability within blocks.
In this context, this would mean schools may have different average test scores but there is relatively less variability in average effects across schools than variability within schools.

In both the BIBD and IBD, the within-block variance estimators are slightly larger than the between-block variance estimators in the ``big'' block analysis.
Based on the form of the biases of these estimators, this may indicate larger within-block than between-block heterogeneity of outcomes.
Additionally, the H\'ajek estimator has consistently smaller estimated variance than the HT estimator for the individual effects.
This may indicate that there is heterogeneity of potential outcomes across blocks and that these effects are not negatively correlated with school size.

Comparing the design-based estimates in the BIBD and IBD, the IBD has a smaller estimated variance.
There are a few likely reasons for this.
First, the IBD in this case is able to use more of the sample in estimation, using three-fourths of the sample whereas the BIBD uses two-thirds of the sample.
This will not always be the case; if comparing the two regular classroom treatments, the BIBD would still use two-thirds of the sample but the IBD would only use half the sample in estimation.
Second, blocks with both treatments present in a pairwise comparison contribute differently to the variance than blocks with only one treatment present.
For both within-block and between-block variability components, there will be more terms related to the treatment effect heterogeneity as the number of blocks with two treatments increases.
In the IBD, half the blocks have both treatments whereas in the BIBD only one-third of blocks do.
Again, this changes if comparing the two regular classroom treatments, where the IBD would have no blocks with both treatments.
Thus, it is important to align both the goals of inference and considerations of increasing participation when designing a BIBD or IBD.
In fact, our proposed variance estimator with within-block bias is not feasible when comparing the two regular classrooms under the proposed IBD, because there are no schools assigned both types of regular classes.

The change in the estimates of the treatment effect in both the BIBD and IBD when only ``big'' blocks are used is also notable.
This indicates that there is heterogeneity across schools.
In particular, it appears that the treatment of smaller classrooms has a larger impact in bigger schools.
One hypothetical explanation for this could be that resources in bigger schools might, in general, be shared by more students compared to smaller schools, making small classrooms with more focused attention and resources even more effective.

\section{Discussion}\label{sec:disc}

Although we have developed a solid framework for analysis of IBDs under the design-based perspective, there are many avenues of further development.
We focus throughout on inference for the finite population using the design-based approach.
However, it is often of interest to generalize findings to a larger population.
One may consider that a model-based approach makes this task more natural.
Indeed, \cite{mccullagh2002statistical} argues that in a sensible (parametric) model, the parameters of the models, in this case the causal effect, should be the same across appropriately defined populations, with the goal of extending inference beyond the units at hand.
Fortunately, the design-based approach can be used for this task as well, with some extensions.
First, the properties of our estimators should be examined under different sampling schemes, such as random sampling of blocks or random sampling of units within blocks \citep[see, e.g.,][]{pashley2020block}.
Second, approaches for generalizability, such as weighting, should be considered \citep[see, e.g.,][]{degtiar2023review}.
Third, covariate adjustment can be incorporated to both improve efficiency and generalizability of results \citep[see, e.g.,][for an example of design-based regression adjustment]{zhao2022reconciling}.
It should also be noted that although the treatment effect is not typically considered constant under the design-based approach, the definition of estimands using potential outcomes makes the population for which inference is being made explicit and the estimands are stable under different randomizations.

\bibliographystyle{apalike}
\bibliography{inc_ref}{}

\begin{appendices}
\newpage

\setcounter{page}{1}
\begin{center}
{\bf \Large  Supplementary Material\\for\\``Design-based Causal Inference for Incomplete Block Designs''}
\end{center}
\renewcommand{\theequation}{S\arabic{equation}}

These supplementary materials are organized as follows:
Section~\ref{append:useful_exp} provides useful expressions and expansions that are used in the proofs.
Section~\ref{supp_matt:add_var_res} provides additional variance estimation results, including finite-sample bias of the variance estimators for the Horvitz-Thompson estimator and variance estimation for the adjusted estimators.
Section~\ref{append:proofs} provides proofs of the results in the main paper.
Section~\ref{append:proof_lemmas} provides proofs of lemmas used in the proofs in Section~\ref{append:proofs}.
Section~\ref{supp_mat:add sim} provides additional simulation results.

\section{Useful expressions}\label{append:useful_exp}
\subsection{Probabilities for treatment assignments and block membership}\label{append:prob_relations}
For a IBD, as defined in the main paper, we provide the following simplifications of probabilities related to assigning treatment to the blocks.

\noindent For $z \in \{1,...,T\}$, \[\prob(z \in \Rk) = \sum_{j=1}^{t}\prob(z = R_{k,j}) = \frac{L_z}{K}.\]
For $k=1,...,K$,  we denote the ``membership'' type of block $k$ based on treatments assigned to it as
\begin{align*}
A_k=\I[z_1,z_2 \in \Rk], \quad B_k=\I[z_1 \in \Rk, z_2 \notin \Rk], \quad C_k = \I[z_1 \notin \Rk, z_2 \in \Rk].
\end{align*}
Then
\begin{align*}
&\E [ A_k] = \prob(z_1,z_2 \in \Rk) = \frac{l_{z_1,z_2}}{K},\quad 
\E [B_k]  = \prob(z_1\in \Rk, z_2 \notin \Rk) = \frac{L_{z_1}-l_{z_1,z_2}}{K},\\
&\E [C_k] = \prob(z_1 \notin \Rk,z_2 \in \Rk) = \frac{L_{z_2}-l_{z_1,z_2}}{K}.
\end{align*}
This is discussed more in Supplementary Materials~\ref{supp_mat:theorem:bias_fe}.

For the rest of this section, we assume BIBD. That is, $l_{z_1,z_2} = l$, $L_{z_1} = L_{z_2} = L$ for all $z_1,z_2 \in \{1,...,T\}$. Define the proportion of blocks containing $z_1$ or $z_2$ (not both) that also contain another treatment $z$ as
\[p(z) =\mathbb{P}(z\in \Rk|z_1 \in R_k, z_2 \notin R_k)=\mathbb{P}(z\in \Rk|z_2 \in R_k, z_1 \notin R_k)=\mathbb{P}(z\in \Rk|B_k=1)=\mathbb{P}(z\in \Rk|C_k=1). \]
Define the a pair of treatments probability that a pair of treatments $z$ and $z'$ are assigned to a block with either $z_1$ or $z_2$ (but not both), for every $z,z' \in \z^c\ z\neq z'$
\[p_{z_1}(z,z') = \mathbb{P}(z,z'\in \Rk|z_1 \in R_k, z_2 \notin R_k),\quad p_{z_2}(z,z') =  \mathbb{P}(z,z'\in \Rk|z_2 \in R_k, z_1 \notin R_k)= \mathbb{P}(z,z'\in \Rk|B_k=1),\quad p_{z_2}(z,z') =  \mathbb{P}(z,z'\in \Rk|C_k=1)  \quad \text{for all } z,z' \in \z^c, \ z\neq z'.\]
For example, referring to the design in Table~\ref{tab:(5,3,6,3) BIBD},  $p(3) = 2/3$ and $p_{1}(3,4) = 1/3$.
If $T=3$, there is only one treatment in each subset of $\tset$ that does not belong to $\z$, so we define $p_{z_1}(z,z')=p_{z_2}(z,z')=0$.

Further, let
\begin{align*}
    p_{z_1z_1}(z) = \mathbb{P}\left(z \in \Rk,\R_j| B_k=B_j=1\right), \quad p_{z_1z_1}(z,z')=\mathbb{P}\left(z \in \Rk, z'\in \R_j| B_k=B_j=1\right).
\end{align*}

Let $n(z)$ be the number of blocks containing $\tilde{z}$ and $z$, but not $\z \setminus\{\tilde{z}\}$, and $n_{\tilde{z}}(z,z')$ be the number of blocks containing $\tilde{z}$, $z$, and $z'$, but not $\z \setminus\{\tilde{z}\}$ for $\tilde{z} \in \z$.
Then, because $L-l$ blocks contain $\tilde{z}$ and not $\z \setminus\{\tilde{z}\}$, $n(z)=p(z)(L-l)$ and $n_{\tilde{z}}(z,z') = p_{\tilde{z}}(z,z')(L-l)$.
So,
\begin{align*}
&p_{z_1z_1}(z) = \frac{n(z)(n(z)-1)}{(L-l)(L-l-1)} = \frac{(L-l)p(z)^2}{L-l-1}-\frac{p(z)}{L-l-1}\\
&\Rightarrow p_{z_1z_1}(z)-p(z)^2 = \frac{p(z)^2-p(z)}{L-l-1}\\
    &p_{z_1z_1}(z,z') = \frac{n(z)n(z')-n_{z_1}(z,z')}{(L-l)(L-l-1)} = \frac{(L-l)p(z)p(z')}{L-l-1}-\frac{p_{z_1}(z,z')}{L-l-1}\\
    &\Rightarrow p_{z_1z_1}(z,z')-p(z)p(z') = \frac{p(z)p(z')-p_{z_1}(z,z')}{L-l-1},
\end{align*}
Analogous expressions hold for conditioning on $C_k$ by replacing $z_1$ with $z_2$.

For block $k$ and $j$ such that $B_k=C_j=1$, 
\begin{align*}
    \mathbb{P}(z\in \Rk,\R_j|B_k=C_j=1)-p(z)^2=\mathbb{P}(z\in \Rk,z' \in \R_j|B_k=C_j=1)-p(z)p(z')=0.
\end{align*}

\subsection{Variance and covariance of indicators of block membership}\label{subsec: var blk mem}
For $j \neq k$,
\begin{align*}
&\Var\left(A_k\right)=\frac{l_{z_1,z_2}}{K}\left(1-\frac{l_{z_1,z_2}}{K}\right),\quad \Var\left(B_k\right)=\frac{L_{z_1}-l_{z_1,z_2}}{K}\left(1-\frac{L_{z_1}-l_{z_1,z_2}}{K}\right)\\ &\Cov\left(A_k, B_k\right)=-\frac{l_{z_1,z_2}(L_{z_1}-l_{z_1,z_2})}{K^2},\quad  \Cov\left(B_k, C_k\right)=-\frac{(L_{z_1}-l_{z_1,z_2})(L_{z_2}-l_{z_1,z_2})}{K^2}, \\
&\Cov\left(A_k, A_j\right)=-\frac{l_{z_1,z_2}(K-l_{z_1,z_2})}{K^2(K-1)}, \\
&\Cov\left(A_k, B_j\right)=\frac{l_{z_1,z_2}(L_{z_1}-l_{z_1,z_2})}{K^2(K-1)}, \\
&\Cov\left(B_k, B_j\right)=\frac{(L_{z_1}-l_{z_1,z_2})(L_{z_1}-l_{z_1,z_2}-K)}{K^2(K-1)}, \\
&\Cov\left(B_k, C_j\right)=\frac{(L_{z_1}-l_{z_1,z_2})(L_{z_2}-l_{z_1,z_2})}{K^2(K-1)}.
\end{align*}
$\Var(C_k)$, $\Cov(A_k,C_k)$, $\Cov(A_k,C_j)$, and $\Cov(C_k,C_j)$ can be computed as the similar above.

\subsection{Useful simplifications of quadratic sums}\label{append:subsec:quad_sums}

Derivation of the finite-population variances involves quadratic sums, so we provide a couple of simplifications of these types of expressions below:
\begin{align*}
&\frac{1}{n-1}\sum_{i=1}^n(x_i-\bar{x})^2=\frac{1}{n-1}\sum_{i=1}^n x_i^2-\frac{n}{n-1}\bar{x}^2=\frac{1}{n}\sum_{i=1}^nx_i^2-\frac{1}{n(n-1)}\sum_{i=1}^n\sum_{j \neq i}x_ix_j,\\
&\frac{1}{K-1}\sum_{k=1}^K\left(x_k-w_k-(\bar{x}-\bar{w})\right)^2
=\frac{1}{K}\sum_{k=1}^K\left(x_k-w_k\right)^2-\frac{1}{K(K-1)}\sum_{k=1}^K\sum_{j\neq k}\left(x_k-w_k\right)\left(x_j-w_j\right)\\
&=\frac{1}{K-1}\left[\sum_{k=1}^K (x_k-\bar{x})^2 + \sum_{k=1}^K(w_k-\bar{w})^2\right]-\frac{2}{K}\left[\sum_{k=1}^Kx_kw_k-\frac{1}{K-1}\sum_{k=1}^K\sum_{j \neq k}x_kw_j\right].
\end{align*}

\subsection{Relationship between the variances of the Horvitz-Thompson and H\'ajek estimator}\label{append:rel ht haj}
Let $\gamma_k = Kw_k$, $\overline{\gamma^d} = K^{-1}\sum_{k=1}^K\gamma_k^d$, and $\overline{\gamma\tilde{Y}(z)} = K^{-1}\sum_{k=1}^K\gamma_k^{2}\ybark(z)$ for $d\in\mathbb{N}$.
Then the following equation represents the relationship between the variances of the Horvitz-Thompson and H\'ajek estimators:
\begin{align}
&(1-K^{-1})(\Shaj(z)-\SHT(z)) =\left[\ybarzw^2\left(\overline{\gamma^2}+1\right)-2\ybarzw\overline{\gamma\tilde{Y}(z)}\right],\label{eq:haj and ht single}\\
&2^{-1}(1-K^{-1})[\Shaj(z)+\Shaj(z')-\Shaj(\tau(z,z'))-(\SHT(z)+\SHT(z')-\SHT(\tau(z,z')))]\nonumber\\
&=\left[\ybarzw\ybar(z';\bm{w})\left(\overline{\gamma^2}+1\right)-\ybarzw\overline{\gamma\tilde{Y}(z')}-\ybar(z;\bm{w})\overline{\gamma\tilde{Y}(z')}\right].\label{eq:haj and ht double}
\end{align}

\subsection{Alternative form of variance expressions for $\tauestfe(z_1,z_2)$}\label{append:alt_var_form}

We can simplify the variance expression of $\tauestfe(z_1,z_2)$ in Section~\ref{subsec:fixed-effect} using some additional notation.

We can write for within-block variance components
\[\bar{S}^2_k(\tsettilde) = \frac{1}{W_{\tilde{z}} }\sum_{\mathcal{C} \in \tset_{\tilde{z}}}\sum_{z\in \mathcal{C}\setminus\{\tilde{z}\}}\frac{1}{t-1} S_k^2(z)=\sum_{z \in \z^c}\frac{p(z)}{t-1}S^2_k(z)\]
and
\[\bar{S}^2_k(\tau(\tilde{z}, \tsettilde)) = \frac{1}{W_{\tilde{z}} }\sum_{\tsetone \in \tsettilde}S^2_k(\tau(\tilde{z}, \tsetone)) = \sum_{z \in \z^c}\frac{p(z)}{t-1}S^2_k(\tau(\tilde{z}, z)) - \sum_{z \in \z^c }\sum_{z' \in \z^c , z' \neq z}\frac{p_{\tilde{z}}(z, z')}{2(t-1)^2}S^2_k(\tau(z, z')).\]

For between-block variance components, we similarly have
\begin{align*}
\bar{S}^2_{bb}(\tau(\tilde{z}, \tsettilde)) &= \frac{1}{W_{\tilde{z}} }\sum_{\tsetone \in \tset_{\tilde{z}}}S_{bb}^2(\tau(\tilde{z},\tsetone)) 
= \sum_{z \in \z^c}\frac{p(z)}{t-1}S^2_{bb}(\tau(\tilde{z}, z)) - \sum_{z \in \z^c }\sum_{z' \in \z^c , z' \neq z}\frac{p_{\tilde{z}}(z, z')}{2(t-1)^2}S^2_{bb}(\tau(z, z')).
\end{align*}

We can also write for the variance estimators,
\begin{align*}
\bar{s}_{bb}^2(\tau(\tilde{z},\tsettilde)) = \frac{1}{W_{\tilde{z}} }\sum_{\tsetone \in \tset_{\tilde{z}}}s_{bb}^2(\tau(\tilde{z},\tsetone)) = \sum_{z \in \z^c}\frac{p(z)}{t-1}s^2_{bb}(\tau(\tilde{z}, z)) - \sum_{z \in \z^c }\sum_{z' \in \z^c , z' \neq z}\frac{p_{\tilde{z}}(z, z')}{2(t-1)^2}s^2_{bb}(\tau(z, z')).
\end{align*}

It's worth noting that pieces of the variance expression simplify nicely in the unreduced design.
An \emph{unreduced} BIBD is one in which all possible combinations of selecting $t$ out of $T$ treatments are assigned across the blocks \citep{oehlert2010first}.
When $T>3$ and we use an unreduced design,
\begin{align*}
\bar{S}^2_k(\tau(\tilde{z}, \tsettilde)) &= \binom{T-1}{t-2}^{-1}\sum_{\tsetone \in \tsettilde}S^2_k(\tau(\tilde{z}, \tsetone))\\
& = \frac{1}{T-2}\sum_{z \in \z^c}S^2_k(\tau(\tilde{z}, z))-\frac{t-2}{2(t-1)(T-2)(T-3)}\sum_{z \in z^c}\sum_{z' \in \z^c , z' \neq z}S^2_k(\tau(z, z'))
\end{align*}
and
\begin{align*}
\bar{S}^2_{bb}(\tau(\tilde{z}, \tsettilde)) &= \binom{T-1}{t-2}^{-1}\sum_{\tsetone \in \tsettilde}S_{bb}^2(\tau(\tilde{z},\tsetone)) \\
&= \frac{1}{T-2}\sum_{z \in \z^c}S^2_{bb}(\tau(\tilde{z}, z)) - \frac{t-2}{2(t-1)(T-2)(T-3)}\sum_{z \in \z^c }\sum_{z' \in \z^c , z' \neq z}S^2_{bb}(\tau(z, z')).
\end{align*}

\section{Additional variance estimation results}\label{supp_matt:add_var_res}
\subsection{The Horvitz-Thompson estimator}\label{supp_mat:finite_bias_ht}
In this section, we provide the finite-sample bias of the variance estimators for the Horvitz-Thompson (HT) and the adjusted estimator. Recall that two variance estimators for the HT estimator:
For $z,z'\in\{1,...,T\}$,
\begin{align*}
&\Shatbbht(z,z') = \frac{l_{z,z'}\I[l_{z,z'}\geq 2]}{2L_zL_{z'}}(s_{\HT}^2(z)+s_{\HT}^2(z')-s_{\HT}^2(\tau(z,z'))),\\
&\Shatwbht(z,z') = \I(l_{z,z'}\geq2)\left(1-\frac{L_zL_{z'}}{l_{z,z'}}\right)\Shatbbht(z,z') +\frac{\I[z=z']}{p_z}\left(\frac{1}{K^2}\sum_{k=1}^K\frac{\I[z\in\Rk](Kw_k)^2s_k^2(z)}{n_k/t}\right),
\end{align*}
and $\Shatbbht = (\Shatbbht(z,z'))_{1\leq z,z'\leq T}$, and $\Shatwbht= (\Shatwbht(z,z'))_{1\leq z,z'\leq T}$.
\begin{theorem}\label{theorem:finite_bias_ht}
The finite-sample bias of $\bm{g}\tr\Shatbbht\bm{g}$ and $\bm{g}\tr\Shatwbht\bm{g}$ is
\begin{align*}
    &\E[\bm{g}\tr\Shatbbht\bm{g}]-\Var(\tauhtg)\\
    &=\frac{\SHT(\tauwg)}{K}-\sum_{z,z':l_{z,z'}\leq1}\frac{g_zg_{z'}l_{z,z'}}{2L_{z}L_{z'}}\left[S^2_{\HT}(z)+S^2_{\HT}(z')-S^2_{\HT}(\tau(z,z'))\right.\\
    &-\left.\left\{\frac{1}{K}\sum_{k=1}^K(Kw_k)^2\left(\frac{S_k^2(z)}{n_k}+\frac{S_k^2(z')}{n_k}-\frac{S_k^2(\tau(z,z'))}{n_k}\right)\right\}\right],\\
    &\E[\bm{g}\tr\Shatwbht\bm{g}]-\Var(\tauhtg)\\
    &=\frac{1}{K}\sum_{k=1}^K\frac{S_k^2(\tauwg)}{n_k}+\sum_{z,z':l_{z,z'}\leq1}\frac{g_zg_{z'}}{2}\left(\frac{1}{K}-\frac{l_{z,z'}}{L_zL_{z'}}\right)\left[S^2_{\HT}(z)+S^2_{\HT}(z')-S^2_{\HT}(\tau(z,z'))\right.\\
    &-\left.\left\{\frac{1}{K}\sum_{k=1}^K(Kw_k)^2\left(\frac{S_k^2(z)}{n_k}+\frac{S_k^2(z')}{n_k}-\frac{S_k^2(\tau(z,z'))}{n_k}\right)\right\}\right].
\end{align*}
\end{theorem}
The proof is provided in Section~\ref{pf:finite_bias_ht}
\subsection{The adjusted estimator}\label{supp_mat:var_est_fe}
We now turn to the variances estimators for the adjusted estimator. In addition to the sample estimators defined in Section~\ref{subsec:est_cov}, we define 
\begin{align*}
\bar{s}_{bb}^2(\tau(\tilde{z},\tsettilde)) =\sum_{z\in \z^c}\frac{p(z)}{t-1}s_{bb}^2(\tau(\tilde{z},z))-\sum_{z\in\mathbf{z}^c}\sum_{z'\in\z^c,z'\neq z}\frac{p_{\tilde{z}}(z,z')}{2(t-1)^2}s_{bb}^2(\tau(z,z'))\
\end{align*}
 where $p(z)$ and $p_{\tilde{z}}(z,z')$ are defined in Section~\ref{append:prob_relations}. 
 We remark that $\bar{s}_{bb}^2(\tau(\tilde{z},\tsettilde))$ is well-defined provided that $s_{bb}^2(\tau(\tilde{z},z))$ and $s_{bb}^2(\tau(z,z'))$ are well-defined for $\tilde{z} \in \z$ and $z,z'\in\z^c$. 
Further, let 
\[\tilde{\sigma}^2_{bb}(z_1, z_2) = \frac{1}{K}\left[s_{bb}^2(\tau(z_1,z_2))+(T-1)\left(\frac{\bar{s}_{bb}^2(\tau(z_1,\tset_{z_1}))}{t/(t-1)}+\frac{\bar{s}_{bb}^2(\tau(z_2,\tset_{z_2}))}{t/(t-1)}\right)\right].\]

We propose the following two variance estimators for the adjusted estimator:
\begin{align*}
\varestwbfe &= \frac{T-t}{T(t-1)}\tilde{\sigma}^2_{bb}(z_1, z_2) 
+ \frac{1}{K^2}\sum_{k=1}^K\left(\frac{\I[z_1 \in \Rk]s_k^2(z_1)}{n_k/T}+\frac{\I[z_2 \in \Rk]s_k^2(z_2)}{n_k/T}\right),\\
    \varestbbfe &= \frac{T-t}{T(t-1)}\tilde{\sigma}^2_{bb}(z_1, z_2)
    + \frac{s_{bb}^2(\tau(z_1,z_2))}{K}.
\end{align*}

\begin{theorem}\label{theorem:bias_fe_var}
The bias of $\varestwbfe$ and $\varestbbfe$ are $K^{-2}\sum_{k=1}^K{S_k^2(\tau(z_1,z_2))}/{n_k}$ and ${S_{bb}^2(\tau(z_1,z_2))}/{K}$, respectively.
Thus, these are conservative estimators.
\end{theorem}
The proof is provided in Section~\ref{supp_mat:proof of bias_fe_var}.

These biases have the same form of (asymptotic) bias as the variance estimators in Theorem~\ref{thm:var est}.
If treatment effects are homogenous within blocks, $\varestwbfe$ is unbiased.
If average treatment effects are homogenous across blocks, $\varestbbfe$ is unbiased.
Both estimators require there to be at least two blocks assigned each pair of treatments $z, z'$ that occur together in the set of treatment sets for this design, $\tset$, in order to estimate each $s_{bb}^2(\tau(z,z'))$.
Further, $\varestwbfe$ requires at least two units assigned to $z \in \z$ in block $k$ if $z \in \Rk$.

\section{Proofs}\label{append:proofs}

\subsection{Proof of Theorem~\ref{theorem:var_ht}}\label{append:proof of var_ht}
\begin{proof}
We define a new potential outcome $\tilde{Y}_i(z)$ for each unit $i$ with $b_i = k$ as $\tilde{Y}_i(z)=Kw_kY_i(z)$. This enables us to define similar definitions corresponding to $\ybark(z)$ and $\ybar(z)$ as
\begin{align*}
\overline{\tilde{Y}}_k(z) = \frac{1}{n_k}\sum_{i:b_i=k}\tilde{Y}_i(z)=Kw_k\ybark(z),\quad \overline{\tilde{Y}}(z) = \frac{1}{K}\sum_{k=1}^K\overline{\tilde{Y}}_k(z) = \sum_{k=1}^Kw_k\ybark(z) = \ybarzw.
\end{align*}
We also define sample estimators for $\overline{\tilde{Y}}_k(z)$ and $\overline{\tilde{Y}}(z)$ as
\begin{align*}
\widehat{\tilde{Y}}_k(z) = \frac{1}{n_k/t}\sum_{i:b_i=k}\I[Z_i = z]\tilde{Y}_i(z)=Kw_k\yestk(z),\quad \widehat{\tilde{Y}}(z) = \frac{1}{L_z}\sum_{k=1}^K\I[z\in\Rk]\widehat{\tilde{Y}}_k(z)=\yht(z).
\end{align*} 

To make $\yestk(z)$ always well defined, take $\yestk(z) = 0$ if $z \notin \Rk$.
Conditional on $z$ being assigned to block $k$,
\begin{align*}
\E\left[\yestk(z)\given z \in \Rk\right]&=\E\left[\frac{1}{n_k/t}\sum_{i:b_i=k}\I[Z_i=z]Y_i(z)\given z \in \Rk\right]\\
&=\frac{1}{n_k/t}\sum_{i:b_i=k}\prob(Z_i=z|z \in \Rk)Y_i(z) \\
&=\frac{t}{n_k}\frac{n_k/t}{n_k}\sum_{i:b_i=k}Y_i(z) \ \left(\because \prob(Z_i=z|z \in \Rk)=\frac{n_k/t}{n_k} \ \forall \ i \text{ in block } k \right) \\ 
&=\ybark(z).
\end{align*}

Using the fact that $ \overline{\tilde{Y}}(z)=\ybar(z)$ and $\yht(z) = \widehat{\tilde{Y}}(z)$, we establish the unbiasedness of $\bm{\yht}$ by the law of iterated expectation because
\begin{align*}
\E[\yht(z)] = \E[ \widehat{\tilde{Y}}(z)] &= \E\left[ \frac{1}{L_z}\sum_{k=1}^K\I[z\in\Rk]\widehat{\tilde{Y}}_k(z)\right] = \E\left[ \frac{1}{L_z}\sum_{k=1}^K\I[z\in\Rk]Kw_k\yestk(z)\right]\\
& = \E\left[ \frac{1}{L_z}\sum_{k=1}^K\I[z\in\Rk]Kw_k\ybark(z)\right] = \ybarzw
\end{align*}
for all $z=1,...,T$.

Now we provide the proof of the covariance matrix of the Horvitz-Thompson estimator. We define the following quantities:
\begin{align*}
&\tilde{\bm{B}}(z,z) =\left(\frac{1}{L_z/K}-1\right)\tilde{S}_{bb}^2(z),\quad \tilde{\bm{B}}(z,z')=\frac{1}{2}\left(\frac{Kl_{z,z'}}{L_zL_{z'}}-1\right)(\tilde{S}_{bb}^2(z)+\tilde{S}_{bb}^2(z')-\tilde{S}_{bb}^2(\tau(z,z'))),\\
&\tilde{\bmw}(z,z) = \frac{1}{L_z}\sum_{k=1}^K\frac{\tilde{S}_k^2(z)}{n_k/(t-1)},\quad \tilde{\bmw}(z,z')=-\frac{l_{z,z'}}{2L_zL_{z'}}\sum_{k=1}^K\left(\frac{\tilde{S}_k^2(z)}{n_k}+\frac{\tilde{S}_k^2(z')}{n_k}-\frac{\tilde{S}_k^2(\tau(z,z'))}{n_k}\right),
\end{align*}
with
\begin{align*}
&\tilde{S}^2_k(z) = \frac{1}{n_k-1}\sum_{i: b_i = k}(\tilde{Y}_i(z) - \overline{\tilde{Y}}_k(z))^2, \quad \tilde{S}^2_{bb}(z) = \frac{1}{K-1}\sum_{k=1}^K(\overline{\tilde{Y}}_k(z) - \ybarzw)^2,\\
&\tilde{S}^2_k(\tau(z, z')) = \frac{1}{n_k-1}\sum_{i: b_i = k}[(\tilde{Y}_i(z) - \tilde{Y}_i(z') -( \overline{\tilde{Y}}_k(z) - \overline{\tilde{Y}}_k(z'))]^2,\quad
\text{and}\\
&\tilde{S}^2_{bb}(\tau(z,z')) = \frac{1}{K-1}\sum_{k=1}^K[\overline{\tilde{Y}}_k(z) - \overline{\tilde{Y}}_k(z') -( \ybarzw - \ybar(z';\bm{w}))]^2.
\end{align*}
We further simplify the above quantities as
\begin{align*}
&\tilde{S}^2_k(z) = \frac{1}{n_k-1}\sum_{i: b_i = k}(\tilde{Y}_i(z) - \overline{\tilde{Y}}_k(z))^2=(Kw_k)^2S_k^2(z),\\
&\tilde{S}^2_{bb}(z) = \frac{1}{K-1}\sum_{k=1}^K(\overline{\tilde{Y}}_k(z) -  \ybarzw)^2=\SHT(z),\\
&\tilde{S}^2_k(\tau(z, z')) = \frac{1}{n_k-1}\sum_{i: b_i = k}[(\tilde{Y}_i(z) - \tilde{Y}_i(z') -( \overline{\tilde{Y}}_k(z) - \overline{\tilde{Y}}_k(z'))]^2=(Kw_k)^2S^2_k(\tau(z, z')),
\\
&\tilde{S}^2_{bb}(\tau(z,z')) = \frac{1}{K-1}\sum_{k=1}^K[\overline{\tilde{Y}}_k(z) - \overline{\tilde{Y}}_k(z') -(  \ybarzw - \ybar(z';\bm{w}))]^2 = \SHT(\tau(z,z')).
\end{align*}

We first compute $\bm{S}_{\HT}(z,z)$. By the law of total variance,
\begin{align*}
\Var(\yht(z)) = \E\left[\Var\left(\yht(z) \given \bfR \right)\right] +\Var\left(\E\left[\yht(z) \given \bfR\right]\right).
\end{align*}

For the first term,
\begin{align*}
&\E\left[\Var\left(\yht(z) \given \bfR \right)\right] = \E\left[\frac{1}{L_z^2}\sum_{k=1}^K(Kw_k)^2\I(z\in \Rk)\Var\left(\yestk(z) \given \Rk \right)\right]\\
&=\frac{1}{L_z^2}\sum_{k=1}^K\prob(z\in \Rk)(Kw_k)^2\Var\left(\yestk(z) \given \Rk \right) = \frac{1}{L_z^2}\sum_{k=1}^K\left(\frac{L_z}{K}\right)\frac{(Kw_k)^2S_k^2(z)}{n_k/(t-1)} = \frac{1}{KL_z}\sum_{k=1}^K\frac{(Kw_k)^2S_k^2(z)}{n_k/(t-1)}.
\end{align*}
For the second term,
\begin{align*}
&\Var\left(\E\left[\yht(z) \given \bfR\right]\right) = \Var\left(\frac{1}{L_z}\sum_{k=1}^K\I(z\in \Rk)Kw_k\ybark(z)\right)\\
&=\frac{1}{L_z^2}\sum_{k=1}^K\left(Kw_k)^2(\ybark(z)\right)^2\Var(\I[z\in \Rk])+\frac{1}{L_z^2}\sum_{k=1}^K\sum_{j\neq k}(Kw_k)(Kw_j)\ybark(z)\ybarj(z)\Cov(\I[z\in\Rk],\I[z\in\Rj])\\
&=\frac{1}{L_z^2}\sum_{k=1}^K\left(\overline{\tilde{Y}}_k(z)\right)^2\left[\frac{L_z}{K}\left(1-\frac{L_z}{K}\right)\right]+\frac{1}{L_z^2}\sum_{k=1}^K\sum_{j\neq k}\overline{\tilde{Y}}_k(z)\overline{\tilde{Y}}_j(z)\left[\frac{L_z}{K}\left(\frac{L_z-1}{K-1}-\frac{L_z}{K}\right)\right]\\
&=\frac{1}{KL_z}\left[\sum_{k=1}^K\left(\overline{\tilde{Y}}_k(z)\right)^2\left(1-\frac{L_z}{K}\right)+(L_z-K)\left(\frac{1}{K}\sum_{k=1}^K\left(\overline{\tilde{Y}}_k(z)\right)^2-\tilde{S}_{bb}^2(z)\right)\right] = \left(\frac{1}{L_z}-\frac{1}{K}\right)\tilde{S}_{bb}^2(z).
\end{align*}
Thus, 
\begin{align*}
\bm{S}_{\HT}(z,z) = \left(\frac{1}{L_z}-\frac{1}{K}\right)\tilde{S}_{bb}^2(z)+\frac{1}{KL_z}\sum_{k=1}^K\frac{\tilde{S}_k^2(z)}{n_k/(t-1)}=\frac{1}{K}\left(\tilde{\bm{B}}(z,z)+\tilde{\bm{W}}(z,z)\right).
\end{align*}

Now we turn to $\bm{S}_{\HT}(z,z')$ for $z\neq z'$. By the law of total covariance,
\begin{align*}
\Cov(\yht(z),\yht(z')) = \E\left[\Cov\left(\yht(z),\yht(z') \given \bfR \right)\right] +\Cov\left(\E\left[\yht(z) \given \bfR\right],\E\left[\yht(z') \given \bfR\right]\right).
\end{align*}
For the first term,
\begin{align*}
&\E\left[\Cov\left(\yht(z),\yht(z') \given \bfR \right)\right]\\
&=\E\left[\Cov\left( \frac{1}{L_z}\sum_{k=1}^K\mathbb{I}[z \in \Rk]Kw_k\yestk(z), \frac{1}{L_{z'}}\sum_{k=1}^K\mathbb{I}[z' \in \Rk]Kw_k\yestk(z') \given \bfR \right)\right]\\
&=\E\left[\frac{1}{L_zL_{z'}}\sum_{k=1}^K\I[z,z'\in \Rk](Kw_k)^2\Cov\left(\yestk(z),\yestk(z')|\I[z,z'\in \Rk]\right)\right]\\
&=\frac{1}{L_zL_{z'}}\sum_{k=1}^K\prob(z,z'\in\Rk)(Kw_k)^2\left[-\frac{1}{2}\left(\frac{S_k^2(z)}{n_k}+\frac{S_k^2(z')}{n_k}-\frac{S_k^2(\tau(z,z'))}{n_k}\right)\right]\\
&=-\frac{l_{z,z'}}{2KL_zL_{z'}}\sum_{k=1}^K(Kw_k)^2\left(\frac{S_k^2(z)}{n_k}+\frac{S_k^2(z')}{n_k}-\frac{S_k^2(\tau(z,z'))}{n_k}\right)\\
&=-\frac{l_{z,z'}}{2KL_zL_{z'}}\sum_{k=1}^K\left(\frac{\tilde{S}_k^2(z)}{n_k}+\frac{\tilde{S}_k^2(z')}{n_k}-\frac{\tilde{S}_k^2(\tau(z,z'))}{n_k}\right).
\end{align*}
For the second term,
\begin{align*}
&\Cov\left(\E\left[\yht(z) \given \bfR\right],\E\left[\yht(z') \given \bfR\right]\right) = \frac{1}{L_zL_{z'}}\Cov\left(\sum_{k=1}^K\I[z\in \Rk]Kw_k\ybark(z),\sum_{k=1}^K\I[z\in \Rk]Kw_k\ybark(z')\right)\\
&=\frac{1}{L_zL_{z'}}\sum_{k=1}^K\overline{\tilde{Y}}_k(z)\overline{\tilde{Y}}_k(z')\Cov\left(\I[z\in\Rk],\I[z'\in\Rk]\right)+\frac{1}{L_zL_{z'}}\sum_{k=1}^K\sum_{j\neq k}\overline{\tilde{Y}}_k(z)\overline{\tilde{Y}}_j(z')\Cov\left(\I[z\in\Rk],\I[z'\in\Rj]\right).
\end{align*}
For the first covariance term above, we have
\begin{align*}
\Cov\left(\I[z\in\Rk],\I[z'\in\Rk]\right) = \prob(z,z'\in \Rk)-\prob(z\in\Rk)\prob(z'\in\Rk) = \frac{Kl_{z,z'}-L_zL_{z'}}{K^2}.
\end{align*}
For the second covariance term above, we have
\begin{align*}
&\Cov\left(\I[z\in\Rk],\I[z'\in\Rj]\right) = \prob(z\in\Rk,z'\in\Rj) - \prob(z\in\Rk)\prob(z'\in\Rj) \\
&= \prob(z\in\Rk,z'\in\Rj,z'\in\Rk)+ \prob(z\in\Rk,z'\in\Rj,z'\notin\Rk)-\frac{L_zL_{z'}}{K^2}\\
&=\prob(z'\in\Rj\;|\;z,z'\in\Rk)\prob(z,z'\in\Rk)+\prob(z'\in\Rj\;|\;z\in\Rk,z'\notin\Rk)\prob(z\in\Rk,z'\notin\Rk)-\frac{L_zL_{z'}}{K^2}\\
&=\frac{L_{z'}-1}{K-1}\left(\frac{l_{z,z'}}{K}\right)+\frac{L_{z'}}{K-1}\left(\frac{L_z-l_{z,z'}}{K}\right)-\frac{L_zL_{z'}}{K^2}=\frac{L_zL_{z'}-Kl_{z,z'}}{K^2(K-1)}.
\end{align*}
Thus,
\begin{align*}
&\Cov\left(\E\left[\yht(z) \given \bfR\right],\E\left[\yht(z') \given \bfR\right]\right) \\
&=\frac{1}{L_zL_{z'}}\sum_{k=1}^K\overline{\tilde{Y}}_k(z)\overline{\tilde{Y}}_k(z')\Cov\left(\I[z\in\Rk],\I[z'\in\Rk]\right)+\frac{1}{L_zL_{z'}}\sum_{k=1}^K\sum_{j\neq k}\overline{\tilde{Y}}_k(z)\overline{\tilde{Y}}_j(z')\Cov\left(\I[z\in\Rk],\I[z'\in\Rj]\right)\\
&=\frac{Kl_{z,z'}-L_zL_{z'}}{K^2}\left(\frac{1}{L_zL_{z'}}\sum_{k=1}^K\overline{\tilde{Y}}_k(z)\overline{\tilde{Y}}_k(z')\right)+\frac{L_zL_{z'}-Kl_{z,z'}}{K^2(K-1)}\left(\frac{1}{L_zL_{z'}}\sum_{k=1}^K\sum_{j\neq k}\overline{\tilde{Y}}_k(z)\overline{\tilde{Y}}_j(z')\right)\\
&=\frac{Kl_{z,z'}-L_zL_{z'}}{KL_zL_{z'}}\left[\frac{1}{K}\sum_{k=1}^K\overline{\tilde{Y}}_k(z)\overline{\tilde{Y}}_k(z')-\frac{1}{K(K-1)}\sum_{k=1}^K\sum_{j\neq k}\overline{\tilde{Y}}_k(z)\overline{\tilde{Y}}_j(z')\right]\\
&=\frac{Kl_{z,z'}-L_zL_{z'}}{2L_zL_{z'}}\left[\frac{\tilde{S}_{bb}^2(z)}{K}+\frac{\tilde{S}_{bb}^2(z')}{K}-\frac{\tilde{S}_{bb}^2(\tau(z,z'))}{K}\right]
\end{align*}
To summarize,
\begin{align*}
&\Cov(\yht(z),\yht(z')) = \E\left[\Cov\left(\yht(z),\yht(z') \given \bfR \right)\right] +\Cov\left(\E\left[\yht(z) \given \bfR\right],\E\left[\yht(z') \given \bfR\right]\right)\\
&=-\frac{l_{z,z'}}{2KL_zL_{z'}}\sum_{k=1}^K\left(\frac{\tilde{S}_k^2(z)}{n_k}+\frac{\tilde{S}_k^2(z')}{n_k}-\frac{\tilde{S}_k^2(\tau(z,z'))}{n_k}\right)+\frac{Kl_{z,z'}-L_zL_{z'}}{2L_zL_{z'}}\left[\frac{\tilde{S}_{bb}^2(z)}{K}+\frac{\tilde{S}_{bb}^2(z')}{K}-\frac{\tilde{S}_{bb}^2(\tau(z,z'))}{K}\right]\\
&=\frac{1}{2L_zL_{z'}}\left[\frac{Kl_{z,z'}-L_zL_{z'}}{K}\left(\tilde{S}_{bb}^2(z)+\tilde{S}_{bb}^2(z')-\tilde{S}_{bb}^2(\tau(z,z'))\right)-\frac{l_{z,z'}}{K}\sum_{k=1}^K\left(\frac{\tilde{S}_k^2(z)}{n_k}+\frac{\tilde{S}_k^2(z')}{n_k}-\frac{\tilde{S}_k^2(\tau(z,z'))}{n_k}\right)\right]\\
&=\frac{1}{K}\left(\tilde{\bm{B}}(z,z')+\tilde{\bm{W}}(z,z')\right).
\end{align*}
Because $p_z = \prob(z\in\Rk) = L_z/K$ and $q_{z,z'} = \prob(z,z'\in\Rk) = l_{z,z'}/K$, we can easily check that $\tilde{\bm{B}} = \bm{B}_{HT}$ and $\tilde{\bmw} = \bmw$. This concludes the proof of Theorem~\ref{theorem:var_ht}.

\end{proof}

\subsection{Proof of Theorem~\ref{thm:clt}}\label{append:proof of clt}
\begin{proof}
Following \cite{zhao2022reconciling}, we decompose $\yht(z)$ for $z \in \{1,...,T\}$ into two components as follows:
\begin{align*}
\yht(z) &=\sum_{k=1}^K \frac{w_k\I[z\in\Rk]\yestk(z)}{\prob(z\in\Rk)}\\
&=\sum_{k=1}^K \frac{w_k\I[z\in\Rk]\ybark(z)}{\prob(z\in\Rk)}+ \sum_{k=1}^K \frac{w_k\I[z\in\Rk]}{\prob(z\in\Rk)}(\yestk(z)-\ybark(z))\\
& = \mu(z) + \sum_{k=1}^K\delta_k(z)
\end{align*}
where
\begin{align*}
\mu(z) = \sum_{k=1}^K \frac{w_k\I[z\in\Rk]\ybark(z)}{\prob(z\in\Rk)},\quad \delta_k(z)= \frac{w_k\I[z\in\Rk]}{\prob(z\in\Rk)}(\yestk(z)-\ybark(z)).
\end{align*}
We remark that the randomness of $\mu(z)$ only comes from the first stage randomization.
By taking vectorization form, we have
\begin{align*}
\bm{\yest}_{\HT} = \bmmu+ \bmdelta,
\end{align*}
where $\bmmu = (\mu(1),...,\mu(T))\tr$, $\bmdelta = \sum_{k=1}^K\bmdelta_k$, and $\bmdelta_k = (\delta_k(1),...,\delta(T))\tr$. 

Letting $ \mathcal{A}$ be the $\sigma$-algebra generated by $\{\Rk\}_{k=1}^K$, we can easily check that
\begin{align*}
&\E[\bmdelta|\mathcal{A}]=0,\quad \Cov(\bmdelta|\mathcal{A}) = \Cov(\sum_{k=1}^{K}\bmdelta_k|\mathcal{A})= \sum_{k=1}^{K}\Cov(\bmdelta_k|\mathcal{A}),\\
& \Cov(\bmdelta) = \sum_{k=1}^{K}\Cov(\bmdelta_k),\quad\Cov(\bmmu,\bmdelta) = \E[\Cov(\bmmu,\bmdelta|\mathcal{A})]+\Cov[\E(\bmmu|\mathcal{A}),\E(\bmdelta|\mathcal{A})] = \E[\bmmu\E[\bmdelta\tr|\mathcal{A}]]-0=0.
\end{align*}
In the above, we use the fact that $\bmmu$ is deterministic for given $\{\Rk\}_{k=1}^K$. By Theorem~\ref{theorem:var_ht} and $\Cov(\bmmu|\mathcal{A}) = \bm{0}_{T\times T}$, we have
\begin{align*}
\Cov(\bmmu) = \frac{1}{K}\bht,\quad \Cov(\bmdelta) = \frac{1}{K}\bmw.
\end{align*}

We introduce the following lemma to establish the asymptotic normality of $\bm{\yest}_{\HT}$:
\begin{Lemma}[Theorem A.1 of \citet{ohlsson1989asymptotic}, Lemma S2 of \citet{zhao2022reconciling}]\label{lemma:clt}
For $K=1,2,...$, let $\{\xi_{K,k}:k=1,...,K\}$ be a martingale difference sequence relative to the filtration $\{\mathcal{F}_{K,k}:k=0,...,K\}$, and let $X_K$ be an $\mathcal{F}_{K,0}$-measurable random variable. Set $\xi_{K} = \sum_{k=1}^K\xi_{K,k}$. Suppose that the following three conditions are fulfilled as $K\rightarrow \infty$.
\begin{enumerate}[label=\upshape(\roman*)]
\item $\sum_{k=1}^K\E[\xi_{K,k}^4]=o(1).$
\item For some sequence of non-negative real numbers $\{\beta_K: K=1,2,...\}$ with $\sup_{K\geq1}\beta_K<\infty$, we have $\E[(\sum_{k=1}^K\E(\xi_{K,k}^2 | \mathcal{F}_{K,k-1})-\beta_K^2)^2]=o(1)$.
\item $\mathcal{L}(X_K)*\mathcal{N}(0,\beta_K^2)\overset{d}{\rightarrow}\mathcal{L}_0$ for some probability distribution $\mathcal{L}_0$, where $\mathcal{L}(X)$ denotes a distribution of $X$ and $*$ denotes convolution.
\end{enumerate}
Then, $\mathcal{L}(X_K+\xi_K)\overset{d}{\rightarrow}\mathcal{L}_0$ as $K\rightarrow \infty$.
\end{Lemma}

Because $\sqrt{K}(\bm{\yht}-\bm{\ybar})= \sqrt{K}(\bmmu-\bm{\ybar}+\bmdelta)$, we show $\sqrt{K}(\bm{\yht}-\bm{\ybar})\overset{d}{\rightarrow}\mathcal{N}(0,\Sigma_{\HT})$ for $\Sigma_{\HT}=\bht+\bmw$ by showing 
\begin{align*}
\bmeta\tr\sqrt{K}(\bmmu-\bm{\ybar}+\bmdelta)\overset{d}{\rightarrow}\mathcal{N}(0,\bmeta\tr\Sigma_{\HT}\bmeta)
\end{align*}
for any deterministic unit vector $\bmeta\in\mathbb{R}^{T}$.

Let $X = \bmeta\tr\sqrt{K}(\bmmu-\bm{\ybar})$, $\xi_k = \bmeta\tr\sqrt{K}\bmdelta_k$, and $\xi = \sum_{k=1}^K\xi_k$.
We define $\mathcal{F}_{K,0} = \mathcal{A}$ where $ \mathcal{A}$ is the $\sigma$-algebra generated by $\{\Rk\}_{k=1}^K$, and $\mathcal{F}_{K,k}$ be the $\sigma$-algebra generated by $\{\Rk\}_{k=1}^K$ and $\{(Z_i)_{i:b_i=l}\}_{l=1,...,k}$ for $k=1,...,K$.
Then $\mathcal{F}_{K,0}\subset\mathcal{F}_{K,1}\subset\cdots\subset\mathcal{F}_{K,K}$ such that $\{\mathcal{F}_{K,k}:k=0,...,K\}$ is a filtration.
We verify below by checking that $(\xi_{k})_{k=1}^K$ and $X$ satisfy the three conditions of Lemma~\ref{lemma:clt} with $\beta_K^2 = \bmeta\tr\bmw\bmeta$ with regard to filtration $\{\mathcal{F}_{K,k}:k=0,...,K\}$.

\underline{Condition $(i)$.} We have 
\begin{align}\label{eq:xi_delta}
\xi_k^4 = K^2(\bmeta\tr\bmdelta_k)^4 \leq K^2\|\bmeta\|_2^4\|\bmdelta_k\|_2^4 = K^2\|\bmdelta_k\|_2^4.
\end{align}
In the last inequality, we use Cauchy-Schwartz inequality. In the last equality, we used the fact that $\bmeta$ is a unit vector. Thus, $\sum_{k=1}^K\E\xi_k^4 \leq K^2\sum_{k=1}^K\E\|\bmdelta_k\|_2^4$
and condition (i) holds if $ K^2\sum_{k=1}^K\E\|\bmdelta_k\|_2^4 = o(1)$.
To ensure this, we introduce a lemma.

\begin{Lemma}\label{lem:fourth moment}
Assume the randomization of incomplete block designs and Assumption~\ref{assump:clt}\ref{itm:fourth moment}. Then,
\begin{enumerate}[label=\upshape(\roman*)]
\item \label{itm:yestk fourth}$K^2\sum_{k=1}^Kw_k^4\E[\yestk(z)^2\yestk(z')^2| z,z'\in\Rk] = o(1)$ for $z,z' = 1,...,T$.
\item \label{itm:delta fourth}$K^2\sum_{k=1}^K\E[\|\bmdelta_k\|_2^4|\Rk = \tsetone] = o(1)$. 
\item \label{itm:sk fourth} $K^2\sum_{k=1}^Kw_k^4\E[(s_k^2(z))^2|z\in \Rk] = o(1)$ for $z,z' = 1,...,T$.
\end{enumerate}
\end{Lemma}
We provide the proof of Lemma~\ref{lem:fourth moment} in Section~\ref{append:proof of fourth moment}.
Thus, under Assumption~\ref{assump:clt}\ref{itm:fourth moment}, we establish $K^2\sum_{k=1}^K\E\|\bmdelta_k\|_2^4 = o(1)$ by Lemma~\ref{lem:fourth moment}.

\underline{Condition $(ii)$.} Let $\sigma = \Var(\xi|\mathcal{F}_{K,0})$, and $\sigma_k =  \Var(\xi_k|\mathcal{F}_{K,k}) =  \Var(\xi_k|\Rk)$ with  $\sigma = \sum_{k=1}^K\sigma_k$.
We verify that  $\E[\xi|\mathcal{F}_{K,0}] = \E[\sum_{k=1}^K\xi_k|\mathcal{F}_{K,0}]=\sqrt{K}\bmeta\tr\E[\sum_{k=1}^K\bmdelta_k|\mathcal{F}_{K,0}]=0$, $\Var(\xi) = \Var(\bmeta\tr\sqrt{K}\bmdelta) = \bmeta\tr\bmw\bmeta = \beta_K^2$ and $\E[\xi_k^2|\mathcal{F}_{K,k-1}] =  \E[\xi_k^2|\mathcal{F}_{K,0}] = \Var(\xi_k|\mathcal{F}_{K,0})=\sigma_k$, then this leads to
\begin{align*}
\sum_{k=1}^K\E[\xi_k^2|\mathcal{F}_{K,k-1}]=\sum_{k=1}^K\sigma_k = \sigma,\quad \beta_K^2 = \Var(\xi) = \E[\Var(\xi|\mathcal{F}_{K,0})] = \E[\sigma].
\end{align*}
Thus, $\E[(\sum_{k=1}^K\E(\xi_{k}^2 | \mathcal{F}_{K,k-1})-\beta_K^2)^2] = \Var(\sigma).$
To satisfy condition (ii) of Lemma~\ref{lemma:clt}, we need $\Var(\sigma) = o(1)$.

To verify this, we view $\sigma_k = \Var(\xi_k|\Rk)$ as the observed value of 
\begin{align*}
\sigma_k(\tsetone) =  \Var(\xi_k|\Rk=\tsetone) = K\bmeta\tr\Cov(\bmdelta|\Rk=\tsetone)\bmeta
\end{align*}
with mean $\bar{\sigma}(\tsetone)=K^{-1}\sum_{k=1}^K\sigma_k(\tsetone) $ and variance $S_{\sigma(\tsetone)}^2=(K-1)^{-1}\sum_{k=1}^K[\sigma_k(\tsetone)-\bar{\sigma}(\tsetone)]^2$ and sample mean $\widehat{\sigma}(\tsetone) = r_{\tsetone}^{-1}\sum_{k:\Rk=\tsetone}\sigma_k(\tsetone)$.
By standard results, $\Var(\widehat{\sigma}(\tsetone))=\frac{K-r_{\tsetone}}{Kr_{\tsetone}}S_{\sigma(\tsetone)}^2$ such that, with $\sigma=\sum_{\tsetone\in\tset}r_{\tsetone}\widehat{\sigma} (\tsetone)$, we establish
\begin{align*}
&\Var(\sigma) = \Var\left(\sum_{\tsetone\in\tset}r_{\tsetone}\widehat{\sigma}(\tsetone)\right)=\sum_{\tsetone\in\tset}\Var\left(r_{\tsetone}\widehat{\sigma}(\tsetone)\right)+\sum_{\tsetone\in\tset}\sum_{\tsetone'\in\tset,\tsetone'\neq \tsetone}\Cov(r_{\tsetone}\widehat{\sigma}(\tsetone),r_{\tsetone'}\widehat{\sigma}(\tsetone'))\\
&\leq \sum_{\tsetone\in\tset}\Var\left(r_{\tsetone}\widehat{\sigma}(\tsetone)\right) + \sum_{\tsetone\in\tset}\sum_{\tsetone'\in\tset,\tsetone'\neq \tsetone}\sqrt{\Var\left(r_{\tsetone}\widehat{\sigma}(\tsetone)\right)}\sqrt{\Var\left(r_{\tsetone'}\widehat{\sigma}(\tsetone')\right)}\\
&=K\sum_{\tsetone\in\tset}\frac{r_{\tsetone}}{K}\left(1-\frac{r_{\tsetone}}{K}\right)S_{\sigma(\tsetone)}^2 + K\sum_{\tsetone\in\tset}\sum_{\tsetone'\in\tset,\tsetone'\neq \tsetone}\sqrt{\frac{r_{\tsetone}}{K}\left(1-\frac{r_{\tsetone}}{K}\right)S_{\sigma(\tsetone)}^2}\sqrt{\frac{r_{\tsetone'}}{K}\left(1-\frac{r_{\tsetone'}}{K}\right)S_{\sigma(\tsetone')}^2}
\end{align*}
by Cauchy-Schwarz inequality. Because
\begin{align*}
\sum_{\tsetone'\in\tset,\tsetone'\neq \tsetone}\sqrt{\frac{r_{\tsetone'}}{K}\left(1-\frac{r_{\tsetone'}}{K}\right)S_{\sigma(\tsetone')}^2}\leq \sum_{\tsetone\in\tset}\sqrt{\frac{r_{\tsetone}}{K}\left(1-\frac{r_{\tsetone}}{K}\right)S_{\sigma(\tsetone)}^2},
\end{align*}
 the upper bound of $\Var(\sigma)$ above is further bounded as
\begin{align*}
&K\sum_{\tsetone\in\tset}\frac{r_{\tsetone}}{K}\left(1-\frac{r_{\tsetone}}{K}\right)S_{\sigma(\tsetone)}^2 + K\sum_{\tsetone\in\tset}\sum_{\tsetone'\in\tset,\tsetone'\neq \tsetone}\sqrt{\frac{r_{\tsetone}}{K}\left(1-\frac{r_{\tsetone}}{K}\right)S_{\sigma(\tsetone)}^2}\sqrt{\frac{r_{\tsetone'}}{K}\left(1-\frac{r_{\tsetone'}}{K}\right)S_{\sigma(\tsetone')}^2}\\
&\leq K\left[\sum_{\tsetone\in\tset}\frac{r_{\tsetone}}{K}\left(1-\frac{r_{\tsetone}}{K}\right)S_{\sigma(\tsetone)}^2 +\left(\sum_{\tsetone\in\tset}\sqrt{\frac{r_{\tsetone}}{K}\left(1-\frac{r_{\tsetone}}{K}\right)S_{\sigma(\tsetone)}^2}\right)^2\right]\\
&\leq K(1+|\tset|)\sum_{\tsetone\in\tset}\frac{r_{\tsetone}}{K}\left(1-\frac{r_{\tsetone}}{K}\right)S_{\sigma(\tsetone)}^2\leq  4^{-1} K(1+|\tset|)\sum_{\tsetone\in\tset}S_{\sigma(\tsetone)}^2.
\end{align*}
In the second and last inequalities, we use Cauchy-Schwarz inequality and $\max_{\tsetone\in\tset}\left|\frac{r_{\tsetone}}{K}\left(1-\frac{r_{\tsetone}}{K}\right)\right|\leq1/4$.

 Provided that $|\tset|=O(1)$, $\Var(\sigma)=o(1)$ is implied by $K\sum_{\tsetone\in\tset}S_{\sigma(\tsetone)}^2 = o(1)$. With $(K-1)S_{\sigma(\tsetone)}^2 = \sum_{k=1}^K\sigma_k^2(\tsetone)-K\bar{\sigma}^2(\tsetone)$, this is in turn guaranteed by 
\begin{align}\label{eq:two sufficient for ii}
\sum_{\tsetone\in\tset}\sum_{k=1}^K\sigma_k^2(\tsetone) = o(1),\quad K\sum_{\tsetone\in\tset}\bar{\sigma}^2(\tsetone) = o(1).
\end{align}
Because $\sigma_k(\tsetone) = \E[\xi_k^2|\Rk=\tsetone]$,
\begin{align*}
\sigma_k^2(\tsetone) = ( \E[\xi_k^2|\Rk=\tsetone])^2\leq \E[\xi_k^4|\Rk = \tsetone]\leq K^2\E[\|\bmdelta_k\|_2^4|\Rk = \tsetone]
\end{align*}
by Jensen's inequality and \eqref{eq:xi_delta}. The first equality in \eqref{eq:two sufficient for ii} then follows from
\begin{align*}
\sum_{\tsetone\in\tset}\sum_{k=1}^K\sigma_k^2(\tsetone)\leq K^2\sum_{\tsetone\in\tset}\sum_{k=1}^K\E[\|\bmdelta_k\|_2^4|\Rk = \tsetone] = o(1)
\end{align*} 
by Lemma~\ref{lem:fourth moment} for $\tset$ with $|\tset|  <\infty$.

The second inequality in \eqref{eq:two sufficient for ii} follows from
\begin{align*}
\bar{\sigma}(\tsetone) = \bmeta\tr\sum_{k=1}^K\Cov(\bmdelta_k|\Rk = \tsetone)\bmeta = \frac{1}{K}\sum_{z\in\tsetone}\sum_{z'\in\tsetone}\eta_z\eta_{z'}\bmw(z,z') = O(1/K).
\end{align*}
This verifies \eqref{eq:two sufficient for ii} for $\tset$ with $|\tset|  <\infty$ and thus condition (ii) in Lemma~\ref{lemma:clt}.

\underline{Condition $(iii)$.} The following lemma establishes asymptotic normality of $\bmmu$, whose proof is provided in Section~\ref{append:proof of clt of mu}.
\begin{Lemma}\label{lem:clt of mu}
Under Assumption~\ref{assump:clt}, $\sqrt{K}(\bmmu-\ybarw)\overset{d}{\rightarrow}\mathcal{N}(0,\bht)$.
\end{Lemma}

The convolution of $\sqrt{K}(\bmmu-\ybarw)$ with $\mathcal{N}(0,\bmeta\tr\bmw\bmeta)$ thus converges in distribution to $\mathcal{N}(0,\bmeta\tr\Sigma_{\HT}\bmeta)$ by the convergence of the characteristic function.

This verifies that $(\xi_{k})_{k=1}^K$ and $X$ satisfy the three conditions in Lemma~\ref{lemma:clt}. The sufficient
condition follows from Lemma~\ref{lemma:clt} and ensures the result for $\bm{\yht}$.

 To facilitate the discussion, it is useful to introduce the intermediate quantity
\begin{align*}
\yht'(z) = \sum_{k=1}^K\frac{w_k\I[z\in\Rk]}{\prob(z\in\Rk)}(\yestk(z)-\ybarzw)=\sum_{k=1}^K\frac{w_k\I[z\in\Rk]}{\prob(z\in\Rk)}\left[\frac{1}{n_k/t}\sum_{i:b_i=k}\I[Z_i=z](Y_i(z)-\ybarzw)\right],
\end{align*}
as the Horvitz-Thompson estimator defined on the centered potential outcomes $Y_i(z)-\ybarzw$. The difference between the H\'ajek estimator and the true finite-population average equals
\begin{align*}
\yhaj(z)-\ybarzw = \frac{\yht(z)-\onehatz\ybarzw}{\onehatz} = \frac{\yht'(z) }{\onehatz}.
\end{align*}
This indicates us the relationship between the asymptotic distribution of the Horvitz-Thompson and H\'ajek estimators: the centered H\'ajek estimator is the ratio estimator between the centered Horvitz-Thompson estimator and estimator of one.
For the asymptotic normality of $\bm{\yhaj}$, we recall that $\bm{\yhaj} -\ybarw = \bm{\onehat}^{-1}\bm{\yht}'$ with $\bm{\onehat} = \text{diag}(\onehatz)_{z=1}^{T}$ and $\bm{\yht}' = (\yht'(1),...,\yht'(T))\tr$. The asymptotic normality of $\bm{\yht}$ extends to $\bm{\yht}'$ as $\sqrt{K}(\bm{\yht}'-\ybarw)\overset{d}{\rightarrow}\mathcal{N}(0,\Sigma_{\haj})$. The result for $\bm{\yhaj}$ follows from Slutsky’s theorem with $\bm{\onehat}^{-1} \overset{p}{\rightarrow} \bm{I}_{T}$ by the following lemma, whose proof is given in Section~\ref{append:proof of consistency}.
\begin{Lemma}\label{lem:consistency}
Let $z,z'\in\{1,...,T\}$. Recall that $\yest_A(z) = l_{z,z'}^{-1}\sum_{k=1}^KKw_k\I[z,z'\in\Rk]\yestk(z)$ and $\yest_{A'}(z) = \sum_{k=1}^K\frac{w_k\I[z,z'\in\Rk]\yestk(z)}{\sum_{k=1}^Kw_k\I[z,z'\in\Rk]}$. Let $\bm{\yest}_A = (\yest_A(1),...,\yest_A(T))\tr$ and $\bm{\yest}_{A'} = (\yest_{A'}(1),...,\yest_{A'}(T))\tr$.
Under Assumption~\ref{assump:clt}\ref{itm:p conv} and \ref{itm:kw conv}, 
$\bm{\yht}\overset{p}{\rightarrow} \ybarw$, $\bm{\yhaj}\overset{p}{\rightarrow} \ybarw$, $\bm{\yest}_A\overset{p}{\rightarrow}\ybarw$, $\bm{\yest}_{A'}\overset{p}{\rightarrow}\ybarw$, and $\bm{\onehat}\overset{p}{\rightarrow} \bm{I}_{T}$.
\end{Lemma}
\end{proof}

\subsection{Proof of Corollary~\ref{cor:var comp}}\label{append:proof of var comp}
\begin{proof}
By \eqref{eq:haj and ht single}, 
\begin{align}\label{eq:ht and haj comp}
\SHT(z)-\Shaj(z) = -K(K-1)^{-1}\left[\ybarzw^2\left(\overline{\gamma^2}+1\right)-2\ybarzw\overline{\gamma\tilde{Y}(z)}\right].
\end{align}
If $Kw_k=1$, then $\ybarzw = K^{-1}\sum_{k=1}^K\ybark(z)$, $\overline{\gamma^2}=1$, and $\overline{\gamma\tilde{Y}(z)}=K^{-1}\sum_{k=1}^K\ybark(z)$. Thus, $\SHT(z)-\Shaj(z)=0$.

If $\ybark(z) = c$ for some constant $c$ over all $k$, then $\ybarzw = c$ and $\overline{\gamma\tilde{Y}(z)}=c\overline{\gamma^2}$. Thus, \eqref{eq:ht and haj comp} can be reduced as
\begin{align*}
 &-K(K-1)^{-1}\left[\ybarzw^2\left(\overline{\gamma^2}+1\right)-2\ybarzw\overline{\gamma\tilde{Y}(z)}\right]\\
 &=-K(K-1)^{-1}\left[c^2\left(\overline{\gamma^2}+1\right)-2c^2\overline{\gamma^2}\right] = -c^2K(K-1)^{-1}(1-\overline{\gamma^2}).
\end{align*}
By Cauchy-Schwarz inequality, $1 = (K^{-1}\sum_{k=1}^Kw_k)^2\leq K^{-1}\sum w_k^2 = \overline{\gamma^2}$. This ensures $\SHT(z)-\Shaj(z) \geq 0$; the equality holds if and only if $w_k = 1/K$ or $c=0$.

Finally, if $Kw_k\ybark(z) = c$ for some constant $c$ over $k$, then $\ybarzw = c$, and $\overline{\gamma\tilde{Y}(z)} = c$. Thus, \eqref{eq:ht and haj comp} can be reduced as
\begin{align*}
&-K(K-1)^{-1}\left[\ybarzw^2\left(\overline{\gamma^2}+1\right)-2\ybarzw\overline{\gamma\tilde{Y}(z)}\right]\\
&=-K(K-1)^{-1}\left[c^2\left(\overline{\gamma^2}+1\right)-2c^2\right]=-K(K-1)^{-1}c^2[\overline{\gamma^2}-1].
\end{align*}
Because $\overline{\gamma^2}-1 = K^{-1}\sum_{k=1}^K(Kw_k-1)^2 \geq 0$, the above term is nonnegative; the equality holds if and only if $c=0$ or $w_k = 1/K$. This concludes the proof.
\end{proof}

\subsection{Proof of Theorem~\ref{thm:var est}}\label{append:proof of var est}
\begin{proof}
We first provide the results for the Horvitz-Thompson estimator's variance estimators. Recall that
\begin{align*}
\sHT(\tau(z,z')) &= (l_{z,z'}-1)^{-1}\sum_{k=1}^K\I[z,z'\in\Rk][Kw_k(\yestk(z)-\yestk(z'))-(\yest_A(z)-\yest_A(z'))]^2\\
&=(l_{z,z'}-1)^{-1}\sum_{k=1}^K\I[z,z'\in\Rk][Kw_k(\yestk(z)-\yestk(z')))]^2-\frac{l_{z,z'}}{l_{z,z'}-1}\left(\yest_A(z)-\yest_A(z')\right)^2.
\end{align*}
Likewise, $\sHT(z)$ can be written as
\begin{align*}
\sHT(z) = (L_z-1)^{-1}\sum_{k=1}^K\I[z\in\Rk](Kw_k\yest_k(z))^2 - \frac{L_z}{L_z-1}\left(\yht(z)-\yht(z')\right)^2.
\end{align*}
We first derive the asymptotic expressions of $(l_{z,z'}-1)^{-1}\sum_{k=1}^K\I[z,z'\in\Rk][(Kw_k)^2\yestk(z)\yestk(z')]$ and $ (L_z-1)^{-1}\sum_{k=1}^K\I[z\in\Rk](Kw_k\yest_k(z))^2$ by  introducing the following lemma, whose proof is provided in Section~\ref{append:proof of sht control}.
\begin{Lemma}\label{lem:sht control}
Under Assumption~\ref{assump:clt}\ref{itm:fourth moment}, it holds that
\begin{align*} 
&(l_{z,z'}-1)^{-1}\sum_{k=1}^K\I[z,z'\in\Rk][Kw_k(\yestk(z)-\yestk(z')))]^2\nonumber\\
& = p_z\bmw(z,z)+p_{z'}\bmw(z',z')-\frac{2p_zp_{z'}}{q_{z,z'}}\bmw(z,z')+\SHT(\tau(z,z'))+(\ybarzw-\ybar(z';\bm{w}))^2+o_p(1),\\
&(L_z-1)^{-1}\sum_{k=1}^K\I[z\in\Rk](Kw_k\yest_k(z))^2 = p_z\bmw(z,z)+\SHT(z)+\ybarzw^2+o_p(1),\\
&\frac{1}{L_z}\sum_{k=1}^K\frac{\I[z\in\Rk](Kw_k)^2s_k^2(z)}{n_k/t} = \frac{1}{K}\sum_{k=1}^K\frac{(Kw_k)^2S_k^2(z)}{n_k/t}+o_p(1).
\end{align*}
\end{Lemma}

Lemma~\ref{lem:consistency} ensures 
\begin{align*}
\frac{l_{z,z'}}{l_{z,z'}-1}\left(\yest_A(z)-\yest_A(z')\right)^2 \overset{p}{\rightarrow} (\ybarzw-\ybar(z';\bm{w}))^2
\end{align*}
as $K$ goes to infinity. Therefore, by Lemma~\ref{lem:sht control}, we prove that
\begin{align*}
\sHT(\tau(z,z')) =  p_z\bmw(z,z)+p_{z'}\bmw(z',z')-\frac{2p_zp_{z'}}{q_{z,z'}}\bmw(z,z')+\SHT(\tau(z,z'))+o_p(1).
\end{align*}
In the same way, we establish that
\begin{align*}
\sHT(z) = p_z\bmw(z,z)+\SHT(z)+o_p(1)
\end{align*}
and this leads to
\begin{align*}
&\Shatbbht(z,z') = \frac{l_{z,z'}}{2L_zL_{z'}}(\sHT(z)+\sHT(z')-\sHT(\tau(z,z')))\\
&=\frac{1}{K}\left(\bmw(z,z')+\bht(z,z')+\frac{1}{2}(\SHT(z)+\SHT(z')-\SHT(\tau(z,z')))\right)+o_p(1)\\
&=\frac{1}{K}\Sigma_{\HT}(z,z')+\frac{1}{2K}(\SHT(z)+\SHT(z')-\SHT(\tau(z,z')))+o_p(1)
\end{align*}
for $l_{z,z'}\geq 2$. Thus, we ensure
\begin{align*}
    &\bm{g}\tr(K \Shatbbht-\bm{\Sigma}_{\HT})\bm{g} = \sum_{l_{z,z'}\geq 2}g_zg_{z'}\left(K \Shatbbht(z,z')-\bm{\Sigma}_{\HT}(z,z')\right)
+\sum_{l_{z,z'}<2}g_zg_{z'}\left(K \Shatbbht(z,z')-\bm{\Sigma}_{\HT}(z,z')\right)\\
&=\frac{1}{2}\sum_{l_{z,z'}\geq 2}g_zg_{z'}(\SHT(z)+\SHT(z')-\SHT(\tau(z,z')))+o_p(1)+\sum_{l_{z,z'}<2}g_zg_{z'}(-\bm{\Sigma}_{*}(z,z')).
\end{align*}
We remark that if $l_{z,z'}<2$, and $q_{z,z'}=l_{z,z'}/K\to 0$ as $K\to\infty$ (as assumed), then 
\begin{align*}
    &\bm{\Sigma}_{*}(z,z')+\frac{1}{2}(S_{*}^2(z)+S_{*}^2(z')-S_{*}^2(\tau(z,z')))\\
    &= \bmw(z,z')+\bm{B}_{*}(z,z')+\frac{1}{2}(S_{*}^2(z)+S_{*}^2(z')-S_{*}^2(\tau(z,z')))\to 0,
\end{align*}
since $l_{z,z'}<2$ only occurs for off-diagonal element and 
\begin{align*}
    &\bmw(z,z')=-\frac{q_{z,z'}}{2p_z p_{z'}}\left[\frac{1}{K}\sum_{k=1}^K(Kw_k)^2\left(\frac{S_k^2(z)}{n_k}+\frac{S_k^2(z')}{n_k}-\frac{S_k^2(\tau(z,z'))}{n_k}\right)\right],\\
    &\bm{B}_{*}(z,z')=\frac{1}{2}\left[\left(\frac{q_{z,z'}}{p_z p_{z'}}-1\right)\left(S_{*}^2(z)+S_{*}^2(z')-S_{*}^2(\tau(z,z'))\right)\right],
\end{align*}
for $* = \HT$ and $\haj$.
Thus, we can write
\begin{align}\label{eq: Sigma small l}
    -\bm{\Sigma}_{*}(z,z')=\frac{1}{2}(S_{*}^2(z)+S_{*}^2(z')-S_{*}^2(\tau(z,z')))+o(1),\quad \text{for $l_{z,z'}<2$ with $* = \HT$ and $\haj$},
\end{align}
and we ensure
\begin{align*}
    &\bm{g}\tr(K \Shatbbht-\bm{\Sigma}_{\HT})\bm{g}\\
    &=\frac{1}{2}\sum_{l_{z,z'}\geq 2}g_zg_{z'}(\SHT(z)+\SHT(z')-\SHT(\tau(z,z')))+o_p(1)\\
    &+\frac{1}{2}\sum_{l_{z,z'}<2}g_zg_{z'}(\SHT(z)+\SHT(z')-\SHT(\tau(z,z')))+o(1)\\
    &=\frac{1}{2}\sum_{z=1}^T\sum_{z'=1}^Tg_zg_{z'}(\SHT(z)+\SHT(z')-\SHT(\tau(z,z')))+o_p(1)+o(1)=\SHT(\tauwg)+o_p(1).
\end{align*}

For $\Shatwbht(z,z')$ with $l_{z,z'}\geq 2$, we can show using the same argument for $\Shatwbht(z,z')$ that
\begin{align*}
K\Shatwbht(z,z') = \bm{\Sigma}_{\HT}(z,z')+\frac{1}{2K}\sum_{k=1}^K\frac{(Kw_k)^2}{n_k}\left(S_{k}^2(z)+S_{k}^2(z')-S_{k}^2(\tau(z,z'))\right)+o_p(1).
\end{align*}
For the diagonal element of $\widehat{\bm{S}}_{\HT}^{wb}$, we have
\begin{align*}
K\Shatwbht(z,z) &= \left(\frac{1}{p_z}-1\right)\sHT(z)+\frac{1}{L_z}\sum_{k=1}^K\frac{\I[z\in\Rk](Kw_k)^2s_k^2(z)}{n_k/t}\\
&= \bm{\Sigma}_{\HT}(z,z)-p_z\bmw(z,z)+\frac{1}{L_z}\sum_{k=1}^K\frac{\I[z\in\Rk](Kw_k)^2s_k^2(z)}{n_k/t}+o_p(1).
\end{align*}
By Lemma~\ref{lem:sht control},
\begin{align*}
K\Shatwbht(z,z) 
&=\bm{\Sigma}_{\HT}(z,z)-p_z\bmw(z,z)+\frac{1}{K}\sum_{k=1}^K\frac{(Kw_k)^2S_k^2(z)}{n_k/t}+o_p(1)\\
&=\bm{\Sigma}_{\HT}(z,z)+\frac{1}{K}\sum_{k=1}^K\frac{(Kw_k)^2S_k^2(z)}{n_k}+o_p(1).
\end{align*}

Then $\bm{g}\tr\left(K\Shatwbht- \bm{\Sigma}_{\HT}\right)\bm{g}$ can be written as
\begin{align*}
&\bm{g}\tr\left(K\Shatwbht-\bm{\Sigma}_{\HT}\right)\bm{g}\\
&=\sum_{z=1}^{T}g_z^2(K\Shatwbht(z,z)- \bm{\Sigma}_{\HT}(z,z))+\sum_{z=1}^{T}\sum_{z'\neq z}g_zg_{z'}(K\Shatwbht(z,z')-\bm{\Sigma}_{\HT}(z,z'))\\
&=\sum_{z=1}^{T}g_z^2\left(\frac{1}{K}\sum_{k=1}^K\frac{(Kw_k)^2S_k^2(z)}{n_k}\right)+\sum_{z=1}^{T}\sum_{z'\neq z}g_zg_{z'}\left[\frac{1}{2K}\sum_{k=1}^K\frac{(Kw_k)^2}{n_k}\left(S_k^2(z)+S_k^2(z')-S_k^2(\tau(z,z'))\right)\right]+o_p(1)\\
&=\frac{1}{K}\sum_{k=1}^K\frac{S_k^2(\tauwg)}{n_k}+o_p(1).
\end{align*}

Now we provide the result for the H\'ajek estimator's variance estimators, first starting by decomposition of $\shaj(z)$ and $\shaj(\tau(z,z'))$, defined as
\begin{align*}
&\shaj(z) = (L_z-1)^{-1}\sum_{k=1}^K\I[z\in\Rk](Kw_k)^2(\yestk(z)-\yhaj(z))^2,\\
&\shaj(\tau(z,z')) = (l_{z,z'}-1)^{-1}\sum_{k=1}^K\I[z,z'\in\Rk](Kw_k)^2[\yestk(z)-\yestk(z')-(\yest_{A'}(z)-\yest_{A'}(z'))]^2\\
&\text{with}\quad \yest_{A'}(z) = \sum_{k=1}^K\frac{w_k\I[z,z'\in\Rk]\yestk(z)}{\sum_{k=1}^Kw_k\I[z,z'\in\Rk]}.
\end{align*}

By Lemma~\ref{lem:sht control}, direct algebra shows that
\begin{align*}
\shaj(z) &= (L_z-1)^{-1}\sum_{k=1}^K\I[z\in\Rk](Kw_k)^2\yestk(z)^2+\left(\yhaj(z)\right)^2(L_z-1)^{-1}\sum_{k=1}^K\I[z\in\Rk](Kw_k)^2\\
&-2\yhaj(z)(L_z-1)^{-1}\sum_{k=1}^K\I[z\in\Rk](Kw_k)^2\yestk(z),\\
&=p_z\bmw(z,z)+\SHT(z)+\ybarzw^2+\left(\yhaj(z)\right)^2(L_z-1)^{-1}\sum_{k=1}^K\I[z\in\Rk](Kw_k)^2\\
&-2\yhaj(z)(L_z-1)^{-1}\sum_{k=1}^K\I[z\in\Rk](Kw_k)^2\yestk(z)+o_p(1),
\end{align*}
and
\begin{align*}
\shaj(\tau(z,z')) &= (l_{z,z'}-1)^{-1}\sum_{k=1}^K\I[z,z'\in\Rk](Kw_k)^2\left(\yestk(z)-\yestk(z')\right)^2\\
&+(\yest_{A'}(z)-\yest_{A'}(z'))^2(l_{z,z'}-1)^{-1}\sum_{k=1}^K\I[z,z'\in\Rk](Kw_k)^2\\
&-2(\yest_{A'}(z)-\yest_{A'}(z'))(l_{z,z'}-1)^{-1}\sum_{k=1}^K\I[z,z'\in\Rk](Kw_k)^2\left(\yestk(z)-\yestk(z')\right)\\
&=p_z\bmw(z,z)+p_{z'}\bmw(z',z')-\frac{2p_zp_{z'}}{q_{z,z'}}\bmw(z,z')+\SHT(\tau(z,z'))+(\ybarzw-\ybar(z';\bm{w}))^2\\
&+(\yest_{A'}(z)-\yest_{A'}(z'))^2(l_{z,z'}-1)^{-1}\sum_{k=1}^K\I[z,z'\in\Rk](Kw_k)^2\\
&-2(\yest_{A'}(z)-\yest_{A'}(z'))(l_{z,z'}-1)^{-1}\sum_{k=1}^K\I[z,z'\in\Rk](Kw_k)^2\left(\yestk(z)-\yestk(z')\right)+o_p(1).
\end{align*}
We introduce the following lemma controls the asymptotic behavior of the remaining terms above.
\begin{Lemma}\label{lem:shaj control}
Recall the notations of $\overline{\gamma\tilde{Y}(z)}$ and $\overline{\gamma^2}$ in Section~\ref{append:rel ht haj}:  $\gamma_k = Kw_k$, $\overline{\gamma^2} = K^{-1}\sum_{k=1}^K\gamma_k^2$, and $\overline{\gamma\tilde{Y}(z)} = K^{-1}\sum_{k=1}^K\gamma_k^{2}\ybark(z)$. Then we have
\begin{enumerate}[label=\upshape(\roman*)]
\item \label{itm:gamma y single}$(L_z-1)^{-1}\sum_{k=1}^K\I[z\in\Rk](Kw_k)^2\yestk(z)-\overline{\gamma\tilde{Y}(z)}=o_p(1).$
\item \label{itm:gamma y double}$(l_{z,z'}-1)^{-1}\sum_{k=1}^K\I[z,z'\in\Rk](Kw_k)^2\yestk(z)-\overline{\gamma\tilde{Y}(z)}=o_p(1)$.
\item \label{itm:gamma2 single}$(L_z-1)^{-1}\sum_{k=1}^K\I[z\in\Rk](Kw_k)^2-\overline{\gamma^2}=o_p(1)$
\item \label{itm:gamma2 double}$(l_{z,z'}-1)^{-1}\sum_{k=1}^K\I[z,z'\in\Rk](Kw_k)^2-\overline{\gamma^2}=o_p(1).$
\end{enumerate}
\end{Lemma}
$\overline{\gamma\tilde{Y}(z)}$ is O(1) because
$\left|\overline{\gamma\tilde{Y}(z)}-\ybarzw\right|^2 = K^{-2}\left|\sum_{k=1}^KKw_k(Kw_k\ybark(z)-\ybarzw)\right|^2\leq\overline{\gamma^2}\SHT(z)+o(1) = O(1)$ by the Cauchy-Schwarz inequality and Assumption~\ref{assump:clt}\ref{itm:kw conv}
 and \ref{itm:bb conv}. Combined with Lemma~\ref{lem:shaj control} and Lemma~\ref{lem:consistency}, this ensures
 \begin{align*}
 \yhaj(z)(L_z-1)^{-1}\sum_{k=1}^K\I[z\in\Rk](Kw_k)^2\yestk(z) = \ybarzw\overline{\gamma\tilde{Y}(z)} +o_p(1).
 \end{align*}
 In the similar way, we have 
  \begin{align*}
 \left(\yhaj(z)\right)^2(L_z-1)^{-1}\sum_{k=1}^K\I[z\in\Rk](Kw_k)^2 = \ybarzw^2\overline{\gamma^2} +o_p(1).
 \end{align*}
 With the above expressions, we establish
 \begin{align*}
 \shaj(z) &=p_z\bmw(z,z)+\SHT(z)-2\ybarzw\overline{\gamma\tilde{Y}(z)}+\ybarzw^2(\overline{\gamma^2}+1) +o_p(1).
 \end{align*}
 Using \eqref{eq:haj and ht single}, the above equations can be written as
 \begin{align*}
  \shaj(z) &= p_z\bmw(z,z)+\SHT(z)+(1-K^{-1})(\Shaj(z)-\SHT(z))+o_p(1)\\
  &= p_z\bmw(z,z)+\Shaj(z)-K^{-1}(\Shaj(z)-\SHT(z))+o_p(1)=p_z\bmw(z,z)+\Shaj(z)+o_p(1).
 \end{align*}
 by Assumption~\ref{assump:clt}\ref{itm:bb conv}. Likewise, for $l_{z,z'}\geq 2$,
 \begin{align*}
 \shaj(\tau(z,z')) &= p_z\bmw(z,z)+p_{z'}\bmw(z',z')-\frac{2p_zp_{z'}}{q_{z,z'}}\bmw(z,z')+\SHT(\tau(z,z'))\\
 &+(\ybarzw-\ybar(z';\bm{w}))^2(\overline{\gamma^2}+1)-2(\ybarzw-\ybar(z';\bm{w}))\left(\overline{\gamma\tilde{Y}(z)}-\overline{\gamma\tilde{Y}(z')}\right)\\
 &= p_z\bmw(z,z)+p_{z'}\bmw(z',z')-\frac{2p_zp_{z'}}{q_{z,z'}}\bmw(z,z')+\SHT(\tau(z,z'))\\
 &+(1-K^{-1})\left(\Shaj(\tau(z,z'))-\SHT(\tau(z,z'))\right)+o_p(1)\\
 &=p_z\bmw(z,z)+p_{z'}\bmw(z',z')-\frac{2p_zp_{z'}}{q_{z,z'}}\bmw(z,z')+\Shaj(\tau(z,z'))+o_p(1).
 \end{align*}
 by \eqref{eq:haj and ht double}. 
 This implies
 \begin{align*}
 \Shatbbhaj(z,z') &= \frac{\I[l_{z,z'}\geq 2]}{K}\left[\bm{\Sigma}_{\haj}(z,z')+2^{-1}\left(\Shaj(z)+\Shaj(z')-\Shaj(\tau(z,z'))\right)+o_p(1)\right]. \end{align*}
 By \eqref{eq: Sigma small l}, we ensure
 \begin{align*}
 \bm{g}\tr\left(K\widehat{\bm{S}}_{\haj}^{bb}-\bm{\Sigma}_{\haj}\right)\bm{g} = \sum_{z=1}^T\sum_{z'=1}^T\frac{g_zg_{z'}}{2}\left(\Shaj(z)+\Shaj(z')-\Shaj(\tau(z,z'))\right)+o_p(1),
 \end{align*}
 and thus we establish
 \begin{align*}
  \bm{g}\tr\left(K\widehat{\bm{S}}_{\haj}^{bb}-\bm{\Sigma}_{\haj}\right)\bm{g}-\Shaj(\tauwg)\overset{p}{\rightarrow} 0.
  \end{align*}

 For $\Shatwbhaj$, since we assume $l_{z,z'}\geq 2$ for all $z$, $z'$ and this implies
 \begin{align*}
 &\Shatwbhaj(z,z') = \frac{1}{K}\left[\bm{\Sigma}_{\haj}(z,z')+\frac{1}{2K}\sum_{k=1}^K\frac{(Kw_k)^2}{n_k}\left(S_{k}^2(z)+S_{k}^2(z')-S_{k}^2(\tau(z,z'))\right)+o_p(1)\right]
 \end{align*} 
 and thus we establish
 \begin{align*}
\bm{g}\tr\left(K\widehat{\bm{S}}_{\haj}^{wb}-\bm{\Sigma}_{\haj}\right)\bm{g}-\frac{1}{K}\sum_{k=1}^K\frac{S_k^2(\tauwg)}{n_k}\overset{p}{\rightarrow} 0.
 \end{align*}
This concludes the proof.

\end{proof}

\subsection{Proof of Corollary~\ref{theorem:var_bibd}}
\begin{proof}
By Theorem~\ref{theorem:var_ht}, $\Var\left(\tauestg\right)$ under BIBD can be expressed as
\begin{align*}
&\Var\left(\tauestg\right) \\
&=\sum_{z=1}^{T}g_z^2\left[\left(\frac{1}{L}-\frac{1}{K}\right)S_{bb}^2(z)\right]+\frac{1}{2}\sum_{z=1}^{T}\sum_{z'\neq z}g_zg_{z'}\left[\left(\frac{l}{L^2}-\frac{1}{K}\right)\left(S_{bb}^2(z)+S_{bb}^2(z')-S_{bb}^2(\tau(z,z'))\right)\right]\\
&+\sum_{z=1}^{T}g_z^2\left[\frac{1}{KL}\sum_{k=1}^K\frac{S_k^2(z)}{n_k/(t-1)}\right]-\frac{1}{2}\sum_{z=1}^{T}\sum_{z'\neq z}g_zg_{z'}\left[\frac{l}{L^2}\frac{1}{K}\sum_{k=1}^K\left(\frac{S_k^2(z)}{n_k}+\frac{S_k^2(z')}{n_k}-\frac{S_k^2(\tau(z,z'))}{n_k}\right)\right].
\end{align*}
By setting $g_{z_1} = 1$, $g_{z_2} = -1$, and $g_{z} = 0$ for $z \neq z_1,z_2$, $\tauestg = \tauestbibd(z_1,z_2) $ and $\Var(\tauestbibd(z_1,z_2))$ can be written as
\begin{align*}
&\Var\left(\tauestbibd(z_1,z_2)\right)\\
&=\left(\frac{1}{L}-\frac{1}{K}\right)\left(S_{bb}^2(z_1)+S_{bb}^2(z_2)\right)-\left(\frac{l}{L^2}-\frac{1}{K}\right)\left(S_{bb}^2(z_1)+S_{bb}^2(z_2)-S_{bb}^2(\tau(z_1,z_2))\right)\\
&+\frac{1}{KL}\sum_{k=1}^K\frac{S_k^2(z_1)+S_k^2(z_2)}{n_k/(t-1)}+\frac{l}{KL^2}\sum_{k=1}^K\left(\frac{S_k^2(z_1)}{n_k}+\frac{S_k^2(z_2)}{n_k}-\frac{S_k^2(\tau(z_1,z_2))}{n_k}\right).
\end{align*}
Under BIBD, because $Kt = TL$ and $l(T-1) = L(t-1)$,
\begin{align*}
&\frac{1}{L}-\frac{1}{K} = \frac{1}{K}\left(\frac{T-t}{t}\right),\frac{l}{L^2}-\frac{1}{K} = \left(\frac{1}{L}\right)\frac{t-1}{T-1}-\frac{1}{K} = \frac{1}{K}\left(\frac{T(t-1)}{t(T-1)}-1\right) = -\frac{T-t}{Kt(T-1)},\\
&\frac{1}{KL} = \frac{1}{K^2}\cdot\frac{T}{t},\quad \frac{l}{KL^2} = \frac{1}{KL}\cdot\frac{t-1}{T-1} = \frac{1}{K^2}\cdot\frac{T(t-1)}{t(T-1)}.
\end{align*}
This lead to the simplification of $\Var\left(\tauestbibd(z_1,z_2)\right)$ as
\begin{align*}
\Var\left(\tauestbibd(z_1,z_2)\right)&=\frac{T-t}{t(T-1)}\left(\frac{S_{bb}^2(z_1)}{K/T}+\frac{S_{bb}^2(z_2)}{K/T}-\frac{S_{bb}^2(\tau(z_1,z_2))}{K}\right)\\
&+\frac{T(t-1)}{t(T-1)}\cdot\frac{1}{K^2}\sum_{k=1}^K\left(\frac{S_k^2(z_1)}{n_k/T}+\frac{S_k^2(z_2)}{n_k/T}-\frac{S_k^2(\tau(z_1,z_2))}{n_k}\right),
\end{align*}
which concludes the proof of Corollary~\ref{theorem:var_bibd}.
\end{proof}

\subsection{Proof of Theorem~\ref{theorem:bias_fe}}\label{supp_mat:theorem:bias_fe}
Here we prove unbiasedness of the adjusted estimator: $\mathbb{E}[\tauestfe(z_1,z_2) ] = \tau(z_1, z_2)$.
\begin{proof}
Using the notation $A_k$, $B_k$, and $C_k$ defined previously and the notation in Supplementary Material~\ref{append:alt_var_form}, we can write
\begin{align*}
    &\tauestfe(z_1,z_2)\\
    &=\frac{t}{lT}\left\{\sum_{k=1}^K\sum_{j=1}^t \I[z_1 = R_{k,j}]\left[\yestk(z_1) -\frac{1}{t}\sum_{m=1}^{t}\yestk(R_{k,m})\right]-\sum_{k=1}^K\sum_{j=1}^t \I[z_2 = R_{k,j}]\left[\yestk(z_2) -\frac{1}{t}\sum_{m=1}^{t}\yestk(R_{k,m})\right]\right\}\\
    &= \frac{t}{lT}\left[\sum_{k=1}^KA_k \left(\yestk(z_1)-\yestk(z_2)\right)+B_k\left(\yestk(z_1)-\frac{1}{t}\sum_{m=1}^{t}\yestk(R_{k,m})\right) - C_k\left(\yestk(z_2)-\frac{1}{t}\sum_{m=1}^{t}\yestk(R_{k,m})\right)\right].
\end{align*}
Its expectation is
\begin{align*}
    &\E[\tauestfe(z_1,z_2)]=\E\left[\E[\tauestfe(z_1,z_2)|\mathbf{R}]\right]\\
     =&\frac{t}{lT}\E\left[\sum_{k=1}^K A_k \left(\ybar_k(z_1)-\ybar_k(z_2)\right)+B_k\left(\left(\frac{t-1}{t}\right)\ybar_k(z_1)-\frac{1}{t}\sum_{z\neq z_1}\I[z \in \Rk]\ybar_k(z)\right)\right.\\
    &\qquad \qquad \left.- C_k\left(\left(\frac{t-1}{t}\right)\ybar_k(z_2)-\frac{1}{t}\sum_{z\neq z_2}\I[z \in \Rk]\ybar_k(z)\right)\right].
\end{align*}

The first term is 
\begin{align*}
\E\left[\sum_{k=1}^K A_k \left(\ybar_k(z_1)-\ybar_k(z_2)\right)\right] = \frac{l}{K}\sum_{k=1}^K  \left(\ybar_k(z_1)-\ybar_k(z_2)\right).
\end{align*}
For the second term (and similarly the third term),
\begin{align*}
    \E\left[\frac{1}{t}\sum_{z\neq z_1}\I[z \in \Rk]\ybar_k(z)\given B_k=1\right] = \frac{1}{t}\sum_{z\neq z_1}\prob(z\in \Rk | B_k=1)\ybar_k(z)=\frac{1}{t}\sum_{z \in \z^c}p(z)\ybar_k(z)
\end{align*}
because $\prob(z_2\in \Rk | B_k=1)=0$.
Thus,
\begin{align*}
&\frac{lT}{t}\E[\tauestfe(z_1,z_2)]  \\
    &=\E\left[\sum_{k=1}^K A_k \left(\ybar_k(z_1)-\ybar_k(z_2)\right)+B_k\left(\ybar_k(z_1)-\frac{1}{t}\sum_{z=1}^T\I[z \in \Rk]\ybar_k(z)\right) - C_k\left(\ybar_k(z_2)-\frac{1}{t}\sum_{z=1}^T\I[z \in \Rk]\ybar_k(z)\right)\right]\\
    &=\E\left[\sum_{k=1}^K A_k \left(\ybar_k(z_1)-\ybar_k(z_2)\right)+\frac{t-1}{t}\left(B_k\ybar_k(z_1)-C_k\ybar_k(z_2)\right)\right] -\sum_{k=1}^K\sum_{z \in \z^c}\left[\E [B_k] p(z)-\E[C_k]p(z)\right]\ybar_k(z)\\
    &=\left(l+\frac{(t-1)(L-l)}{t}\right)\frac{1}{K}\sum_{k=1}^K\left(\ybar_k(z_1)-\ybar_k(z_2)\right)\\
    &=\frac{lT}{Kt}\sum_{k=1}^K\left(\ybar_k(z_1)-\ybar_k(z_2)\right).
\end{align*}
The last two lines simplify because $(t-1)(L-l) = l(T-t)$ and $ \E [B_k]=\E [C_k] = \frac{L-l}{K}$.
Therefore $\tauestfe(z_1,z_2)$ is an unbiased estimator of $\tau(z_1,z_2)$.
\end{proof}

\subsection{Proof of Theorem~\ref{thrm:adj_var}}\label{append:subsec_fe_proof}
We show here the derivation of the variance of $\tauestfe(z_1,z_2)$,
\begin{align*}
&\Var\left(\tauestfe(z_1,z_2)\right)\\
&=\frac{T-t}{T(t-1)}\left[\frac{1}{K}\left\{S_{bb}^2(\tau(z_1,z_2))+(T-1)\left(\frac{\bar{S}_{bb}^2(\tau(z_1,\tset_{z_1}))}{t/(t-1)}+\frac{\bar{S}_{bb}^2(\tau(z_2,\tset_{z_2}))}{t/(t-1)}\right)\right\}\right]\\
&+\frac{T-1}{T(t-1)}\left[\frac{1}{K^2}\sum_{k=1}^Kt\left(\frac{S_k^2(z_1)}{n_k/t}+\frac{S_k^2(z_2)}{n_k/t}-\frac{S_k^2(\tau(z_1,z_2))}{n_k}\right)+(T-t)\left(\frac{\V{z_1}}{t/(t-1)}+\frac{\V{z_2}}{t/(t-1)}\right)\right].
    \end{align*}
    
\begin{proof}
Before getting to the variance of $\tauestfe(z_1,z_2)$, we examine the definition of $\tauestfe(z_1,z_2)$ further. 

If a block is assigned both $z_1$ and $z_2$, then its adjustment cancels in $\yestadj(z_1)-\yestadj(z_2)$. 
If a block is only assigned one of the two treatments under comparison, for example, if $z_1$ is assigned to the block but $z_2$ is not, this block will be used to estimate $\yestadj(z_1)$ but not $\yestadj(z_2)$. 
If neither $z_1$ nor  $z_2$ are assigned to a block, that block will not contribute to $\tauestfe(z_1,z_2)$. 
Thus, there are three general ways a block may contribute to $\tauestfe(z_1,z_2)$ and we define group memberships to classify blocks according to contribution.
A block with group membership (i) $A$ is a block for which the assigned treatment set contains both $z_1$ and $z_2$
(ii) $B$ is a block for which the assigned treatment set contains $z_1$ but not $z_2$ 
(iii) $C$ is a block for which the assigned treatment set contains  $z_2$ but not $z_1$.
We denote indicators for group membership $A$, $B$, and $C$ for block $k$ as 
$$A_k = \I[z_1\in \Rk,z_2 \in \Rk], \quad B_k = \I[z_1 \in \Rk, z_2 \notin \Rk], \quad C_k = \I[z_1 \notin \Rk, z_2 \in \Rk].$$

The above indicators are random due to the first randomization. 
Because we take the set of treatment subsets used in the BIBD design as fixed, however, the number of blocks with a certain group membership is invariant under the randomization. 
In particular, we observe that there are $l$ blocks whose treatment includes $\z$, meaning there are $l$ blocks with membership $A$.
Further, there are $L$ blocks total with $z_1$, so there are $L-l$ blocks with membership $B$. 
Similarly, there are $L-l$ blocks with membership $C$.
For example, consider a design from  \cite{oehlert2010first} with parameters $\bm{\lambda} = (K,T,t,L,l) = (10,5,3,6,3)$, with the treatment sets (comprising $\tset$) given in given in Table~\ref{tab:(5,3,6,3) BIBD} and set $z_1=1$, $z_2=2$.
If there are $r$ blocks assigned to each treatment set (column), then there are $3r$ blocks with membership $A$, $3r$ blocks with membership $B$, and $3r$ blocks with membership $C$.

We will use the same decomposition of variance as in Section~\ref{subsec: finite_ht}, into the expectation of the conditional variance and the variance of the conditional expectation.
We start with the conditional variance of $\tauestfe(z_1,z_2)$, which is
\begin{align*}
    &\Var\left(\tauestfe(z_1,z_2)|\bfR\right)\\
    &=\left(\frac{t}{lT}\right)^2\sum_{k=1}^K\left[A_k\Var\left(\yestk(z_1)-\yestk(z_2)|\mathbf{R}\right)+B_k\Var\left(\yestk(z_1)-\frac{1}{t}\sum_{m=1}^{t}\yestk(R_{k,m})\given\bfR\right)\right.\\
    &\qquad \qquad \qquad \qquad \left.+C_k\Var\left(\yestk(z_2)-\frac{1}{t}\sum_{m=1}^{t}\yestk(R_{k,m})\given\bfR\right)\right].
\end{align*}
We know that
\begin{align*}
    \Var\left(\yestk(z_1)-\yestk(z_2)\given\mathbf{R}\right)=\frac{S_k^2(z_1)}{n_k/t}+\frac{S_k^2(z_2)}{n_k/t}-\frac{S_k^2(\tau(z_1,z_2))}{n_k}.
\end{align*}
For $\tilde{z} \in \z = \{z_1,z_2\}$ and $\tsetone \in \tsettilde$, consider
\begin{align*}
\Var\left(\yestk(\tilde{z})-\frac{1}{t}\sum_{m=1}^{t}\yestk(R_{k,m})\given \Rk = \tsetone \right) = \left(\frac{t-1}{t}\right)^2\Var\left(\tauestk(\tilde{z}, \tsetone)\given \Rk = \tsetone \right)
\end{align*}
where $\tauestk(\tilde{z}, \tsetone) = \yestk(\tilde{z})-\frac{1}{t-1}\sum_{z \in  \tsetone\setminus\{\tilde{z}\}}\yestk(z)$.
By Theorem 3 of \cite{li2017general}, 
\begin{align*}
&\Var\left(\tauestk(\tilde{z}, \tsetone)\given \Rk = \tsetone \right)= \frac{S^2_k(\tilde{z})}{n_k/t} + \frac{1}{(t-1)^2}\sum_{z \in \tsetone\setminus\{\tilde{z}\}} \frac{S^2_k(z)}{n_k/t} - \frac{S^2_k(\tau(\tilde{z}, \tsetone)) }{n_k}
\end{align*}
where 
\begin{align*}
&S^2_k(\tau(\tilde{z}, \tsetone)) =\frac{1}{n_k-1}\sum_{i:b_i=k}\left(Y_i(\tilde{z})-\frac{1}{t-1}\sum_{z \in \tsetone\setminus\{\tilde{z}\}}Y_i(z)-\left(\overline{Y}_k(\tilde{z})-\frac{1}{t-1}\sum_{z \in \tsetone\setminus\{\tilde{z}\} }\overline{Y}_k(z)\right)\right)^2\\
&= \frac{1}{(t-1)^2}\frac{1}{n_k-1}\sum_{z \in \tsetone\setminus\{\tilde{z}\}}\sum_{i:b_i=k}\left(\tau_i(\tilde{z}, z) - \tau(\tilde{z}, z)\right)^2 \\
& \qquad+ \frac{1}{(t-1)^2}\frac{1}{n_k-1}\sum_{z \in \tsetone\setminus\{\tilde{z}\}}\sum_{z' \in \tsetone\setminus\{\tilde{z}\}, z' \neq z}\sum_{i:b_i=k}\left(\tau_i(\tilde{z}, z) - \tau(\tilde{z}, z)\right)\left(\tau_i(\tilde{z}, z') - \tau(\tilde{z}, z')\right)\\
&= \frac{1}{(t-1)^2}\sum_{z \in \tsetone\setminus\{\tilde{z}\}}S^2_k(\tau(\tilde{z}, z)) + \frac{1}{2(t-1)^2}\sum_{z \in \tsetone\setminus\{\tilde{z}\}}\sum_{z' \in \tsetone\setminus\{\tilde{z}\}, z' \neq z}\left[S^2_k(\tau(\tilde{z}, z)) + S^2_k(\tau(\tilde{z}, z')) - S^2_k(\tau(z, z'))\right]\\
&=\sum_{z \in  \tsetone\setminus\{\tilde{z}\}}\frac{1}{t-1}S^2_k(\tau(\tilde{z}, z)) - \sum_{z \in  \tsetone\setminus\{\tilde{z}\} }\sum_{z' \in  \tsetone\setminus\{\tilde{z}\}, z' \neq z}\frac{1}{2(t-1)^2}S^2_k(\tau(z, z')).
\end{align*}
By taking the expectation over $\tsetone \in \tsettilde$,
\begin{align*}
&\E\left[\Var\left(\tauestk(\tilde{z}, \tsetone)\given \Rk = \tsetone \right)\right]=\frac{S^2_k(\tilde{z})}{n_k/t} + \frac{1}{W_{\tilde{z}} }\sum_{\mathcal{C} \in \tset_{\tilde{z}}}\sum_{z\in \mathcal{C}\setminus\{\tilde{z}\}}\frac{t}{(t-1)^2} \frac{S_k^2(z)}{n_k}-\frac{1}{W_{\tilde{z}} }\sum_{\mathcal{C} \in \tset_{\tilde{z}}}\frac{S^2_k(\tau(\tilde{z}, \tsetone)) }{n_k}=\V{\tilde{z}}.
\end{align*}

Thus,
\begin{align}
    &\E\left[\Var\left(\tauestfe(z_1,z_2)|\bfR\right)\right]\\
    &=\left(\frac{t}{lT}\right)^2\sum_{k=1}^K\left[\E [A_k] \left(\frac{S_k^2(z_1)}{n_k/t}+\frac{S_k^2(z_2)}{n_k/t}-\frac{S_k^2(\tau(z_1,z_2))}{n_k}\right)+\left\{\E [B_k]\frac{\V{z_1}}{(t/(t-1))^2}+\E [C_k]\frac{\V{z_2})}{(t/(t-1))^2} \right\}\right]\nonumber\\
    &=\frac{T-1}{T(t-1)}\left[\frac{1}{K^2}\sum_{k=1}^K\left\{t\left(\frac{S_k^2(z_1)}{n_k/t}+\frac{S_k^2(z_2)}{n_k/t}-\frac{S_k^2(\tau(z_1,z_2))}{n_k}\right)+(T-t)\left(\frac{\V{z_1}}{t/(t-1)}+\frac{\V{z_2}}{t/(t-1)}\right)\right\}\right].\label{eq:fe_exp_cond_var}
    \end{align}

Now consider the variance of the conditional expectation:
\begin{align*}
    &\Var\left(\E\left[\tauestfe(z_1,z_2)\given \mathbf{R}\right]\right)\\
    &=\left(\frac{t}{lT}\right)^2\Var\left[\sum_{k=1}^KA_k(\ybar_k(z_1)-\ybar_k(z_2))+\frac{B_k(t-1)}{t}\left(\ybar_k(z_1)-\frac{1}{t-1}\sum_{z \in \z^c}\I[z \in \Rk]\ybar_k(z)\right)\right.\nonumber\\
    &\qquad\qquad\qquad\qquad \left.-\frac{C_k(t-1)}{t}\left(\ybar_k(z_2)-\frac{1}{t-1}\sum_{z \in \z^c}\I[z \in \Rk]\ybar_k(z)\right)\right]\\
    &=\left(\frac{t}{lT}\right)^2\Var\left[\sum_{k=1}^KA_k(\ybar_k(z_1)-\ybar_k(z_2))+\sum_{\tsetone \in \tset_{z_1}}\frac{B_k\I[\Rk = \tsetone](t-1)}{t}\left(\ybar_k(z_1)-\frac{1}{t-1}\sum_{z \in \tsetone \setminus z_1}\ybar_k(z)\right)\right. \\
    &\qquad\qquad\qquad\qquad \left.-\sum_{\tsetone \in \tset_{z_2}}\frac{C_k\I[\Rk = \tsetone](t-1)}{t}\left(\ybar_k(z_2)-\frac{1}{t-1}\sum_{z \in \tsetone \setminus z_2}\ybar_k(z)\right)\right].
\end{align*}

Now we have a format for the conditional variance to apply Theorem 3 of \citet{li2017general}.
To help see the connection to that result better, we can conceptualize a hypothetical experiment.
Consider an experiment with potential outcomes
\begin{align*}
X_k(a) = l\left(\ybar_k(z_1)-\ybar_k(z_2)\right) & \text{ if } A_k = 1,\\
X_k(b, \tsetone) = \frac{L-l}{W_{z_1}}\frac{(t-1)}{t}\left(\ybar_k(z_1)-\frac{1}{t-1}\sum_{z \in \tsetone \setminus z_1}\ybar_k(z)\right) & \text{ if } B_k = 1 \text{ and } \I[\Rk = \tsetone]= 1,\\
X_k(c, \tsetone)  = \frac{L-l}{W_{z_2}}\frac{(t-1)}{t}\left(\ybar_k(z_2)-\frac{1}{t-1}\sum_{z \in \tsetone \setminus z_1}\ybar_k(z)\right) & \text{ if } C_k = 1 \text{ and } \I[\Rk = \tsetone]= 1,\\
X_k(d)  =0 & \text{ if } A_k = B_k C_k = 0.
\end{align*}
Note that the $X_k$'s are fixed functions of potential outcomes (i.e., not random).
Keep the probability of a block having $A_k = 1$, $B_k = 1$, $ \I[\Rk = \tsetone]= 1$, etc. the same as our original BIBD.
Then we can write the conditional variance above as 

\begin{align*}
    &\left(\frac{lT}{t}\right)^2\Var\left(\E\left[\tauestfe(z_1,z_2)\given \mathbf{R}\right]\right)\\
    &=\Var\left(\sum_{k=1}^K\frac{A_k}{l}X_k(a)+\sum_{\tsetone \in \tset_{z_1}}\frac{B_k\I[\Rk = \tsetone]}{(L-l)/W_{z_1}}X_k(b, \tsetone) -\sum_{\tsetone \in \tset_{z_2}}\frac{C_k\I[\Rk = \tsetone]}{(L-l)/W_{z_2}}X_k(c, \tsetone)\right).
\end{align*}
This is a the variability of a contrast of ``observed'' means from the hypothetical experiment described above, making Theorem 3 of \citet{li2017general} directly applicable.

By Theorem 3 of \citet{li2017general},
\begin{align*}
    &\left(\frac{lT}{t}\right)^2\Var\left(\E\left[\tauestfe(z_1,z_2)\given \mathbf{R}\right]\right)\\
    &=\frac{1}{K-1}\sum_{k=1}^Kl(\ybar_k(z_1)-\ybar_k(z_2) - \ybar(z_1)-\ybar(z_2))^2\\
    &\qquad +\frac{1}{K-1}\sum_{k=1}^K\sum_{\tsetone \in \tset_{z_1}}\frac{(t-1)^2}{t^2}\frac{L-l}{W_{z_1}}\left(\ybar_k(z_1)-\frac{1}{t-1}\sum_{z \in \tsetone \setminus z_1}\ybar_k(z) - \left[\ybar(z_1)-\frac{1}{t-1}\sum_{z \in \tsetone \setminus z_1}\ybar(z)\right]\right)^2\\
    &\qquad +\frac{1}{K-1}\sum_{k=1}^K\sum_{\tsetone \in \tset_{z_2}}\frac{(t-1)^2}{t^2}\frac{L-l}{W_{z_2}}\left(\ybar_k(z_2)-\frac{1}{t-1}\sum_{z \in \tsetone \setminus z_2}\ybar_k(z)-\left[ \ybar(z_2)-\frac{1}{t-1}\sum_{z \in \tsetone \setminus z_2}\ybar(z)\right]\right)^2\\
    &\qquad - \frac{1}{K-1}\frac{1}{K}\sum_{k=1}^K \left(\left[l + (L-l)\frac{t-1}{t}\right](\ybar_k(z_1) -\ybar(z_1) ) - \left[l + (L-l)\frac{t-1}{t}\right](\ybar_k(z_2) -\ybar(z_2)) \right.\\
    &\qquad \left. + \frac{L-l}{W_{z_1}}\frac{1}{t}\sum_{\tsetone \in \tset_{z_1}}\sum_{z \in \tsetone \setminus z_1}(\ybar_k(z) - \ybar(z)) + \frac{L-l}{W_{z_2}}\frac{1}{t}\sum_{\tsetone \in \tset_{z_2}}\sum_{z \in \tsetone \setminus z_2}(\ybar_k(z) - \ybar(z)) \right)^2
    \end{align*}
    Note that there are there are an equal number of $z \notin \z$ across type B and type C blocks so portion of the last term is 0.
    Further, recall definitions
\[S_{bb}^2(\tau(\tilde{z},\tsetone)) = \frac{1}{K-1}\sum_{k=1}^K\left(\overline{Y}_k(\tilde{z}) - \frac{1}{t-1}\sum_{z \in \tsetone\setminus\{\tilde{z}\}}\overline{Y}_k(z) - \left[\overline{Y}(\tilde{z}) - \frac{1}{t-1}\sum_{z \in \tsetone\setminus\{\tilde{z}\}}\overline{Y}(z)\right]\right)^2,\]
\begin{align*}
\bar{S}^2_{bb}(\tau(\tilde{z}, \tsettilde)) &= \frac{1}{W_{\tilde{z}} }\sum_{\tsetone \in \tset_{\tilde{z}}}S_{bb}^2(\tau(\tilde{z},\tsetone)) 
= \sum_{z \in \z^c}\frac{p(z)}{t-1}S^2_{bb}(\tau(\tilde{z}, z)) - \sum_{z \in \z^c }\sum_{z' \in \z^c , z' \neq z}\frac{p_{\tilde{z}}(z, z')}{2(t-1)^2}S^2_{bb}(\tau(z, z')).
\end{align*}

Then we can simplify further,
\begin{align*}
&\left(\frac{lT}{t}\right)^2\Var\left(\E\left[\tauestfe(z_1,z_2)\given \mathbf{R}\right]\right)\\
        &=\frac{1}{K-1}\sum_{k=1}^Kl(\ybar_k(z_1)-\ybar_k(z_2) - \ybar(z_1)-\ybar(z_2))^2 +\frac{(t-1)^2}{t^2}(L-l)\bar{S}^2_{bb}(\tau(z_1, \tsettilde))^2 \\
    &\qquad+\frac{(t-1)^2}{t^2}(L-l)\bar{S}^2_{bb}(\tau(z_2, \tsettilde))^2 - \frac{1}{K-1}\frac{1}{K}\sum_{k=1}^K\frac{l^2T^2}{t^2} \left(\ybar_k(z_1) -\ybar(z_1) - (\ybar_k(z_2) -\ybar(z_2)) \right)^2\\
    &=\frac{(t-1)^2}{t^2}(L-l)\bar{S}^2_{bb}(\tau(z_1, \tsettilde))^2 +\frac{(t-1)^2}{t^2}(L-l)\bar{S}^2_{bb}(\tau(z_2, \tsettilde))^2 + \left(\frac{l^2T^2}{t^2K}-l\right)S_{bb}^2(\tau(z_1,z_2))\\
    &=\frac{(t-1)K}{t}(L-l)\frac{\bar{S}^2_{bb}(\tau(z_1, \tsettilde))^2}{t/(t-1)K} +\frac{(t-1)K}{t}(L-l)\frac{\bar{S}^2_{bb}(\tau(z_2, \tsettilde))^2}{t/(t-1)K} + \frac{1}{K}\frac{lT^2}{t^2}\frac{T-t}{T(t-1)}S_{bb}^2(\tau(z_1,z_2))\\
    &=\frac{l^2T^2}{t^2}\frac{T-t}{T(t-1)}\left[(T-1)\left(\frac{\bar{S}^2_{bb}(\tau(z_1, \tsettilde))^2}{t/(t-1)K} +\frac{\bar{S}^2_{bb}(\tau(z_2, \tsettilde))^2}{t/(t-1)K}\right) +\frac{S_{bb}^2(\tau(z_1,z_2))}{K}\right].
\end{align*}

We provide some simplification of the constants below using the relations $TL=tK$ and $l=\frac{L(t-1)}{T-1}$:
\begin{align*}
l+(L-l)\frac{t-1}{t} & = l+l\frac{T-t}{t} = l\frac{T}{t},\\
\frac{(t-1)K}{t}(L-l) &= \frac{L(T-t)}{T-1}\frac{(t-1)K}{t}= \frac{T-t}{T-1}\frac{LT}{t^2}L(t-1)= L(T-t)\frac{lT}{t^2}= \frac{l^2T^2}{t^2}\frac{(T-t)(T-1)}{T(t-1)}\\
\frac{l^2T^2}{t^2K} - l&=-\frac{1}{K}\frac{l^2}{t^2}\left(\frac{LTt}{l} -T^2\right)=-\frac{1}{K}\frac{l^2T^2}{t^2}\left(\frac{Lt - lT}{lT} \right)=-\frac{1}{K}\frac{l^2T^2}{t^2}\left(\frac{lt(T-1)/(t-1) - lT}{lT} \right)\\
& =-\frac{1}{K}\frac{l^2T^2}{t^2}\left(\frac{t(T-1) - T(t-1)}{T(t-1)} \right)=-\frac{1}{K}\frac{l^2T^2}{t^2}\left(\frac{T-t}{T(t-1)} \right).
\end{align*}

So we have for the variance of the conditional expectation
\begin{align}
&\Var\left(\E\left[\tauestfe(z_1,z_2)\given \mathbf{R}\right]\right) \nonumber\\
    &=\frac{T-t}{T(t-1)}\left[(T-1)\left(\frac{\bar{S}^2_{bb}(\tau(z_1, \tsettilde))^2}{t/(t-1)K} +\frac{\bar{S}^2_{bb}(\tau(z_2, \tsettilde))^2}{t/(t-1)K}\right) +\frac{S_{bb}^2(\tau(z_1,z_2))}{K}\right].\label{eq:fe_var_cond_ex}
\end{align}

The overall variance, adding the variance of the conditional expectation from (\ref{eq:fe_var_cond_ex}) and the expectation of the conditional variance from (\ref{eq:fe_exp_cond_var}), is then
\begin{align*}
&\Var\left(\tauestfe(z_1,z_2)\right) \\
&=\frac{T-t}{T(t-1)}\left[\frac{1}{K}\left\{(T-1)\left(\frac{\bar{S}_{bb}^2(\tau(z_1,\tset_{z_1}))}{t/(t-1)}+\frac{\bar{S}_{bb}^2(\tau(z_2,\tset_{z_2}))}{t/(t-1)}\right) + S_{bb}^2(\tau(z_1,z_2))\right\}\right]\\
&+\frac{T-1}{T(t-1)}\frac{1}{K^2}\sum_{k=1}^K\left[(T-t)\left(\frac{\V{z_1}}{t/(t-1)}+\frac{\V{z_2}}{t/(t-1)}\right)+ t\Vtau\right]
    \end{align*}
        
\end{proof}
\subsection{Expectation of variability over random blocking}\label{append:random_block}
In this section we derive the expectation of variance components if blocking is done at random (i.e., outcomes are uncorrelated with blocks).
Consider if we blocked units into $K$ blocks at random, fixing the block size $n_k$.
Let $B_i(k)$ be the indicator that unit $i$ was assigned to block $k$ ($B_i(k) = \mathbb{I}(b_i=k)$).
In this scenario the $B_i(k)$'s are random variables and we assume the probability of each unit being assigned to a particular block $k$ is $n_k/N$.
Let $\tilde{\ybar}_k(z) = \frac{1}{n_k}\sum_{i=1}^NB_i(k)Y_i(z)$ and $\tilde{\ybar}(z) = \frac{1}{K}\sum_{k=1}^K\tilde{\ybar}_k(z)$.
We have
\[S^2_k(z) = \frac{1}{n_k-1}\sum_{i=1}^NB_i(k)(Y_i(z) - \tilde{\ybar}_k(z))^2\]
and
\[S_{bb}^2(z) = \frac{1}{K-1}\sum_{k=1}^K\left(\tilde{\ybar}_k(z) - \tilde{\ybar}(z)\right)^2. \]

It is useful to note that
\[\E[B_i(k)]= \frac{n_k}{N}, \quad \E[B_i(k)B_j(k)]= \frac{n_k(n_k-1)}{N(N-1)}  \text{ for } j\neq i, \quad E[B_i(k)B_j(h)]= \frac{n_kn_h}{N(N-1)}  \text{ for } j\neq i, h \neq k.\]

We first find the expectation of $S^2_k(z)$ over the random blocking:
\begin{align*}
\E[S^2_k(z)] &=\E\left[ \frac{1}{n_k}\sum_{i=1}^NB_i(k)Y_i(z)^2 - \frac{1}{n_k(n_k-1)}\sum_{i=1}^N\sum_{j\neq i}B_i(k)B_j(k)Y_i(z)Y_j(z)\right]\\
&= \frac{1}{n_k}\sum_{i=1}^N \frac{n_k}{N}Y_i(z)^2 - \frac{1}{n_k(n_k-1)}\sum_{i=1}^N\sum_{j\neq i}\frac{n_k(n_k-1)}{N(N-1)}Y_i(z)Y_j(z)\\
&= \frac{1}{N}\sum_{i=1}^NY_i(z)^2 - \frac{1}{N(N-1)}\sum_{i=1}^N\sum_{j\neq i}Y_i(z)Y_j(z)\\
&= \frac{1}{N-1}\sum_{i=1}^N(Y_i(z) - \ybar(z))^2 \equiv S^2(z).
\end{align*}

Next find the expectation of $S_{bb}^2(z)$  over the random blocking:
\begin{align*}
\E[S_{bb}^2(z)] &= \E\left[\frac{1}{K}\sum_{k=1}^K\tilde{\ybar}_k^2(z) - \frac{1}{K(K-1)}\sum_{k=1}^K\sum_{h\neq k}\tilde{\ybar}_k(z)\tilde{\ybar}_h(z)\right]\\
&= \frac{1}{K}\E\Big[\sum_{k=1}^K\frac{1}{n_k^2}\left(\sum_{i=1}^NB_i(k)Y_i(z)^2 + \sum_{i=1}^N\sum_{j\neq i}B_i(k)B_j(k)Y_i(z)Y_j(z)\right)\\
&\qquad \qquad - \frac{1}{K-1}\sum_{k=1}^K\sum_{h\neq k}\frac{1}{n_k}\frac{1}{n_j}\sum_{i=1}^N\sum_{j \neq i}B_i(k)B_j(h)Y_i(z)Y_j(z)\Big]\\
&= \frac{1}{K}\Bigg[\sum_{k=1}^K\frac{1}{n_k}\left(\sum_{i=1}^N\frac{1}{N}Y_i(z)^2 + \sum_{i=1}^N\sum_{j\neq i}\frac{n_k-1}{N(N-1)}Y_i(z)Y_j(z)\right)\\
&\qquad \qquad - \frac{1}{K-1}\sum_{k=1}^K\sum_{h\neq k}\sum_{i=1}^N\sum_{j \neq i}\frac{1}{n_k}\frac{n_k}{N(N-1)}Y_i(z)Y_j(z)\Bigg]\\
&= \frac{1}{K}\sum_{k=1}^K\frac{1}{n_k}\left(\sum_{i=1}^N\frac{1}{N}Y_i(z)^2 - \frac{1}{N(N-1)}\sum_{i=1}^N\sum_{j\neq i}Y_i(z)Y_j(z)\right)\\
&=\frac{S^2(z)}{K}\sum_{k=1}^K\frac{1}{n_k}.
\end{align*}

So we see that
 \[\frac{1}{K}\sum_{k=1}^K\frac{1}{n_k}\E[S^2_k(z)] = \E[S_{bb}^2(z)].\]

\subsection{Equivalence of adjusted estimator and weighted least square estimator}\label{append:equiv_adj}
We assume the model
\begin{align*}
Y_{izk} = \mu +\tau_z +\beta_k +\epsilon_{izk},\quad \epsilon_{izk}\overset{iid}{\sim}N(0,\sigma^2).
\end{align*}
To get the weighted least square estimator where weights are proportional to the inverse of block sizes, we set the loss function as
\begin{align*}
\mathcal{L} = \sum_{k=1}^K\sum_{z=1}^{T}\frac{\I[z\in\Rk]}{n_k}\sum_{i:Z_i=z,b_i=k}(Y_{izk}-\mu-\tau_z-\beta_k)^2.
\end{align*}
Without loss of generality, we assume $\mu = 0$. When $n_k \equiv t$, it reduces to the classical least square estimator.
By taking the derivative with respect to $\beta_k$, we have
\begin{align}
&\frac{\partial \mathcal{L}}{\partial\beta_k} = 0\nonumber\\
&\iff \sum_{z=1}^{T}\frac{\I[z\in\Rk]}{n_k}\sum_{i:Z_i=z,b_i=k} Y_{izk} = \beta_k\left(\sum_{z=1}^{T}\frac{\I[z\in\Rk]}{n_k}\sum_{i:Z_i=z,b_i=k}1\right)+\sum_{z=1}^{T}\tau_z\left(\frac{\I[z\in\Rk]}{n_k}\sum_{i:Z_i=z,b_i=k}1\right)\label{eq: beta k}.
\end{align}

By taking derivative with respect to $\tau_z$, we have
\begin{align}
&\frac{\partial \mathcal{L}}{\partial\tau_z} = 0\nonumber\\
&\iff \sum_{k=1}^K\frac{\I[z\in\Rk]}{n_k}\sum_{i:Z_i=z,b_i=k}Y_{izk} = \tau_z\left(\sum_{k=1}^K\frac{\I[z\in\Rk]}{n_k}\sum_{i:Z_i=z,b_i=k}1\right)+\sum_{k=1}^K\frac{\I[z\in\Rk]}{n_k}\beta_k\sum_{i:Z_i=z,b_i=k}1 \label{eq: tau z}.
\end{align}

We note that 
\begin{align*}
\sum_{i:Z_i=z,b_i=k}1 = (n_k/t)\I[z\in\Rk],\quad \sum_{z=1}^{T}\frac{\I[z\in\Rk]}{n_k}\sum_{i:Z_i=z,b_i=k}1 =\sum_{z=1}^{T}\frac{(n_k/t)\I[z\in\Rk]}{n_k} = 1.
\end{align*}
Then from \eqref{eq: beta k}, $\beta_k$ can be represented as
\begin{align*}
\beta_k = \sum_{z=1}^T\frac{\I[z\in\Rk]}{n_k}\sum_{i:Z_i=z,b_i=k}Y_{izk} -\frac{1}{t}\sum_{z=1}^{T}\tau_z\I[z\in\Rk].
\end{align*}
By substituting the above expression for $\beta_k$ into \eqref{eq: tau z} we establish
\begin{align*}
 &\sum_{k=1}^K\frac{\I[z\in\Rk]}{n_k}\sum_{i:Z_i=z,b_i=k}Y_{izk}\\
 & = \frac{\tau_z}{t}\sum_{k=1}^K\I[z\in\Rk]+\sum_{k=1}^K\frac{\I[z\in\Rk]}{t}\left[ \sum_{z=1}^T\frac{\I[z\in\Rk]}{n_k}\sum_{i:Z_i=z,b_i=k}Y_{izk} -\frac{1}{t}\sum_{z'=1}^{T}\tau_{z'}\I[z'\in\Rk]\right].
\end{align*}

Because $\sum_{k=1}^K\I[z\in\Rk]=L_z
 $, by multiplying $t$ to both sides, the above equation is equivalent to
 \begin{align*}
 &\sum_{k=1}^K\I[z\in\Rk]\left[\frac{1}{n_k/t}\sum_{i:Z_i=z,b_i=k}Y_{izk}-\frac{1}{t} \sum_{z=1}^T\frac{\I[z\in\Rk]}{n_k/t}\sum_{i:Z_i=z,b_i=k}Y_{izk}\right]=L_z\tau_z -\sum_{z'=1}^{T}\sum_{k=1}^K\frac{\I[z,z'\in\Rk]}{t}\tau_{z'}.
 \end{align*}
 Because 
 \begin{align*}
 &\frac{1}{n_k/t}\sum_{i:Z_i=z,b_i=k}Y_{izk} = \yestk(z), \quad \frac{1}{t} \sum_{z=1}^T\frac{\I[z\in\Rk]}{n_k/t}\sum_{i:Z_i=z,b_i=k}Y_{izk} = \frac{1}{t}\sum_{m=1}^{t}\yestk(R_{k,m}),\\
 &\sum_{z'=1}^{T}\sum_{k=1}^K\frac{\I[z,z'\in\Rk]}{t}\tau_{z'} = \frac{1}{t}\sum_{z'=1}^{T}l_{z,z'}\tau_{z'},\quad\text{with} \quad l_{z,z} = L_z,
 \end{align*}
 we further establish that
 \begin{align*}
\sum_{k=1}^K\I[z\in\Rk]\left(\yestk(z)-\frac{1}{t}\sum_{m=1}^{t}\yestk(R_{k,m})\right) = \frac{L_z(t-1)}{t}\tau_z -\sum_{z'\neq z}\frac{l_{z,z'}}{t}\tau_{z'}.
 \end{align*}
In a BIBD, $L_z \overset{z}{\equiv} L$, $l_{z,z'} \overset{z,z'}{\equiv} l = \frac{L(t-1)}{T-1}$, so this leads to
\begin{align*}
\frac{t}{lT}\yestadj(z) = \frac{T-1}{T}\tau_z -\frac{1}{T}\sum_{z'\neq z}\tau_{z'}
\end{align*}
and for $z \neq z'$,
\begin{align*}
\tau_{z}-\tau_{z'} = \frac{T-1}{T}\tau_z -\frac{1}{T}\sum_{z''\neq z}\tau_{z''} - \left(\frac{T-1}{T}\tau_{z'} -\frac{1}{T}\sum_{z''\neq z'}\tau_{z''}\right) = \frac{t}{lT}
\left(\yestadj(z)-\yestadj(z')\right) = \tauestfe(z,z').
\end{align*}

\subsection{Comparison of design-based estimators to adjusted estimators}\label{append:unadj_fe_comp}
\subsubsection{In general}\label{proof:unadj_fe_comp}
The difference between the adjusted-estimator and design-based estimator, in general, is
\begin{align*}
&\Var(\tauestfe(z_1,z_2))-\Var(\tauestbibd(z_1,z_2))\\
&=\frac{T-t}{Kt}\left[\frac{2Tt-T-t}{T(T-1)(t-1)}\left\{S_{bb}^2(\tau(z_1,z_2))+\frac{1}{K}\sum_{k=1}^KV_k(z_1,z_2)\right\}\right.\\
&+\frac{T-1}{T}\left(\bar{S}_{bb}^2(\tau(z_1,\tset_{z_1}))+\bar{S}_{bb}^2(\tau(z_2,\tset_{z_2}))\right)-\frac{T}{T-1}\left(S_{bb}^2(z_1)+S_{bb}^2(z_2)\right)\\
&+\left.\frac{t-1}{K}\sum_{k=1}^K\left\{\frac{T-1}{T}\left(\frac{\V{z_1}+\V{z_2}}{t-1}\right)-\frac{T}{T-1}\left(\frac{S_k^2(z_1)}{n_k}+\frac{S_k^2(z_2)}{n_k}\right)\right\}\right].
\end{align*}

\begin{proof}
Recall that 
\begin{align*}
\Var(\tauestbibd(z_1,z_2))&=\frac{T-t}{t(T-1)}\left(\frac{S_{bb}^2(z_1)}{K/T}+\frac{S_{bb}^2(z_2)}{K/T}-\frac{S_{bb}^2(\tau(z_1,z_2))}{K}\right)\\
&+\frac{T(t-1)}{t(T-1)}\left[\frac{1}{K^2}\sum_{k=1}^K\left(\frac{S_k^2(z_1)}{n_k/T}+\frac{S_k^2(z_2)}{n_k/T}-\frac{S_k^2(\tau(z_1,z_2))}{n_k}\right)\right]
\end{align*}
and
 \begin{align*}
&\Var(\tauestfe(z_1,z_2))\\
&=\frac{T-t}{T(t-1)}\left[\frac{1}{K}\left\{S_{bb}^2(\tau(z_1,z_2))+(T-1)\left(\frac{\bar{S}_{bb}^2(\tau(z_1,\tset_{z_1}))}{t/(t-1)}+\frac{\bar{S}_{bb}^2(\tau(z_2,\tset_{z_2}))}{t/(t-1)}\right)\right\}\right]\\
&+\frac{T-1}{T(t-1)}\left[\frac{1}{K^2}\sum_{k=1}^K\left\{t\Vtau+(T-t)\left(\frac{\V{z_1}}{t/(t-1)}+\frac{\V{z_2}}{t/(t-1)}\right)\right\}\right].
    \end{align*}
We first compare the between-block variability part:
\begin{align*}
    &\frac{T-t}{T(t-1)}\left[\frac{1}{K}\left\{S_{bb}^2(\tau(z_1,z_2))+(T-1)\left(\frac{\bar{S}_{bb}^2(\tau(z_1,\tset_{z_1}))}{t/(t-1)}+\frac{\bar{S}_{bb}^2(\tau(z_2,\tset_{z_2}))}{t/(t-1)}\right)\right\}\right]\\
    &-\frac{T-t}{t(T-1)}\left(\frac{S_{bb}^2(z_1)}{K/T}+\frac{S_{bb}^2(z_2)}{K/T}-\frac{S_{bb}^2(\tau(z_1,z_2))}{K}\right)\\
    &=\frac{T-t}{K}\left[\frac{1}{T(t-1)}\left\{S_{bb}^2(\tau(z_1,z_2))+(T-1)\left(\frac{\bar{S}_{bb}^2(\tau(z_1,\tset_{z_1}))}{t/(t-1)}+\frac{\bar{S}_{bb}^2(\tau(z_2,\tset_{z_2}))}{t/(t-1)}\right)\right\}\right.\\
    &\left.-\frac{1}{t(T-1)}\left(\frac{S_{bb}^2(z_1)}{1/T}+\frac{S_{bb}^2(z_2)}{1/T}-S_{bb}^2(\tau(z_1,z_2))\right)\right]\\
    &=\frac{T-t}{K}\left[\left(\frac{1}{T(t-1)}+\frac{1}{t(T-1)}\right)S_{bb}^2(\tau(z_1,z_2))+\frac{T-1}{Tt}\left(\bar{S}_{bb}^2(\tau(z_1,\tset_{z_1}))+\bar{S}_{bb}^2(\tau(z_2,\tset_{z_2}))\right)\right.\\
    &-\left.\frac{T}{t(T-1)}\left(S_{bb}^2(z_1)+S_{bb}^2(z_2)\right)\right]\\
    &=\frac{T-t}{Kt}\left[\left(\frac{t}{T(t-1)}+\frac{1}{T-1}\right)S_{bb}^2(\tau(z_1,z_2))+\frac{T-1}{T}\left(\bar{S}_{bb}^2(\tau(z_1,\tset_{z_1}))+\bar{S}_{bb}^2(\tau(z_2,\tset_{z_2}))\right)\right.\\
    &-\left.\frac{T}{T-1}\left(S_{bb}^2(z_1)+S_{bb}^2(z_2)\right)\right].
\end{align*}
Since 
\begin{align*}
    \frac{t}{T(t-1)}+\frac{1}{T-1} = \frac{t(T-1)+T(t-1)}{T(T-1)(t-1)} = \frac{2Tt-T-t}{T(T-1)(t-1)},
\end{align*}
the difference of between-block part is
\begin{align}
    &\frac{T-t}{Kt}\left[\frac{2Tt-T-t}{T(T-1)(t-1)}S_{bb}^2(\tau(z_1,z_2))+\frac{T-1}{T}\left(\bar{S}_{bb}^2(\tau(z_1,\tset_{z_1}))+\bar{S}_{bb}^2(\tau(z_2,\tset_{z_2}))\right)-\frac{T}{T-1}\left(S_{bb}^2(z_1)+S_{bb}^2(z_2)\right)\right].\label{eq: bb}
\end{align}
Now we turn to the within-block part:
\begin{align*}
    &\frac{T-1}{T(t-1)}\left[\frac{1}{K^2}\sum_{k=1}^K\left\{t\Vtau+(T-t)\left(\frac{\V{z_1}}{t/(t-1)}+\frac{\V{z_2}}{t/(t-1)}\right)\right\}\right]\\
    &-\frac{T(t-1)}{t(T-1)}\left[\frac{1}{K^2}\sum_{k=1}^K\left(\frac{S_k^2(z_1)}{n_k/T}+\frac{S_k^2(z_2)}{n_k/T}-\frac{S_k^2(\tau(z_1,z_2))}{n_k}\right)\right]\\
    &=\frac{T-1}{T(t-1)}\left[\frac{1}{K^2}\sum_{k=1}^K\left\{t\Vtau+(T-t)\left(\frac{\V{z_1}}{t/(t-1)}+\frac{\V{z_2}}{t/(t-1)}\right)\right\}\right]\\
    &-\frac{T(t-1)}{t(T-1)}\left[\frac{1}{K^2}\sum_{k=1}^K\left(V_k(z_1,z_2)+\frac{S_k^2(z_1)}{n_k/(T-t)}+\frac{S_k^2(z_2)}{n_k/(T-t)}\right)\right]\\
    &=\left(\frac{t(T-1)}{T(t-1)}-\frac{T(t-1)}{t(T-1)}\right)\frac{1}{K^2}\sum_{k=1}^KV_k(z_1,z_2)\\
    &+\frac{T-t}{K^2t}\sum_{k=1}^K\left[\frac{T-1}{T}\left(\V{z_1}+\V{z_2}\right)-\frac{T(t-1)}{T-1}\left(\frac{S_k^2(z_1)}{n_k}+\frac{S_k^2(z_2)}{n_k}\right)\right].
\end{align*}
Since 
\begin{align*}
    \frac{t(T-1)}{T(t-1)}-\frac{T(t-1)}{t(T-1)} = \frac{(t(T-1)+T(t-1))(T-t)}{t(T-1)T(t-1)}=\left(\frac{T-t}{t}\right)\frac{2Tt-T-t}{T(T-1)(t-1)},
\end{align*}
and
\begin{align*}
    &\frac{T-1}{T}\left(\V{z_1}+\V{z_2}\right)-\frac{T(t-1)}{T-1}\left(\frac{S_k^2(z_1)}{n_k}+\frac{S_k^2(z_2)}{n_k}\right)\\
    &=(t-1)\left[\frac{T-1}{T}\left(\frac{\V{z_1}+\V{z_2}}{t-1}\right)-\frac{T}{T-1}\left(\frac{S_k^2(z_1)}{n_k}+\frac{S_k^2(z_2)}{n_k}\right)\right],
\end{align*}
we establish
\begin{align}
    &\left(\frac{t(T-1)}{T(t-1)}-\frac{T(t-1)}{t(T-1)}\right)\frac{1}{K^2}\sum_{k=1}^KV_k(z_1,z_2)\nonumber\\
    &+\frac{T-t}{K^2t}\sum_{k=1}^K\left[\frac{T-1}{T}\left(\V{z_1}+\V{z_2}\right)-\frac{T(t-1)}{T-1}\left(\frac{S_k^2(z_1)}{n_k}+\frac{S_k^2(z_2)}{n_k}\right)\right]\nonumber\\
    &=\frac{T-t}{K^2t}\sum_{k=1}^K\left[\left(\frac{2Tt-T-t}{(T-1)T(t-1)}\right)V_k(z_1,z_2)\right.\nonumber\\
    &+\left.(t-1)\left\{\frac{T-1}{T}\left(\frac{\V{z_1}+\V{z_2}}{t-1}\right)-\frac{T}{T-1}\left(\frac{S_k^2(z_1)}{n_k}+\frac{S_k^2(z_2)}{n_k}\right)\right\}\right].\label{eq: wb}
\end{align}
Combining \eqref{eq: bb} and \eqref{eq: wb}, we have
\begin{align*}
    &\Var(\tauestfe(z_1,z_2))-\Var(\tauestbibd(z_1,z_2))\\
    &=\frac{T-t}{Kt}\left[\frac{2Tt-T-t}{T(T-1)(t-1)}\left(S_{bb}^2(\tau(z_1,z_2))+\frac{1}{K}\sum_{k=1}^KV_k(z_1,z_2)\right)\right.\\
    &+\frac{T-1}{T}\left(\bar{S}_{bb}^2(\tau(z_1,\tset_{z_1}))+\bar{S}_{bb}^2(\tau(z_2,\tset_{z_2}))\right)-\frac{T}{T-1}\left(S_{bb}^2(z_1)+S_{bb}^2(z_2)\right)\\
    &+\left.\frac{t-1}{K}\sum_{k=1}^K\left\{\frac{T-1}{T}\left(\frac{\V{z_1}+\V{z_2}}{t-1}\right)-\frac{T}{T-1}\left(\frac{S_k^2(z_1)}{n_k}+\frac{S_k^2(z_2)}{n_k}\right)\right\}\right].
\end{align*}
\end{proof}
We remark that 
\begin{align*}
    \V{\tilde{z}} = \frac{1}{W_{\tilde{z}}} \sum_{\tsetone \in \tsettilde} \Var(\tauestk(\tilde{z}, \tsetone)\mid\Rk = \tsetone)=\frac{S^2_k(\tilde{z})}{n_k/t} + \frac{1}{t-1}\frac{\bar{S}^2_k(\tsettilde)}{n_k/t} -\frac{\bar{S}^2_k(\tau(\tilde{z}, \tsettilde))}{n_k}.
\end{align*}
This leads to
\begin{align*}
   &n_k\left[ \frac{T-1}{T(t-1)}\V{z_1}-\frac{TS_k^2(z_1)}{n_k(T-1)}\right] \\
   &=  \frac{T-1}{T(t-1)}\left[\frac{S^2_k(z_1)}{1/t} + \frac{1}{t-1}\frac{\bar{S}^2_k(\tset_{z_1})}{1/t} -\bar{S}^2_k(\tau(z_1, \tset_{z_1}))\right]-\frac{TS_k^2(z_1)}{T-1}\\
   &=\left(\frac{t(T-1)}{T(t-1)}-\frac{T}{T-1}\right)S_k^2(z_1)+\left(\frac{t(T-1)}{T(t-1)^2}\right)\bar{S}^2_k(\tset_{z_1})-\frac{T-1}{T(t-1)}\bar{S}^2_k(\tau(z_1, \tset_{z_1})).
\end{align*}
If $S_k^2(z)=S_k^2$ for all $z$ and $S_k^2(\tau(z,z'))=0$ for all $z,z'$, the above term reduces as
\begin{align*}
    &\left(\frac{t(T-1)}{T(t-1)}-\frac{T}{T-1}\right)S_k^2+\left(\frac{t(T-1)}{T(t-1)^2}\right)S^2_k \\
    &= \left(\frac{t^2(T-1)}{T(t-1)^2}-\frac{T}{T-1}\right)S_k^2 = \frac{T}{T-1}\left(\frac{t^2(T-1)^2}{T^2(t-1)^2}-1\right)S_k^2 >0,
\end{align*}
since $t(T-1)>T(t-1)$. Since $V_k(z_1,z_2)\geq 0$, the within-block part is nonnegative if $S_k^2(z)=S_k^2$ for all $z$, which holds if $Y_i(z)-Y_i(z')$ is constant for all $z,z'$, and $S_k^2(\tau(z,z'))=0$ for all $z,z'$.

\subsubsection{Under a linear form}
We first consider the case where the potential outcomes follow the additive linear fixed-effects relationship given in Section~\ref{subsec:fixed and unadj}.
As a reminder, the model is
\begin{align*}
    Y_i(z) = \alpha_z +\beta_k +\epsilon_{zi},
\end{align*}
for unit $i$ with $b_i=k$, and we take $\sum_{i:b_i=k}\epsilon_{zi}=0$.
Recall this implies
\begin{align*}
    \ybark(z) = \alpha_z +\beta_k,\quad \ybar(z) = \alpha_z +\bar{\beta}, \text{ with } \bar{\beta} = \frac{1}{K}\sum_{k=1}^K\beta_k.
\end{align*}
Thus, for $z\neq z'$,
\begin{align*}
    &S_k^2(z) = \frac{1}{n_k-1}\sum_{i:b_i=k}\epsilon_{zi}^2,\quad S_k^2(\tau(z,z')) = \frac{1}{n_k-1}\sum_{i:b_i=k}(\epsilon_{zi}-\epsilon_{z'i})^2,\\
    &S_{bb}^2(z) = \frac{1}{K-1}\sum_{k=1}^K(\beta_k-\bar{\beta})^2\equiv S_{bb}^2,\quad S_{bb}^2(\tau(z,z')) = 0.
\end{align*}

First, consider again the assumption that potential outcomes are independent in the sense that $\sum_{i}\epsilon_{zi}\epsilon_{z'i} = 0$, so that $S_k^2(\tau(z,z')) = 2S_k^2$ for all $z\neq z'$.
We compare the precision of $\tauestbibd(z_1,z_2)$ and $\tauestfe(z_1,z_2)$ in terms of $\Var(\tauestclus(z_1,z_2))$ and $\Var(\tauestcbd(z_1,z_2))$, which correspond to the between-block and within-block variability, respectively, for $\tauestbibd(z_1,z_2)$.
Under the same potential outcomes model assumptions above independent potential outcomes, $\Var(\tauestclus(z_1,z_2)) = \Var(\tauestcbd(z_1,z_2))$ when
\begin{align*}
S_{bb}^2 = \frac{T-1+\rho}{T}\left(\frac{1}{K}\sum_{k=1}^K\frac{S_k^2}{n_k}\right).
\end{align*}
When $\Var(\tauestclus(z_1,z_2)) = \Var(\tauestcbd(z_1,z_2))$, 
\begin{align*}
\Var(\tauestbibd(z_1,z_2))-\Var(\tauestfe(z_1,z_2)) = \frac{2(T-t)\rho}{t-1}\left(\frac{1}{K^2}\sum_{k=1}^K\frac{S_k^2}{n_k}\right).
\end{align*}
Thus, $\Var(\tauestbibd(z_1,z_2))\geq\Var(\tauestfe(z_1,z_2))$ when $\Var(\tauestclus(z_1,z_2)) = \Var(\tauestcbd(z_1,z_2))$, with the equality holding when $\rho=0$.
We see this exact result in Figure~\ref{fig: se_comp_est_S1}.

Alternatively, in this model, without the assumption of independent potential outcomes, the between block component of $\Var(\tauestfe(z_1,z_2))$ vanishes. However, the between block component of $\Var(\tauestbibd(z_1,z_2))$ is
\begin{align*}
    \frac{T-t}{t(T-1)}\frac{2S_{bb}^2}{K/T}=\frac{T-t}{t(T-1)}\Var(\tauestclus(z_1,z_2)).
\end{align*}
If we assume $S_k^2(z)\equiv S_k^2$ for all $z$ and 
  $S_k^2(\tau(z,z')) \equiv 2S_k^2(1-\rho)$ for some constant $\rho \in [0,1]$ for all $z\neq z'$, the within block component of $\Var(\tauestbibd(z_1,z_2))$ is
\begin{align*}
    \frac{T(t-1)}{t(T-1)}\left[\frac{1}{K^2}
    \sum_{k=1}^K\left(\frac{2TS_k^2}{n_k}-\frac{2S_k^2(1-\rho)}{n_k}\right)\right] = \frac{T(t-1)}{t(T-1)}\frac{2(T-1+\rho)}{K^2}\sum_{k=1}^K\frac{S_k^2}{n_k}= \frac{T(t-1)}{t(T-1)}\Var(\tauestcbd(z_1,z_2)).
\end{align*}

Under the same assumption, 
\begin{align*}
   \V{\tilde{z}} = \frac{t^2}{t-1}\frac{S_k^2}{n_k}-\frac{t}{2(t-1)}\frac{2S_k^2(1-\rho)}{n_k} = \frac{t(t-1+\rho)}{t-1}\frac{S_k^2}{n_k},
\end{align*}
so the within block component of $\Var(\tauestfe(z_1,z_2))$ is
\begin{align*}
    &\frac{t(T-1)}{T(t-1)}\left[\frac{1}{K^2}\sum_{k=1}^K\left(\frac{2tS_k^2}{n_k}-\frac{2S_k^2(1-\rho)}{n_k}\right)\right]+\frac{2(T-1)(T-t)}{T(t-1)}\left[\frac{t-1+\rho}{K^2}\sum_{k=1}^K\frac{S_k^2}{n_k}\right]\\
    &=\left(\frac{T-1}{t-1}\right)\frac{2(t-1+\rho)}{K^2}
    \sum_{k=1}^K\frac{S_k^2}{n_k}=\frac{(T-1)(t-1+\rho)}{(t-1)(T-1+\rho)}\Var(\tauestcbd(z_1,z_2)).
\end{align*}
Thus, $\Var(\tauestbibd(z_1,z_2)) \geq\Var(\tauestfe(z_1,z_2))$ holds when
\begin{align*}
\frac{T-t}{t(T-1)}\frac{TS_{bb}^2}{K}+\frac{T(t-1)}{t(T-1)}\frac{T-1+\rho}{K^2}\sum_{k=1}^K\frac{S_k^2}{n_k}-\frac{T-1}{t-1}\frac{t-1+\rho}{K^2}\sum_{k=1}^K\frac{S_k^2}{n_k}\geq0\\
\iff \frac{T(T-t)}{t(T-1)}S_{bb}^2\geq\frac{1}{K}\sum_{k=1}^K\frac{S_k^2}{n_k}\left[\frac{T-1(t-1+\rho)}{t-1}-\frac{T(t-1)(T-1+\rho)}{t(T-1)}\right].
    \end{align*}
    Because
    \begin{align*}
   & \frac{T-1(t-1+\rho)}{t-1}-\frac{T(t-1)(T-1+\rho)}{t(T-1)} = T-1-\frac{T(t-1)}{t}+\rho\left[\frac{T-1}{t-1}-\frac{T(t-1)}{t(T-1)}\right]\\
    &=\frac{T-t}{t}+\rho\left[\frac{t(T-1)^2-T(t-1)^2}{t(t-1)(T-1)}\right]=\frac{T-t}{t}+\rho\left[\frac{(Tt-1)(T-t)}{t(t-1)(T-1)}\right]=\frac{T-t}{t}\left[1+\rho\frac{Tt-1}{(t-1)(T-1)}\right],
    \end{align*}
    the above inequality can be reduced as
    \begin{align*}
    \frac{T}{T-1}S_{bb}^2\geq\frac{1}{K}\sum_{k=1}^K\frac{S_k^2}{n_k}\left[1+\frac{\rho(Tt-1)}{(t-1)(T-1)}\right]
    \iff \frac{\Var(\tauestclus(z_1,z_2))}{\Var(\tauestcbd(z_1,z_2))}\geq \frac{(T-1)(t-1)+\rho(Tt-1)}{(T-1)(t-1)+\rho(t-1)}.
    \end{align*}
    So, we can say that  $\Var(\tauestbibd(z_1,z_2)) \geq\Var(\tauestfe(z_1,z_2))$ holds when
    \begin{align*}
    S_{bb}^2\geq \frac{c(\rho;T,t)}{K}\sum_{k=1}^K\frac{S_k^2}{n_k}  \quad \text{with}\quad c(\rho;T,t) = \frac{T-1}{T}\left(1+\frac{\rho}{t-1}\right)+\rho.
    \end{align*}

We remark that $c(\rho;T,t)$ is an increasing function of $\rho$ and $c(\rho;T,t) \in \left[\frac{T-1}{T},\frac{t(T-1)}{T(t-1)}+1\right]$.
Also, $c(\rho;T,t) = 1$ when $\rho = (t-1)/(Tt-1)$. When $c(\rho;T,t) = 1$, we can directly compare the precision of $\tauestbibd(z_1,z_2)$ and $\tauestfe(z_1,z_2)$ with the ratio of between block variance to within block variance. In particular, when random blocking occurs, then $S_{bb}^2\approx \frac{1}{K}\sum_{k=1}^K\frac{S_k^2}{n_k}$ as we have shown, so we expect that the precision of two estimators will be similar. However, when $c(\rho;T,t) \neq 1$, 
unlike the comparisons discussed in Section~\ref{subsec:fixed and unadj}, this means that even if random blocking occurs, the relationship between the precision of the unadjusted estimator and the adjusted estimator can change depending on the correlation of potential outcomes, $\rho$.
For example, $c(\rho;T,t) = \frac{T-1}{T}$ when $\rho=0$, so we should expect that random blocking would lead to higher precision of the adjusted estimator compared to the unadjusted estimator.

We can also consider model misspecification of the adjusted estimator.
For example, we can assume of the following interactive fixed-effects model:
\begin{align*}
    Y_i(z) = \alpha_z +\beta_k + \gamma_z\delta_k+ \epsilon_{zi},
\end{align*}
for unit $i$ with $b_i=k$.
We make the same assumptions on the correlation of potential outcomes as before.
Then, for $z\neq z'$, $S_k^2(z)$ and $S_k^2(\tau(z,z'))$ are the same as above, but
\begin{align*}
    &S_{bb}^2(z) = \frac{1}{K-1}\sum_{k=1}^K(\beta_k-\bar{\beta}+\gamma_z(\delta_k-\bar{\delta}))^2,\quad S_{bb}^2(\tau(z,z')) = (\gamma_z-\gamma_{z'})^2\left(\frac{1}{K-1}\sum_{k=1}^K(\delta_k-\bar{\delta})^2\right).
\end{align*}
As discussed in the main text, this means that the between block component of the variance of the adjusted estimator increases because the $S_{bb}^2(\tau(z,z'))$ terms are no longer zero under this assumption.
In this model,
\begin{align*}
    &\Var(\tauestfe(z_1,z_2))=\frac{(T-1)(t-1+\rho)}{(t-1)(T-1+\rho)}\Var(\tauestcbd(z_1,z_2))\\
    &+\frac{T-t}{T(t-1)}\left[\frac{\Var(\delta)}{K}\left((\gamma_{z_1}-\gamma_{z_2})^2+\frac{(t-1)(T-1)}{t}\left\{\sum_{z\in \z^c}\frac{p(z)}{t-1}\left((\gamma_{z_1}-\gamma_{z})^2+(\gamma_{z_2}-\gamma_{z})^2\right)\right.\right.\right.\\
    &\left.\left.\left.-\sum_{z\in \z^c}\sum_{z'\in \z^c,z'\neq z}\frac{p_{z_1}(z,z')+p_{z_2}(z,z')}{2(t-1)^2}(\gamma_{z}-\gamma_{z'})^2\right\}\right)\right],
\end{align*}
where $\Var(\delta) = \frac{1}{K-1}\sum_{k=1}^K(\delta_k-\bar{\delta})^2$. Thus, $\Var(\tauestfe(z_1,z_2))$ does not depend on $\{\beta_k\}_{k=1}^K$.
\subsection{Proof of Theorem~\ref{theorem:finite_bias_ht}}\label{pf:finite_bias_ht}
\begin{proof}
We start with the conditional expectation of $\sHT(z)$,\begin{align*}
&(L_z-1)\E\left[\sHT(z)\given\bfR\right]\\
&= \E\left[\sum_{k=1}^K\I[z \in \Rk]\left(Kw_k\yestk(z)-\yht(z)\right)^2\given\bfR\right]\\
&=\sum_{k=1}^K\I[z \in \Rk]\left[\Var\left(Kw_k\yestk(z)\given\bfR\right)+\left(\E\left[Kw_k\yestk(z)\given\bfR\right]\right)^2\right]\\
&+\E\left[\sum_{k=1}^K\I[z \in \Rk]\yht(z)\left(\yht(z)-2Kw_k\yestk(z)\right)\given\bfR\right]\\
&=\sum_{k=1}^K\I[z \in \Rk]\left(\frac{(Kw_k)^2S_k^2(z)}{n_k/(t-1)}+(Kw_k)^2\ybar_k(z)^2\right)-L_z\E\left[\yht(z)^2\given\bfR\right]\\
&=\sum_{k=1}^K\I[z \in \Rk]\left(\frac{(Kw_k)^2S_k^2(z)}{n_k/(t-1)}+(Kw_k)^2\ybar_k(z)^2\right)\\
&-\frac{1}{L_z}\left[\sum_{k=1}^K\I[z\in \Rk]\frac{(Kw_k)^2S_k^2(z)}{n_k/(t-1)}+\left(\sum_{k=1}^K\I[z \in \Rk]Kw_k\ybar_k(z)\right)^2\right]\\
&=\sum_{k=1}^K\I[z \in \Rk]\frac{L_z-1}{L_z}\frac{(Kw_k)^2S_k^2(z)}{n_k/(t-1)}+\sum_{k=1}^K\I[z\in \Rk](Kw_k)^2\ybar_k(z)^2 - \frac{1}{L_z}\left(\sum_{k=1}^K\I[z \in \Rk]Kw_k\ybar_k(z)\right)^2.
\end{align*}
By the law of iterated expectation, taking the expectation over $\bfR$,
\begin{align*}
\E[\sHT(z)]&=\frac{1}{K}\sum_{k=1}^K\frac{(Kw_k)^2S_k^2(z)}{n_k/(t-1)}+\frac{1}{K}\sum_{k=1}^K(Kw_k)^2\ybar_k(z)^2-\frac{1}{K(K-1)}\sum_{k=1}^K\sum_{j\neq k}K^2w_k\ybar_k(z)w_j\ybar_j(z)\\
&=\frac{1}{K}\sum_{k=1}^K\frac{(Kw_k)^2S_k^2(z)}{n_k/(t-1)} + \SHT(z).
\end{align*}

Similarly, provided that $l_{z,z'}\geq 2$,
\begin{align*}
\E\left[s_{bb}^2(\tau(z,z'))\right]=\frac{1}{K}\sum_{k=1}^K(Kw_k)^2\left(\frac{S_k^2(z)}{n_k/t} + \frac{S_k^2(z')}{n_k/t}-\frac{S_k^2(\tau(z,z;))}{n_k}\right) + \SHT(\tau(z,z')).
\end{align*}

For $s_k^2(z)$, conditional on $z \in \Rk$, this is a standard variance estimator with known expectation, $\E\left[s_k^2(z)|z \in \Rk\right] =S_k^2(z)$.
Then it's easy to see that $\E[\I[z \in \Rk]s_k^2(z)]=\frac{L_z}{K}S_k^2(z)$.

To summarize, we have the following results for the expectation of the sample estimators:
\begin{align}
    &\E [\sHT(z)] = \SHT(z) + \frac{1}{K}\sum_{k=1}^{K}\frac{(Kw_k)^2S_k^2(z)}{n_k/(t-1)},\label{eq:exp of sht}\\
    &\E [\sHT(\tau(z,z'))] = \SHT(\tau(z,z')) + \frac{1}{K}\sum_{k=1}^{K}(Kw_k)^2\left(\frac{S_k^2(z)}{n_k/t}+\frac{S_k^2(z')}{n_k/t}-\frac{S_k^2(\tau(z,z'))}{n_k}\right),\label{eq:exp of sht tau}\\
    &\E [\I[z \in \Rk]s_{k}^2(z) ]= \frac{L_z}{K}S_k^2(z).\label{eq:exp of sk}
\end{align}

For $z\neq z'$, the expectation of $\Shatbbht(z,z')$ is
\begin{align*}
    &\E[\Shatbbht(z,z')] = \E\left[\frac{l_{z,z'}\I[l_{z,z'}\geq 2]}{2L_zL_{z'}}(\sHT(z)+\sHT(z')-\sHT(\tau(z,z')))\right]\\
    &= \frac{l_{z,z'}\I[l_{z,z'}\geq 2]}{2L_zL_{z'}}\left[ \SHT(z)+ \SHT(z')-\SHT(\tau(z,z'))\right]\\
    &-\frac{l_{z,z'}\I[l_{z,z'}\geq 2]}{2KL_zL_{z'}}\sum_{k=1}^K\frac{(Kw_k)^2}{n_k}\left[S_k^2(z)+S_k^2(z')-S_k^2(\tau(z,z'))\right].
\end{align*}
Note that
\begin{align*}
    \bm{S}_{\HT}(z,z')& = \frac{1}{K}(\bht(z,z')+\bmw(z,z'))\\
    &= \frac{1}{2}\left[\left(\frac{l_{z,z'}}{L_z L_{z'}}-\frac{1}{K}\right)\left(S^2_{\HT}(z)+S^2_{\HT}(z')-S^2_{\HT}(\tau(z,z'))\right)\right]\\
    &-\frac{l_{z,z'}}{2L_z L_{z'}}\left[\frac{1}{K}\sum_{k=1}^K(Kw_k)^2\left(\frac{S_k^2(z)}{n_k}+\frac{S_k^2(z')}{n_k}-\frac{S_k^2(\tau(z,z'))}{n_k}\right)\right].
    \end{align*}
Thus,
\begin{align*}
    &\E[\Shatbbht(z,z')]-\bm{S}_{\HT}(z,z')\\
    &= \frac{1}{2K}\left(S^2_{\HT}(z)+S^2_{\HT}(z')-S^2_{\HT}(\tau(z,z'))\right)\\
    &-\frac{\I[l_{z,z'}<2]l_{z,z'}}{2L_zL_{z'}}\left[S^2_{\HT}(z)+S^2_{\HT}(z')-S^2_{\HT}(\tau(z,z'))-\frac{1}{K}\sum_{k=1}^K(Kw_k)^2\left(\frac{S_k^2(z)}{n_k}+\frac{S_k^2(z')}{n_k}-\frac{S_k^2(\tau(z,z'))}{n_k}\right)\right].
\end{align*}

For the diagonal element, $l_{z,z}=L_z\geq2$ so the expectation of $\Shatbbht(z,z)$ is
\begin{align*}
\E[\Shatbbht(z,z)]=\frac{1}{L_z}\E[\SHT(z)]&= \frac{1}{L_z}\left( \SHT(z) + \frac{1}{K}\sum_{k=1}^{K}\frac{(Kw_k)^2S_k^2(z)}{n_k/(t-1)}\right)\\
&= \frac{1}{K}\left(\bht(z,z)+\bmw(z,z)\right)+\frac{1}{K}S^2_{\HT}(z).
\end{align*}
Thus,
\begin{align*}
    &\E[\Shatbbht(z,z)]-\bm{S}_{\HT}(z,z) = \frac{1}{K}\SHT(z)
\end{align*}

By combining the above, we have
\begin{align*}
    &\bm{g}\tr\left(\E[\Shatbbht]-\bm{S}_{\HT}\right)\bm{g} - \frac{\SHT(\tauwg)}{K}\\
    &=\sum_{z,z':l_{z,z'}\leq1}\frac{g_zg_{z'}l_{z,z'}}{2L_{z}L_{z'}}\left[\frac{1}{K}\sum_{k=1}^K(Kw_k)^2\left(\frac{S_k^2(z)}{n_k}+\frac{S_k^2(z')}{n_k}-\frac{S_k^2(\tau(z,z'))}{n_k}\right)\right]\\
    &-\sum_{z,z':l_{z,z'}\leq1}\frac{g_zg_{z'}l_{z,z'}}{2L_{z}L_{z'}}\left[S^2_{\HT}(z)+S^2_{\HT}(z')-S^2_{\HT}(\tau(z,z'))\right]
\end{align*}

Now we turn to the bias of $\bm{g}\tr\Shatwbht(z,z')\bm{g}$.
Recall that
\begin{align*}
     \Shatwbht(z,z')=\I(l_{z,z'}\geq2)\left(1-\frac{L_zL_{z'}}{Kl_{z,z'}}\right)\Shatbbht(z,z') +\frac{\I[z=z']}{p_z}\left(\frac{1}{K^2}\sum_{k=1}^K\frac{\I[z\in\Rk](Kw_k)^2s_k^2(z)}{n_k/t}\right)
\end{align*}
In a similar way to $\Shatbbht(z,z')$, we divide into two cases: (i) $z\neq z'$; (ii) $z =z'$. For $z\neq z'$, the expectation of $\Shatwbht(z,z')$ is
\begin{align*}
    &\E[\Shatwbht(z,z')] = \E\left[ \left(1-\frac{L_zL_{z'}}{Kl_{z,z'}}\right)\frac{l_{z,z'}\I[l_{z,z'}\geq 2]}{2L_zL_{z'}}(\sHT(z)+\sHT(z')-\sHT(\tau(z,z'))) \right]\\
    &= \I[l_{z,z'}\geq 2]\left(\frac{l_{z,z'}}{2L_zL_{z'}}-\frac{1}{2K}\right)\left[ \SHT(z)+ \SHT(z')-\SHT(\tau(z,z'))\right]\\
    &-\I[l_{z,z'}\geq 2]\left(\frac{l_{z,z'}}{2L_zL_{z'}}-\frac{1}{2K}\right)\sum_{k=1}^K\frac{(Kw_k)^2}{n_k}\left[S_k^2(z)+S_k^2(z')-S_k^2(\tau(z,z'))\right]\\
    &=\bm{S}_{\HT}(z,z')+\frac{1}{2K}\left(S^2_{\HT}(z)+S^2_{\HT}(z')-S^2_{\HT}(\tau(z,z'))\right)\\
    &-\frac{\I[l_{z,z'}<2]l_{z,z'}}{2L_zL_{z'}}\left[S^2_{\HT}(z)+S^2_{\HT}(z')-S^2_{\HT}(\tau(z,z'))-\frac{1}{K}\sum_{k=1}^K(Kw_k)^2\left(\frac{S_k^2(z)}{n_k}+\frac{S_k^2(z')}{n_k}-\frac{S_k^2(\tau(z,z'))}{n_k}\right)\right]\\
    &-\frac{\I[l_{z,z'}\geq 2]}{2K}\left[S^2_{\HT}(z)+S^2_{\HT}(z')-S^2_{\HT}(\tau(z,z'))-\frac{1}{K}\sum_{k=1}^K(Kw_k)^2\left(\frac{S_k^2(z)}{n_k}+\frac{S_k^2(z')}{n_k}-\frac{S_k^2(\tau(z,z'))}{n_k}\right)\right]\\
    &=\bm{S}_{\HT}(z,z')+\frac{1}{2K^2}\sum_{k=1}^K(Kw_k)^2\left(\frac{S_k^2(z)}{n_k}+\frac{S_k^2(z')}{n_k}-\frac{S_k^2(\tau(z,z'))}{n_k}\right)\\
    &+\frac{\I[l_{z,z'}<2]}{2}\left(\frac{1}{K}-\frac{l_{z,z'}}{L_zL_{z'}}\right)\left(S^2_{\HT}(z)+S^2_{\HT}(z')-S^2_{\HT}(\tau(z,z'))\right)\\
    &+\I[l_{z,z'}<2]\left(\frac{l_{z,z'}}{L_zL_{z'}}-\frac{1}{K}\right)\frac{1}{2K}\sum_{k=1}^K(Kw_k)^2\left(\frac{S_k^2(z)}{n_k}+\frac{S_k^2(z')}{n_k}-\frac{S_k^2(\tau(z,z'))}{n_k}\right).
\end{align*}
For the diagonal element, $l_{z,z}=L_z\geq2$ so the expectation of $\Shatwbht(z,z)$ is
\begin{align*}
    &\E[\Shatwbht(z,z)]=\E\left[\left(1-\frac{L_z}{K}\right)\Shatbbht(z,z) +\frac{K}{L_z}\left(\frac{1}{K^2}\sum_{k=1}^K\frac{\I[z\in\Rk](Kw_k)^2s_k^2(z)}{n_k/t}\right)\right]\\
    &=\left(1-\frac{L_z}{K}\right)\frac{1}{L_z}\E[\SHT(z)]+\frac{1}{KL_z}\sum_{k=1}^K\frac{(Kw_k)^2\E[\I[z\in\Rk]s_k^2(z)]}{n_k/t}\\
    &=\left(1-\frac{L_z}{K}\right)\frac{1}{L_z}\left( \SHT(z) + \frac{1}{K}\sum_{k=1}^{K}\frac{(Kw_k)^2S_k^2(z)}{n_k/(t-1)}\right)+\frac{1}{K^2}\sum_{k=1}^K\frac{(Kw_k)^2S_k^2(z)}{n_k/t}\\
    & =\left(\frac{1}{L_z}-\frac{1}{K}\right)\SHT(z)+\frac{1}{K}\bmw(z,z) +\frac{1}{K^2}\sum_{k=1}^K\frac{(Kw_k)^2S_k^2(z)}{n_k}=\bm{S}_{\HT}(z,z)+\frac{1}{K^2}\sum_{k=1}^K\frac{(Kw_k)^2S_k^2(z)}{n_k}.
\end{align*}
By combining the above, we have
\begin{align*}
    &\bm{g}\tr\left(\E[\Shatwbht]-\bm{S}_{\HT}\right)\bm{g}\\
    &=\frac{1}{K}\sum_{k=1}^K\frac{S_k^2(\tauwg)}{n_k}+\sum_{z,z':l_{z,z'}\leq1}\frac{g_zg_{z'}}{2}\left(\frac{1}{K}-\frac{l_{z,z'}}{L_zL_{z'}}\right)\left[S^2_{\HT}(z)+S^2_{\HT}(z')-S^2_{\HT}(\tau(z,z'))\right.\\
    &-\left.\left\{\frac{1}{K}\sum_{k=1}^K(Kw_k)^2\left(\frac{S_k^2(z)}{n_k}+\frac{S_k^2(z')}{n_k}-\frac{S_k^2(\tau(z,z'))}{n_k}\right)\right\}\right].
\end{align*}
\end{proof}

\subsection{Proof of Theorem~\ref{theorem:bias_fe_var}}\label{supp_mat:proof of bias_fe_var}

Here we derive the bias of $\varestwbfe$ and $\varestbbfe$, recalling
\begin{align*}
\varestwbfe &= \frac{T-t}{T(t-1)}\left[\frac{1}{K}\left\{s_{bb}^2(\tau(z_1,z_2))+(T-1)\left(\frac{\bar{s}_{bb}^2(\tau(z_1,\tset_{z_1}))}{t/(t-1)}+\frac{\bar{s}_{bb}^2(\tau(z_2,\tset_{z_2}))}{t/(t-1)}\right)\right\}\right] \\
&+ \frac{1}{K^2}\sum_{k=1}^K\left(\frac{s_k^2(z_1)}{n_k/T}+\frac{s_k^2(z_2)}{n_k/T}\right),\\
    \varestbbfe &= \frac{T-t}{T(t-1)}\left[\frac{1}{K}\left\{s_{bb}^2(\tau(z_1,z_2))+(T-1)\left(\frac{\bar{s}_{bb}^2(\tau(z_1,\tset_{z_1}))}{t/(t-1)}+\frac{\bar{s}_{bb}^2(\tau(z_2,\tset_{z_2}))}{t/(t-1)}\right)\right\}\right] \\
    &+ \frac{s_{bb}^2(\tau(z_1,z_2))}{K}.
    \end{align*}
\begin{proof}

In Section~\ref{pf:finite_bias_ht}, we derived the finite-sample expectation of $\sHT$ and $s_k^2(z)$. Because we consider a BIBD for the adjusted estimator, we have the following result:

\begin{align}
    &\E [s_{bb}^2(z)] = S_{bb}^2(z) + \frac{1}{K}\sum_{k=1}^{K}\frac{S_k^2(z)}{n_k/(t-1)},\label{eq:exp of sbb}\\
    &\E [s_{bb}^2(\tau(z,z'))] = S_{bb}^2(\tau(z,z')) + \frac{1}{K}\sum_{k=1}^{K}\left(\frac{S_k^2(z)}{n_k/t}+\frac{S_k^2(z')}{n_k/t}-\frac{S_k^2(\tau(z,z'))}{n_k}\right).\label{eq:exp of sbb tau}
\end{align}
Applying \eqref{eq:exp of sk}, \eqref{eq:exp of sbb}, and \eqref{eq:exp of sbb tau} to the components of the variance estimators,
\begin{align*}
    &\E\left[\bar{s}_{bb}^2(\tau(\tilde{z},\tsettilde))\right]\\
    &=\E\left[\sum_{z\in \z^c}\frac{p(z)}{t-1}s_{bb}^2(\tau(\tilde{z},z))-\sum_{z\in\mathbf{z}^c}\sum_{z'\in\z^c,z'\neq z}\frac{p_{\tilde{z}}(z,z')}{2(t-1)^2}s_{bb}^2(\tau(z,z'))\right]\\
    &=\sum_{z\in \z^c}\frac{p(z)}{t-1}S_{bb}^2(\tau(\tilde{z},z))-\sum_{z\in\mathbf{z}^c}\sum_{z'\in\z^c,z'\neq z}\frac{p_{\tilde{z}}(z,z')}{2(t-1)^2}S_{bb}^2(\tau(z,z'))\\
    &+\sum_{z\in\z^c}\frac{p(z)}{t-1}\left[\frac{1}{K}\sum_{k=1}^K\left(\frac{S_k^2(\tilde{z})}{n_k/t}+\frac{S_k^2(z)}{n_k/t}-\frac{S_k^2(\tau(\tilde{z},z))}{n_k}\right)\right]\\
    &-\sum_{z\in\mathbf{z}^c}\sum_{z'\in\z^c,z'\neq z}\frac{p_{\tilde{z}}(z,z')}{2(t-1)^2}\left[\frac{1}{K}\sum_{k=1}^K\left(\frac{S_k^2(\tilde{z})}{n_k/t}+\frac{S_k^2(z')}{n_k/t}-\frac{S_k^2(\tau(z,z'))}{n_k}\right)\right]\\
    &=\bar{S}_{bb}^2(\tau(\tilde{z},\tsettilde))+\frac{1}{K}\sum_{k=1}^K\V{\tilde{z}}.
\end{align*}
where $\bar{S}_{bb}^2(\tau(\tilde{z},\tsettilde))$ and $ \V{\tilde{z}}$ are defined in Section~\ref{subsec:fixed-effect}.
Thus, 
\begin{align*}
    &\E\left[ \frac{T-t}{T(t-1)}\left[\frac{1}{K}\left\{s_{bb}^2(\tau(z_1,z_2))+(T-1)\left(\frac{\bar{s}_{bb}^2(\tau(z_1,\tset_{z_1}))}{t/(t-1)}+\frac{\bar{s}_{bb}^2(\tau(z_2,\tset_{z_2}))}{t/(t-1)}\right)\right\}\right]\right]\\
    &= \Var(\tauestfe(z_1,z_2)) - \frac{1}{K^2}\sum_{k=1}^K\left(\frac{S_k^2(z_1)}{n_k/t}+\frac{S_k^2(z_2)}{n_k/t}-\frac{S_k^2(\tau(z_1,z_2))}{n_k}\right).
\end{align*}

We add terms 
\[\frac{1}{K^2}\sum_{k=1}^K\left(\frac{\I[z_1 \in \Rk]s_k^2(z_1)}{n_k/T}+\frac{\I[z_2 \in \Rk]s_k^2(z_2)}{n_k/T}\right)\]
for $\varestwbfe$ or 
\[\frac{s_{bb}^2(\tau(z_1,z_2))}{K}\]
for $\varestbbfe$ so that the biases of variance estimator are nonnegative.
Therefore, the biases of our variance estimators are
\begin{align*}
\E\left[\varestwbfe\right]-\Var(\tauestfe(z_1,z_2)) &=\frac{1}{K^2}\sum_{k=1}^K\frac{S_k^2(\tau(z_1,z_2))}{n_k},\\
\E\left[\varestbbfe\right]-\Var(\tauestfe(z_1,z_2)) &=\frac{S_{bb}^2(\tau(z_1,z_2))}{K}.\end{align*}
\end{proof}


\section{Proof of Lemmas}\label{append:proof_lemmas}
\subsection{Proof of Lemma~\ref{lem:fourth moment}}\label{append:proof of fourth moment}
\begin{proof}
For statement~\ref{itm:yestk fourth}, it follows from $\yestk(z) = (n_k/t)^{-1}\sum_{i:b_i=k}\I[Z_i=z]Y_i(z)$ for $z\in\Rk$ that
\begin{align*}
\yestk(z)^4\leq (n_k/t)^{-1}\sum_{i:b_i=k}\I[Z_i=z]Y_i(z)^4\leq (n_k/t)^{-1}\sum_{i:b_i=k}Y_i(z)^4.
\end{align*}
By Assumption~\ref{assump:clt}\ref{itm:fourth moment}, this ensures
\begin{align*}
K^2\sum_{k=1}^Kw_k^4\I[z\in\Rk]\E[\yestk(z)^4|\Rk]\leq t\cdot K^2\sum_{k=1}^Kw_k^4n_k^{-1}\sum_{i:b_i=k}Y_i(z)^4=o(1).
\end{align*}
The result then follows from $\I[z,z'\in\Rk]\yestk(z)^2\yestk(z')^2\leq 2^{-1}(\I[z\in\Rk]\yestk(z)^4+\I[z'\in\Rk]\yestk(z')^4)$.

For statement~\ref{itm:delta fourth}, note that $\|\bmdelta_k\|_2^2 = \sum_{z=1}^{T}\frac{\I[z\in\Rk]}{\prob(z\in\Rk)^2}w_k^2(\yestk(z)-\ybark(z))^2$. This ensures
\begin{align*}
\|\bmdelta_k\|_2^4 =\left( \sum_{z=1}^{T}\frac{\I[z\in\Rk]}{\prob(z\in\Rk)^2}w_k^2(\yestk(z)-\ybark(z))^2\right)^2\leq T\sum_{z=1}^{T}\frac{\I[z\in\Rk]}{\prob(z\in\Rk)^4}w_k^4(\yestk(z)-\ybark(z))^4
\end{align*}
by Cauchy-Schwarz inequality and hence
\begin{align*}
\E[\|\bmdelta_k\|_2^4|\Rk]&\leq T\E\left[\sum_{z=1}^{T}\frac{\I[z\in\Rk]}{\prob(z\in\Rk)^{4}}w_k^4(\yestk(z)-\ybark(z))^4\given\Rk\right]\\
&= T\sum_{z=1}^{T}\frac{\I[z\in\Rk]}{\prob(z\in\Rk)^{4}}w_k^4\E\left[(\yestk(z)-\ybark(z))^4\given \Rk\right].
\end{align*}
Because $\prob(z\in\Rk)^4 = L_z^4/K^4 \rightarrow (p_z^{\infty})^4\in (0,1)$, a sufficient condition for $K^2\sum_{k=1}^K\E[\|\bmdelta_k\|_2^4|\Rk] = o(1)$ provided that $T$ is constant is
\begin{align*}
K^{2}\sum_{k=1}^Kw_k^4\I[z\in\Rk]\E[(\yestk(z)-\ybark(z))^4|\Rk]=o(1),\quad\text{for}\quad z=1,...,T.
\end{align*}
Because $(u+v)^4 = [(u+v)^2]^2\leq [2u^2+2v^2]^2 \leq 2(4u^4+4v^4)$, this ensures
\begin{align*}
\sum_{k=1}^Kw_k^4\I[z\in\Rk]\E[(\yestk(z)-\ybark(z))^4|\Rk]\leq 8\sum_{k=1}^Kw_k^4\I[z\in\Rk]\E[\yestk(z)^4|\Rk]+8\sum_{k=1}^Kw_k^4\ybark(z)^4,
\end{align*}
and the proof is concluded because by the statement~\ref{itm:yestk fourth} and
\begin{align*}
K^2\sum_{k=1}^Kw_k^4\ybark(z)^4\leq K^2\sum_{k=1}^K \frac{1}{n_k}\sum_{i:b_i=k}(w_kY_{i}(z))^4=o(1)
\end{align*}
which follows from Cauchy-Schwarz inequality and Assumption~\ref{assump:clt}\ref{itm:fourth moment}.

For statement~\ref{itm:sk fourth}, Cauchy-Schwarz inequality ensures that
\begin{align*}
\I[z\in\Rk](s_k^2(z))^2\leq \frac{\I[z\in\Rk]}{(n_k/t-1)^2}\sum_{i:b_i=k}\I[Z_i=z](Y_i(z)-\yestk(z))^4
\end{align*}
for $z\in\Rk$ and this further leads to 
\begin{align*}
&\frac{\I[z\in\Rk]}{(n_k/t-1)^2}\sum_{i:b_i=k}\I[Z_i=z](Y_i(z)-\yestk(z))^4 \leq \frac{8\I[z\in\Rk]}{(n_k/t-1)^2}\sum_{i:b_i=k}\I[Z_i=z](Y_i(z)^4+(\yestk(z))^4)\\
&= \frac{8\I[z\in\Rk]}{(n_k/t-1)^2}\sum_{i:b_i=k}\I[Z_i=z]Y_i(z)^4+ \frac{8n_k/t\I[z\in\Rk]}{(n_k/t-1)^2}(\yestk(z))^4\leq \frac{16}{(n_k/t-1)^2}\sum_{i:b_i=k}Y_i(z)^4.
\end{align*}
By taking conditional expectation over $\Rk$, we have
\begin{align*}
&\frac{1}{K^2}\sum_{k=1}^K\frac{(Kw_k)^4}{(n_k/t)^2}\I[z\in\Rk]\E[(s_k^2(z))^2|\Rk]\leq \frac{1}{K^2}\sum_{k=1}^K\frac{16(Kw_k)^4}{(n_k/t)^2(n_k/t-1)^2}\sum_{i:b_i=k}Y_i(z)^4\\
&\leq \frac{16t^2}{K^2}\sum_{k=1}^K\frac{(Kw_k)^4}{n_k}\sum_{i:b_i=k}Y_i(z)^4 = o(1)
\end{align*}
because $n_k/t\geq n_k/t-1\geq 1$ and $t$ is fixed. This concludes the proof.
\end{proof}

\subsection{Proof of Lemma~\ref{lem:clt of mu}}\label{append:proof of clt of mu}
\begin{proof}
We first show how $\Cov(\bmmu)$ can be computed by Theorem 3 of \cite{li2017general}. We consider the first stage randomization of incomplete block designs as completely randomized experiments with total $K$ units (blocks) in which $r_{\tsetone}$ units (blocks) receives a treatment $\tsetone$ for all $\tsetone\in\tset$. We set new (fixed) potential outcomes for given $\tsetone$ as
\begin{align*}
\ybark(\tsetone) = Kw_k\sum_{z=1}^{T}\frac{r_{\tsetone}}{L_z}\I[z\in\tsetone]\bm{e}_z\ybark(z),
\end{align*}
where $\bm{e}_z \in \mathbb{R}^{T}$ denotes a standard basis vector with $z$-th coordinate 1.
Thus $\ybark(\tsetone)$ is a vector with $z$th entry equal to $ Kw_k\frac{r_{\tsetone}}{L_z}\ybark(z)$ is $z \in \tsetone$ and equal to 0 otherwise. 
We denote $\ybar(\tsetone) = K^{-1}\sum_{k=1}^K\ybark(\tsetone)$ as averaged potential outcomes. Our target estimand is $\sum_{\tsetone\in\tset}\ybar(\tsetone)$ and this equivalent to $\ybarw$ because
\begin{align*}
&\sum_{\tsetone\in\tset}\ybar(\tsetone) = \sum_{\tsetone\in\tset} \frac{1}{K}\sum_{k=1}^K\ybark(\tsetone)=  \sum_{\tsetone\in\tset} \frac{1}{K}\sum_{k=1}^KKw_k\sum_{z=1}^{T}\frac{r_{\tsetone}}{L_z}\I[z\in\tsetone]\bm{e}_z\ybark(z)\\
&=\sum_{z=1}^{T}\sum_{k=1}^K\frac{w_k}{L_z}\bm{e}_z\ybark(z)\sum_{\tsetone\in\tset}r_{\tsetone}\I[z\in\tsetone] = \sum_{z=1}^{T}\sum_{k=1}^K\bm{e}_zw_k\ybark(z)=\ybarw.
\end{align*} 
Here, we used $\sum_{\tsetone\in\tset}r_{\tsetone}\I[z\in\tsetone]=L_z$.
Our estimator for $\ybar(\tsetone)$ is $\yest(\tsetone) = r_{\tsetone}^{-1}\sum_{k=1}^K\I[\Rk = \tsetone]\ybark(\tsetone)$ and estimator for $\sum_{\tsetone\in\tset}\ybar(\tsetone)$ is 
\begin{align*}
&\sum_{\tsetone\in\tset}\yest(\tsetone) = \sum_{\tsetone\in\tset} r_{\tsetone}^{-1}\sum_{k=1}^K\I[\Rk = \tsetone]Kw_k\sum_{z=1}^{T}\frac{r_{\tsetone}}{L_z}\I[z\in\tsetone]\bm{e}_z\ybark(z)\\
&\sum_{z=1}^{T}\sum_{k=1}^K\bm{e}_z\frac{w_k}{L_z/K}\ybark(z)\sum_{\tsetone\in\tset}\I[z\in\tsetone,\Rk=\tsetone]=\sum_{z=1}^{T}\bm{e}_z\sum_{k=1}^K\frac{w_k\I[z\in\Rk]}{L_z/K}\ybark(z) = \bmmu.
\end{align*}
By Theorem 3 of \cite{li2017general}, covariance matrix of $\sum_{\tsetone\in\tset}\yest(\tsetone)=\bmmu$ is
\begin{align*}
\Cov(\bmmu) &= \sum_{\tsetone\in\tset}\frac{1}{r_{\tsetone}}\left[\frac{1}{K-1}\sum_{k=1}^K(\ybark(\tsetone)-\ybar(\tsetone))(\ybark(\tsetone)-\ybar(\tsetone))\tr\right]\\
&-\frac{1}{K}\left[\frac{1}{K-1}\sum_{k=1}^K\left(\sum_{\tsetone\in\tset}(\ybark(\tsetone)-\ybar(\tsetone))\right)\left(\sum_{\tsetone\in\tset}(\ybark(\tsetone)-\ybar(\tsetone))\right)\tr\right].
\end{align*}
For the first term,
\begin{align*}
 &\sum_{\tsetone\in\tset}\frac{1}{r_{\tsetone}}\left[\frac{1}{K-1}\sum_{k=1}^K(\ybark(\tsetone)-\ybar(\tsetone))(\ybark(\tsetone)-\ybar(\tsetone))\tr\right]\\
 =&\sum_{\tsetone\in\tset}\frac{1}{r_{\tsetone}}\left[\frac{1}{K-1}\sum_{k=1}^K\left(\sum_{z=1}^{T}\frac{r_{\tsetone}\I[z\in\tsetone]}{L_z}\bm{e}_z(Kw_K\ybark(z)-\ybar(z))\right)\left(\sum_{z=1}^{T}\frac{r_{\tsetone}\I[z\in\tsetone]}{L_z}\bm{e}_z(Kw_K\ybark(z)-\ybar(z))\right)\tr\right]\\
 =&\sum_{\tsetone\in\tset}\frac{1}{r_{\tsetone}}\left[\frac{1}{K-1}\sum_{k=1}^K\left(\sum_{z=1}^{T}\frac{r_{\tsetone}^2\I[z\in\tsetone]}{L_z^2}(Kw_K\ybark(z)-\ybar(z))^2\bm{e}_z\bm{e}_z\tr\right)\right.\\
 &+\left.\frac{1}{K-1}\sum_{k=1}^K\left(\sum_{z=1}^{T}\sum_{z'\neq z}\frac{r_{\tsetone}^2\I[z,z'\in\tsetone]}{L_zL_{z'}}(Kw_K\ybark(z)-\ybar(z))(Kw_K\ybark(z')-\ybar(z'))\bm{e}_z\bm{e}_{z'}\tr\right)\right]\\
 =&\sum_{z=1}^{T}\frac{\bm{e}_z\bm{e}_z\tr}{L_z^2}S_{\HT}(z)\sum_{\tsetone\in\tset}r_{\tsetone}\I[z\in\tsetone]+\sum_{z=1}^{T}\sum_{z'\neq z}\frac{\bm{e}_z\bm{e}_{z'}\tr}{L_zL_{z'}}\left[\frac{1}{2}(S_{\HT}(z)+S_{\HT}(z')-S_{\HT}(\tau(z,z'))\right]\sum_{\tsetone\in\tset}r_{\tsetone}\I[z,z'\in\tsetone]\\
=&\sum_{z=1}^{T}\frac{\bm{e}_z\bm{e}_z\tr}{L_z}S_{\HT}(z)+\sum_{z=1}^{T}\sum_{z'\neq z}\frac{l_{z,z'}\bm{e}_z\bm{e}_{z'}\tr}{L_zL_{z'}}\left[\frac{1}{2}(S_{\HT}(z)+S_{\HT}(z')-S_{\HT}(\tau(z,z'))\right].
\end{align*}
Here we used $\sum_{\tsetone\in\tset}r_{\tsetone}\I[z,z'\in\tsetone]=l_{z,z'}$.

For the second term, we have
\begin{align*}
\sum_{\tsetone\in\tset}\ybark(\tsetone) = Kw_k\sum_{z=1}^{T}\frac{\bm{e}_z}{L_z}\ybark(z)\sum_{\tsetone\in\tset}r_{\tsetone}\I[z\in\tsetone] = \sum_{z=1}^{T}\bm{e}_zKw_k\ybark(z),\quad \sum_{\tsetone\in\tset}\ybar(\tsetone) = \ybarw.
\end{align*}
This ensures
\begin{align*}
&\frac{1}{K}\left[\frac{1}{K-1}\sum_{k=1}^K\left(\sum_{\tsetone\in\tset}(\ybark(\tsetone)-\ybar(\tsetone))\right)\left(\sum_{\tsetone\in\tset}(\ybark(\tsetone)-\ybar(\tsetone))\right)\tr\right] \\
= &\frac{1}{K}\sum_{z=1}^{T}\sum_{z'=1}^{T}\frac{\bm{e}_z\bm{e}_{z'}\tr}{2}(S_{\HT}(z)+S_{\HT}(z')-S_{\HT}(\tau(z,z')))
\end{align*}
and thus we conclude $\Cov(\bmmu) = K^{-1}\bht$.

We now prove the asymptotic normality of $\bmmu$. We define 
\begin{align*}
m_{\tsetone}(z) = \max_{1\leq k \leq K}[\bm{e}_z\tr(\ybark(\tsetone)-\ybar(\tsetone))]^2.
\end{align*}
for $z=1,...,T$. $m_{\tsetone}(z)$ can be written as
\begin{align*}
m_{\tsetone}(z) &= \max_{1\leq k\leq K}\left[\bm{e}_z\tr\sum_{z'=1}^{T}\frac{r_{\tsetone}\I[z'\in\tsetone]}{L_{z'}}\bm{e}_{z'}(Kw_k\ybark(z')-\ybar(z'))\right]^2 \\
&= \max_{1\leq k\leq K}\left[\frac{r_{\tsetone}\I[z\in\tsetone]}{L_{z}}(Kw_k\ybark(z)-\ybar(z))\right]^2= \max_{1\leq k\leq K}\frac{r_{\tsetone}^2\I[z\in\tsetone]}{L_{z}^2}(Kw_k\ybark(z)-\ybar(z))^2\\
& = \frac{r_{\tsetone}^2\I[z\in\tsetone]}{L_{z}^2}\max_{1\leq k\leq K}(Kw_k\ybark(z)-\ybar(z))^2.
\end{align*}
This ensures
\begin{align*}
\frac{m_{\tsetone}(z)}{r_{\tsetone}^2\Var(\bm{e}_z\tr\bmmu)} &= \frac{\I[z\in\tsetone]}{L_{z}^2/K^2}\cdot\max_{1\leq k\leq K}\frac{(Kw_k\ybark(z)-\ybar(z))^2}{K}\cdot\frac{1}{K\Var(\bm{e}_z\tr\bmmu)}\\
&\leq \frac{1}{L_{z}^2/K^2}\cdot\max_{1\leq k\leq K}\frac{(Kw_k\ybark(z)-\ybar(z))^2}{K}\cdot\frac{1}{K\Var(\bm{e}_z\tr\bmmu)}.
\end{align*}
Provided that 
\begin{align*}
p_z\rightarrow p_z^{\infty} \in (0,1),\quad \max_{1\leq k\leq K}\frac{(Kw_k\ybark(z)-\ybar(z))^2}{K}\rightarrow 0, \quad \bht \rightarrow \bm{V},
\end{align*}
the above term converges to zero. By Theorem 4 of \cite{li2017general} and the argument of the proof of Theorem 5 of \cite{li2017general}, this establishes
\begin{align*}
\sqrt{K}(\bmmu-\ybarw)\overset{d}{\rightarrow} \mathcal{N}(0,\bm{V}).
\end{align*}
This concludes the proof.
\end{proof}

\subsection{Proof of Lemma~\ref{lem:consistency}}\label{append:proof of consistency}
\begin{proof}
Because $\E\bm{\yht}=\ybarw$ and $\Cov(\bm{\yht}) = O(1/K)$, $\bm{\yht}\overset{p}{\rightarrow}\ybarw$ by Markov inequality. 

Because $\onehatz$ is $\yht(z)$ where all potential outcomes are one, $\E\bm{\onehat} = \bm{I}_{T}$ and
\begin{align*}
\Cov(\bm{\onehat}) &= \text{diag}\left(\left((p_z)^{-1}-1\right)\frac{1}{K-1}\sum_{k=1}^K(Kw_k-1)^2\right)_{z=1}^{T}\\
 &= \text{diag}\left(\left((p_z)^{-1}-1\right)\left(\frac{K^2}{K-1}\sum_{k=1}^Kw_k^2-\frac{K}{K-1}\right)\right)_{z=1}^{T}
\end{align*}
by Theorem~\ref{theorem:var_ht}.
Provided that $p_z\rightarrow p_z^{\infty}\in(0,1)$ and $K^{-1}\sum_{k=1}^K(Kw_k)^2 = O(1)$ as $K$ goes to infinity, $\onehat \overset{p}{\rightarrow} \bm{I}_{T}$ by the Markov inequality. 
The above results directly shows $\bm{\yhaj} -\ybarw = \onehat^{-1}\bm{\yht}-\ybarw = \bm{0}+o_p(1)$.

For $\bm{\yest}_A$, we have $\E\yest_A(z) = \ybarzw$ and
\begin{align*}
&\Var(\yest_A(z)) = \Var\left((q_{z,z'})^{-1}\sum_{k=1}^K\I[z,z'\in\Rk]w_k\yestk(z)\right)\\
&=(q_{z,z'})^{-2}\left[\sum_{k=1}^K\Var\left(\I[z,z'\in\Rk]w_k\yestk(z)\right)+\sum_{k=1}^K\sum_{j\neq k}\Cov\left(\I[z,z'\in\Rk]w_k\yestk(z),\I[z,z'\in\Rj]w_j\yest_j(z)\right)\right]
\end{align*}
By the variance of the indicator of block membership in \ref{subsec: var blk mem} and the law of total variance, we establish that
\begin{align*}
&\Var\left(\I[z,z'\in\Rk]w_k\yestk(z)\right) = \E\left[\Var\left(\I[z,z'\in\Rk]w_k\yestk(z)|\Rk\right)\right]+\Var\left(\E\left[\I[z,z'\in\Rk]w_k\yestk(z)|\Rk\right]\right)\\
&=\E\left[\I[z,z'\in\Rk]w_k^2\left(\frac{S_k^2(z)}{n_k/(t-1)}\right)\right]+\Var\left(\I[z,z'\in\Rk]w_k\ybark(z)\right)\\
&=(q_{z,z'})^2w_k^2\left(\frac{S_k^2(z)}{n_k/(t-1)}\right)+q_{z,z'}(1-q_{z,z'})w_k^2\ybark(z)^2,
\end{align*}
and
\begin{align*}
&\Cov\left(\I[z,z'\in\Rk]w_k\yestk(z),\I[z,z'\in\Rj]w_j\yest_j(z)\right)\\
&=\E\left[\Cov\left(\I[z,z'\in\Rk]w_k\yestk(z),\I[z,z'\in\Rj]w_j\yest_j(z)\given \Rk,\Rj\right)\right]\\
&+\Cov\left(\E\left[\I[z,z'\in\Rk]w_k\yestk(z)\given\Rk\right],\E\left[\I[z,z'\in\Rj]w_j\yest_j(z)\given\Rj\right]\right)\\
&=\Cov\left(\E\left[\I[z,z'\in\Rk]w_k\yestk(z)\given\Rk\right],\E\left[\I[z,z'\in\Rj]w_j\yest_j(z)\given\Rj\right]\right)\\
&=\Cov\left(\I[z,z'\in\Rk]w_k\ybark(z),\I[z,z'\in\Rj]w_j\ybar_j(z)\right)\\
&=-w_kw_j\ybark(z)\ybar_j(z)\frac{q_{z,z'}(1-q_{z,z'})}{K-1}.
\end{align*}
By the above simplification, we further simplify $\Var(\yest_A(z))$ as
\begin{align*}
&\Var(\yest_A(z))\\
&=\frac{1}{q_{z,z'}}\left[\sum_{k=1}^K\left(q_{z,z'}w_k^2\left(\frac{S_k^2(z)}{n_k/(t-1)}\right)+(1-q_{z,z'})w_k^2\ybark(z)^2\right)-\frac{1-q_{z,z'}}{K-1}\sum_{k=1}^K\sum_{j\neq k}w_kw_j\ybark(z)\ybar_j(z)\right]\\
&=\frac{1}{K^2}\sum_{k=1}^K\frac{(Kw_k)^2S_k^2(z)}{n_k/(t-1)}+\frac{1-q_{z,z'}}{Kq_{z,z'}}\SHT(z).
\end{align*}
By Assumption~\ref{assump:clt}\ref{itm:p conv}, \ref{itm:bb conv} and \ref{itm:wb conv}, $\Var(\yest_A(z))=o(1)$. By Markov inequality, we establish that $\yest_A(z)\overset{p}{\rightarrow} \ybarzw$ for all $z$ and this leads to $\bm{\yest}_A\overset{p}{\rightarrow} \ybarw$.

Finally, $\bm{\yest}_{A'}-\ybarw = \onehat^{-1}\bm{\yest}_A-\ybarw = \bm{0}+o_p(1)$ because $\bm{\yest}_A\overset{p}{\rightarrow} \ybarw$ and $\onehat^{-1}\overset{p}{\rightarrow} \bm{I}$. This concludes the proof.
\end{proof}

\subsection{Proof of Lemma~\ref{lem:sht control}}\label{append:proof of sht control}

\begin{proof}
For the expectation of $(l_{z,z'}-1)^{-1}\sum_{k=1}^K\I[z,z'\in\Rk][(Kw_k)^2\yestk(z)\yestk(z')]$, we set $X_k=\I[z,z'\in\Rk][(Kw_k)^2\yestk(z)\yestk(z')]$ and $\E[X_k|\Rk] = \mu_k$. Direct calculation shows that
\begin{align*}
\mu_k &= \E[X_k|\Rk] = \E[\I[z,z'\in\Rk][(Kw_k)^2\yestk(z)\yestk(z')]|\Rk] \\
&= \I[z,z'\in\Rk](Kw_k)^2\left(\Cov(\yestk(z),\yestk(z'))|z,z'\in\Rk]+\E[\yestk(z)|z\in\Rk]\E[\yestk(z')|z'\in\Rk]\right)\\
& = \I[z,z'\in\Rk](Kw_k)^2\left[-(2n_k)^{-1}(S_k^2(z)+S_k^2(z')-S_k^2(\tau(z,z')))+\ybark(z)\ybark(z')\right],\\
\E X_k &= \E\mu_k = q_{z,z'}(Kw_k)^2\left[-(2n_k)^{-1}(S_k^2(z)+S_k^2(z')-S_k^2(\tau(z,z')))+\ybark(z)\ybark(z')\right].
\end{align*}
Recalling definition of $\bmw$ and $\bht$ 
\begin{align*}
&\bht(z,z) =\left(\frac{1}{p_z}-1\right)S^2_{\HT}(z),\quad \bht(z,z')=\frac{1}{2}\left[\left(\frac{q_{z,z'}}{p_z p_{z'}}-1\right)\left(S^2_{\HT}(z)+S^2_{\HT}(z')-S^2_{\HT}(\tau(z,z'))\right)\right],\\
&\bmw(z,z) = \frac{1}{p_z^{K}}\left(\frac{1}{K}\sum_{k=1}^K\frac{(Kw_k)^2S_k^2(z)}{n_k/(t-1)}\right),\\
 &\bmw(z,z')=-\frac{q_{z,z'}}{2p_z p_{z'}}\left[\frac{1}{K}\sum_{k=1}^K(Kw_k)^2\left(\frac{S_k^2(z)}{n_k}+\frac{S_k^2(z')}{n_k}-\frac{S_k^2(\tau(z,z'))}{n_k}\right)\right],
\end{align*}
we ensure that
\begin{align*}
&\E\left[(l_{z,z'}-1)^{-1}\sum_{k=1}^K\I[z,z'\in\Rk][(Kw_k)^2\yestk(z)\yestk(z')]\right] = \E\left[(l_{z,z'}-1)^{-1}\sum_{k=1}^KX_k\right]\\
&=(l_{z,z'}-1)^{-1}\sum_{k=1}^K q_{z,z'}(Kw_k)^2\left[-(2n_k)^{-1}(S_k^2(z)+S_k^2(z')-S_k^2(\tau(z,z')))+\ybark(z)\ybark(z')\right]\\
&=\frac{l_{z,z'}}{l_{z,z'}-1}\left(\frac{p_zp_{z'}}{q_{z,z'}}\bmw(z,z')+\frac{1}{K}\sum_{k=1}^K\ybark(z)\ybark(z')\right)\\
&=\frac{l_{z,z'}}{l_{z,z'}-1}\left(\frac{p_zp_{z'}}{q_{z,z'}}\bmw(z,z')-\frac{K-1}{K}\left(1-\frac{q_{z,z'}}{p_zp_{z'}}\right)^{-1}\bht(z,z')+\ybarzw\ybar(z';\bm{w})\right).
\end{align*}

Provided that $z$ and $z'$ belong to $\Rk$, the distribution of $X_k$ does not depend on $\Rk$. Using this fact, we have
\begin{align*}
\Var(X_k) &=\E[\Var(X_k|\Rk)]+\Var(\E[X_k|\Rk]) = q_{z,z'}\Var(X_k|z,z'\in\Rk)+\E[(\E[X_k|z,z'\in\Rk])^2]-(\E X_k)^2\\
&=q_{z,z'}(\E[X_k^2|z,z'\in\Rk]-\mu_k^2)+q_{z,z'}\mu_k^2-(q_{z,z'})^2\mu_k^2 = q_{z,z'}\E[X_k^2|z,z'\in\Rk]-(q_{z,z'})^2\mu_k^2,\\
\Cov(X_k,X_j) & = \E[\Cov(X_k,X_j|\Rk,\Rj)]+\Cov(\E[X_k|\Rk],\E[X_j|\Rj])\\
&= \Cov(\E[X_k|z,z'\in\Rk],\E[X_j|z,z'\in\Rj])\quad ( \Cov(X_k,X_j|z,z'\in \Rk\text{ and }\Rj)=0)\\
&=\E[\mu_k\mu_j]-(q_{z,z'})^2\mu_k\mu_j = q_{z,z'}\left(\frac{l_{z,z'}-1}{K-1}-q_{z,z'}\right)\mu_k\mu_j.
\end{align*}

This leads to
\begin{align*}
&\Var\left((l_{z,z'}-1)^{-1}\sum_{k=1}^KX_k\right)=(l_{z,z'}-1)^{-2}\left[\sum_{k=1}^K\Var(X_k)+\sum_{k=1}^K\sum_{j\neq k}\Cov(X_k,X_j)\right]\\
&=(l_{z,z'}-1)^{-2}\left[\sum_{k=1}^K\left(q_{z,z'}\E[X_k^2|z,z'\in\Rk]-(q_{z,z'})^2\mu_k^2\right)+\sum_{k=1}^K\sum_{j\neq k}q_{z,z'}\left(\frac{l_{z,z'}-1}{K-1}-q_{z,z'}\right)\mu_k\mu_j\right]\\
&= (l_{z,z'}-1)^{-2}\left[\sum_{k=1}^Kq_{z,z'}\E[X_k^2|z,z'\in\Rk]+q_{z,z'}\left(\frac{l_{z,z'}-1}{K-1}-q_{z,z'}\right)\left(\sum_{k=1}^K\mu_k\right)^2-q_{z,z'}\left(\frac{l_{z,z'}-1}{K-1}\right)\sum_{k=1}^K\mu_k^2\right]\\
&\leq K^2(l_{z,z'}-1)^{-2}\left[K^{-2}\sum_{k=1}^Kq_{z,z'}\E[X_k^2|z,z'\in\Rk]-K^{-2}q_{z,z'}\left(\frac{l_{z,z'}-1}{K-1}\right)\sum_{k=1}^K\mu_k^2\right]\\
&\leq  K^2(l_{z,z'}-1)^{-2}K^{-2}\sum_{k=1}^Kq_{z,z'}\E[X_k^2|z,z'\in\Rk].
\end{align*}
In the first inequality, we use the fact that $(K-1)^{-1}(l_{z,z'}-1)<q_{z,z'} = K^{-1}l_{z,z'}$. In the second inequality, we use that $l_{z,z'}-1>0$ and $\sum_{k=1}^K\mu_k^2>0$.

Because $K^2(l_{z,z'}-1)^{-2} = (q_{z,z'})^{-2}+o(1)$, thus, we ensure
\begin{align*}
0\leq\limsup_{K\rightarrow\infty}\Var\left((l_{z,z'}-1)^{-1}\sum_{k=1}^KX_k\right)\leq (q_{z,z'})^{-1}\lim_{K\rightarrow\infty}K^{-2}\sum_{k=1}^K\E[X_k^2|z,z'\in\Rk]=0
\end{align*}
by Lemma~\ref{lem:fourth moment}\ref{itm:yestk fourth} and by Markov inequality,
\begin{align*}
&(l_{z,z'}-1)^{-1}\sum_{k=1}^K\I[z,z'\in\Rk][(Kw_k)^2\yestk(z)\yestk(z')]\\
&=\E\left[(l_{z,z'}-1)^{-1}\sum_{k=1}^K\I[z,z'\in\Rk][(Kw_k)^2\yestk(z)\yestk(z')]\right]+o_p(1)\\
& = \frac{l_{z,z'}}{l_{z,z'}-1}\left(\frac{p_zp_{z'}}{q_{z,z'}}\bmw(z,z')-\frac{K-1}{K}\left(1-\frac{q_{z,z'}}{p_zp_{z'}}\right)^{-1}\bht(z,z')+\ybarzw\ybar(z';\bm{w})\right) +o_p(1)\\
&=\frac{p_zp_{z'}}{q_{z,z'}}\bmw(z,z')-\left(1-\frac{q_{z,z'}}{p_zp_{z'}}\right)^{-1}\bht(z,z')+\ybarzw\ybar(z';\bm{w}) +o_p(1).
\end{align*}
In a similar way, we have 
\begin{align*}
&(l_{z,z'}-1)^{-1}\sum_{k=1}^K\I[z,z'\in\Rk][Kw_k(\yestk(z)-\yestk(z')))]^2\\
&=\frac{Kq_{z,z'}p_z}{l_{z,z'}-1}\bmw(z,z)+\frac{(K-1)q_{z,z'}}{l_{z,z'}-1}\SHT(z)+\frac{(K-1)q_{z,z'}}{l_{z,z'}-1}(\ybarzw)^2\\
&+\frac{Kq_{z,z'}p_z}{l_{z,z'}-1}\bmw(z',z')+\frac{(K-1)q_{z,z'}}{l_{z,z'}-1}\SHT(z')+\frac{(K-1)q_{z,z'}}{l_{z,z'}-1}(\ybar(z';\bm{w}))^2\\
&-2\left( \frac{l_{z,z'}}{l_{z,z'}-1}\left(\frac{p_zp_{z'}}{q_{z,z'}}\bmw(z,z')-\frac{K-1}{K}\left(1-\frac{q_{z,z'}}{p_zp_{z'}}\right)^{-1}\bht(z,z')+\ybarzw\ybar(z';\bm{w})\right)\right)+o_p(1)\\
& = p_z\bmw(z,z)+p_{z'}\bmw(z',z')-\frac{2p_zp_{z'}}{q_{z,z'}}\bmw(z,z')+\SHT(\tau(z,z'))+(\ybarzw-\ybar(z';\bm{w}))^2+o_p(1).
\end{align*}

By \eqref{eq:exp of sk}, we know
\begin{align*}
\E\left[\frac{1}{L_z}\sum_{k=1}^K\frac{\I[z\in\Rk](Kw_k)^2s_k^2(z)}{n_k/t}\right] = \frac{1}{K}\sum_{k=1}^K\frac{(Kw_k)^2S_k^2(z)}{n_k/t}.
\end{align*}
By the law of total variance,
\begin{align*}
&\Var\left(\I[z\in\Rk](Kw_k)^2s_k^2(z)\right) = \E\left[\Var\left(\I[z\in\Rk](Kw_k)^2s_k^2(z)|\Rk\right)\right]+\Var\left(\E\left[\I[z\in\Rk](Kw_k)^2s_k^2(z)|\Rk\right]\right)\\
&=\E\left[\I[z\in\Rk]\E\left[(Kw_k)^4(s_k^2(z))^2|\Rk\right]\right]-\E\left[\left(\E[\I[z\in\Rk](Kw_k)^2s_k^2(z)|\Rk]\right)^2\right]+\Var(\I[z\in\Rk](Kw_k)^2S_k^2(z))\\
&=\E[\I[z\in\Rk]\E[(Kw_k)^4(s_k^2(z))^2|\Rk]]-\E[(\I[z\in\Rk](Kw_k)^2S_k^2(z))^2]+\left[(Kw_k)^2S_k^2(z)\right]^2p_z(1-p_z)\\
&=p_z(Kw_k)^4\E[(s_k^2(z))^2|z\in\Rk]-(p_z)^2\left[(Kw_k)^2S_k^2(z)\right]^2,
\end{align*}
and
\begin{align*}
&\Cov\left(\I[z\in\Rk](Kw_k)^2s_k^2(z),\I[z\in\Rj](Kw_j)^2s_j^2(z))\right)\\
&=\Cov\left(\E[\I[z\in\Rk](Kw_k)^2s_k^2(z)|\Rk],\E[\I[z\in\Rj](Kw_j)^2s_j^2(z)|\Rj]\right)\\
&=\frac{p_z(p_z-1)}{K-1}(Kw_k)^2(Kw_j)^2S_k^2(z)S_j^2(z).
\end{align*}

This ensures 
\begin{align*}
&\Var\left(\frac{1}{L_z}\sum_{k=1}^K\frac{\I[z\in\Rk](Kw_k)^2s_k^2(z)}{n_k/t}\right)\\
&= \frac{1}{L_z^2}\sum_{k=1}^K\frac{\Var(\I[z\in\Rk](Kw_k)^2s_k^2(z))}{(n_k/t)^2}
+\frac{1}{L_z^2}\sum_{k=1}^K\sum_{j\neq k}\frac{\Cov\left(\I[z\in\Rk](Kw_k)^2s_k^2(z)),\I[z\in\Rj](Kw_j)^2s_j^2(z)\right)}{(n_k/t)(n_j/t)}\\
&=\frac{1}{L_z^2}\sum_{k=1}^K\frac{p_z(Kw_k)^4\E[(s_k^2(z))^2|z\in\Rk]-(p_z)^2\left[(Kw_k)^2S_k^2(z)\right]^2}{(n_k/t)^2}\\
&+\frac{1}{L_z^2}\frac{p_z(p_z-1)}{K-1}\sum_{k=1}^K\sum_{j\neq k}\frac{(Kw_k)^2(Kw_j)^2S_k^2(z)S_j^2(z)}{(n_k/t)(n_j/t)}\\
&=\frac{1}{L_z^2}\sum_{k=1}^K\frac{p_z(Kw_k)^4\E[(s_k^2(z))^2|z\in\Rk]}{(n_k/t)^2}+\frac{1}{L_z^2}\frac{p_z(p_z-1)}{K-1}\left(\sum_{k=1}^K\frac{(Kw_k)^2S_k^2(z)}{n_k/t}\right)^2\\
&+\frac{p_z}{L_z^2}\left(\frac{p_z-1}{K-1}-p_z\right)\sum_{k=1}^K\frac{[(Kw_k)^2S_k^2(z)]^2}{(n_k/t)^2}\\
&\leq \frac{1}{p_zK^2}\sum_{k=1}^K\frac{(Kw_k)^4\E[(s_k^2(z))^2|z\in\Rk]}{(n_k/t)^2} = o(1)
\end{align*}
by Lemma~\ref{lem:fourth moment}\ref{itm:sk fourth}. In the inequality, we use the fact that $(K-1)^{-1}(p_z-1)<(K-1)^{-1}(L_z-1)<p_z = K^{-1}L_z$. Thus, $\Var\left(\frac{1}{L_z}\sum_{k=1}^K\frac{\I[z\in\Rk](Kw_k)^2s_k^2(z)}{n_k/t}\right) \overset{p}{\rightarrow} 0$ as $K\rightarrow \infty$. By Markov inequality, we establish that
\begin{align*}
\frac{1}{L_z}\sum_{k=1}^K\frac{\I[z\in\Rk](Kw_k)^2s_k^2(z)}{n_k/t}\overset{p}{\rightarrow}\frac{1}{K}\sum_{k=1}^K\frac{(Kw_k)^2S_k^2(z)}{n_k/t}
\end{align*}
and this concludes the proof.
\end{proof}

\subsection{Proof of Lemma~\ref{lem:shaj control}}\label{append:proof of shaj control}
\begin{proof}
For statement~\ref{itm:gamma y single}, we first take expectation:
\begin{align*}
&\E\left[(L_z-1)^{-1}\sum_{k=1}^K(Kw_k)^2\I[z\in\Rk]\yestk(z)\right] \\
&= L_z(L_z-1)^{-1}K^{-1}\sum_{k=1}^K(Kw_k)^2\ybark(z) = K^{-1}\sum_{k=1}^K(Kw_k)^2\ybark(z)+o(1).
\end{align*}
By the law of total variance and covariance,
\begin{align*}
\Var(\I[z\in\Rk]\yestk(z)) &= \Var\left(\E[\I[z\in\Rk]\yestk(z)|\Rk\right)+\E\left[\Var(\I[z\in\Rk]\yestk(z)|\Rk)\right]\\
&=\Var\left(\I[z\in\Rk]\ybark(z)\right)+\E\left[\I[z\in\Rk](n_k/(t-1))^{-1}S_k^2(z)\right]\\
&=p_z\left[\frac{S_k^2(z)}{n_k/(t-1)}+(1-p_z)\ybark(z)^2\right],\\
\Cov(\I[z\in\Rk]\yestk(z),\I[z\in\Rj]\yest_j(z))& = \Cov\left(\E[\I[z\in\Rk]\yestk(z)|\Rk],\E[\I[z\in\Rj]\yest_j(z)|\Rj]\right)\\
&+\E\left[\Cov(\I[z\in\Rk]\yestk(z),\I[z\in\Rj]\yest_j(z)|\Rk,\Rj)\right]\\
&=\Cov\left(\E[\I[z\in\Rk]\yestk(z)|\Rk],\E[\I[z\in\Rj]\yest_j(z)|\Rj]\right)\\
&=-p_z(1-p_z)(K-1)^{-1}\ybark(z)\ybarj(z).
\end{align*}
This ensures the expression of the variance of $(L_z-1)^{-1}\sum_{k=1}^K(Kw_k)^2\I[z\in\Rk]\yestk(z)$ as
\begin{align*}
&\Var\left((L_z-1)^{-1}\sum_{k=1}^K(Kw_k)^2\I[z\in\Rk]\yestk(z)\right)\\
&=(L_z-1)^{-2}\sum_{k=1}^K(Kw_k)^4\Var(\I[z\in\Rk]\yestk(z))\\
&+(L_z-1)^{-2}\sum_{k=1}^K(Kw_k)^2(Kw_j)^2\Cov\left(\I[z\in\Rk]\yestk(z),\I[z\in\Rj]\yest_j(z)\right)\\
&=(L_z-1)^{-2}p_z\sum_{k=1}^K(Kw_k)^4\left[\frac{S_k^2(z)}{n_k/(t-1)}+(1-p_z)\ybark(z)^2\right]\\
&-(L_z-1)^{-2}p_z(1-p_z)(K-1)^{-1}\sum_{k=1}^K(Kw_k)^2(Kw_j)^2\ybark(z)\ybarj(z)\\
&=\frac{p_zK^2}{(L_z-1)^2}\frac{t-1}{K^2}\sum_{k=1}^K\frac{(Kw_k)^4S_k^2(z)}{n_k}+\frac{p_z(1-p_z)K^2}{(L_z-1)^2}\frac{1}{K^2(K-1)}\sum_{k=1}^K\left((Kw_k)^2\ybark(z)-K^{-1}\sum_{k=1}^K(Kw_k)^2\ybark(z)\right)^2.
\end{align*}
Because 
\begin{align*}
S_k^2(z) &= (n_k-1)^{-1}\sum_{i:b_i=k}(Y_i(z)-\ybark(z))^2\leq2(n_k-1)^{-1}\sum_{i:b_i=k}(Y_i(z)^2+\ybark(z)^2)\\
&=2(n_k-1)^{-1}\sum_{i:b_i=k}Y_i(z)^2+2n_k(n_k-1)^{-1}\ybark(z)^2\leq 4(n_k-1)^{-1}\sum_{i:b_i=k}Y_i(z)^2 \leq 4\sum_{i:b_i=k}Y_i(z)^2
\end{align*}
by the Cauchy-Schwarz inequality and the fact that $n_k-1\geq 1$, we ensure that
\begin{align*}
&K^{-2}\sum_{k=1}^K(Kw_k)^4n_k^{-1}S_k^2(z) \leq 4K^{-2}\sum_{k=1}^K(Kw_k)^4n_k^{-1}\sum_{i:b_i=k}Y_i(z)^2\\
&\leq\left(K^{-2}\sum_{k=1}^K(Kw_k)^4n_k^{-1}\sum_{i:b_i=k}Y_i(z)^4\right)^{1/2}\left(K^{-2}\sum_{k=1}^K(Kw_k)^4\right)^{1/2}.
\end{align*}
By Assumption~\ref{assump:clt}\ref{itm:fourth moment}, the first term in the right hand side above goes to zero. By taking $Y_i(z)=Y_i(z')=1$ for $i:b_i=k$ in Lemma~\ref{lem:fourth moment}\ref{itm:yestk fourth}, we have the second term in right hand side above goes to zero. Thus, we establish 
\begin{align*}
\frac{1}{K^2}\sum_{k=1}^K\frac{(Kw_k)^4S_k^2(z)}{n_k}=o(1).
\end{align*}
On the other hand,
\begin{align*}
&\frac{1}{K^2(K-1)}\sum_{k=1}^K\left((Kw_k)^2\ybark(z)-K^{-1}\sum_{k=1}^K(Kw_k)^2\ybark(z)\right)^2\\
& = \frac{1}{K-1}\sum_{k=1}^K\left(K^{-1}(Kw_k)^2\ybark(z)-K^{-1}\sum_{k=1}^KK^{-1}(Kw_k)^2\ybark(z)\right)^2\\
&=\frac{1}{K-1}\sum_{k=1}^K\left(Kw_k^2\ybark(z)\right)^2-\frac{K}{K-1}\left(K^{-1}\sum_{k=1}^KKw_k^2\ybark(z)\right)^2\\
&\leq \frac{1}{K-1}\sum_{k=1}^K\left(Kw_k^2\ybark(z)\right)^2 = \frac{1}{K-1}\sum_{k=1}^KK^2w_k^4\ybark(z)^2\leq \frac{1}{K-1}\sum_{k=1}^KK^2w_k^4n_k^{-1}\sum_{i:b_i=k}Y_i(z)^2
 \end{align*}
 and we showed that this also goes to zero in the above. Because $K^2(L_z-1)^{-2}= (p_z)^{-2} +o(1)$ by Assumption~\ref{assump:clt}\ref{itm:p conv}, the variance of $(L_z-1)^{-1}\sum_{k=1}^K(Kw_k)^2\I[z\in\Rk]\yestk(z)$ goes to zero and the statement \ref{itm:gamma y single} is proved by Markov inequality.
 
 For statement \ref{itm:gamma y double}, because
 \begin{align*}
 &(l_{z,z'}-1)^{-1}\sum_{k=1}^K\I[z,z'\in\Rk](Kw_k)^2\yestk(z)\yestk(z') \\
 &= \frac{p_zp_{z'}}{q_{z,z'}}\bmw(z,z')-\left(1-\frac{q_{z,z'}}{p_zp_{z'}}\right)^{-1}\bht(z,z')+\ybarzw\ybar(z';\bm{w})+o_p(1)
 \end{align*}
 as shown in the proof of Lemma~\ref{lem:sht control}, by taking $Y_i(z')=1$ for $i:b_i=k$ for all $k$, we have
 $\bmw(z,z') = 0$, $\left(1-\frac{q_{z,z'}}{p_zp_{z'}}\right)^{-1}\bht(z,z') = K^{-1}\sum_{k=1}^K(Kw_k)^2\ybark(z)-\ybarzw+o(1)$ and $\ybar(z';\bm{w})=1$. This ensures
 \begin{align*}
 (l_{z,z'}-1)^{-1}\sum_{k=1}^K\I[z,z'\in\Rk](Kw_k)^2\yestk(z) =  K^{-1}\sum_{k=1}^K(Kw_k)^2\ybark(z)+o_p(1).
 \end{align*}
 
 For statement \ref{itm:gamma2 single}, we take $Y_i(z) = 1$ for $i:b_i=k$ for all $k$ in Lemma~\ref{lem:sht control} and this ensures $\bmw(z,z)=0$, $\SHT(z) = (K-1)^{-1}(Kw_k-1)^2$, and $\ybarzw = 1$, thus we have
\begin{align*}
(L_z-1)^{-1}\sum_{k=1}^K\I[z\in\Rk](Kw_k)^2 &= (K-1)^{-1}\sum_{k=1}^K(Kw_k)^2-\frac{K}{K-1}+1+o_p(1)\\
& = K^{-1}\sum_{k=1}^K(Kw_k)^2+o_p(1).
\end{align*}

 For statement \ref{itm:gamma2 double}, we take $Y_i(z) = 1$ and $Y_i(z') = 0$ for $i:b_i=k$ for all $k$ in Lemma~\ref{lem:sht control} and this ensures $\bmw(z,z)=\bmw(z',z')=\bmw(z,z')=0$, $\SHT(\tau(z,z')) = (K-1)^{-1}(Kw_k-1)^2$, and $\ybarzw = 1$, $\ybar(z';\bm{w})=0$, thus we have
\begin{align*}
(l_{z,z'}-1)^{-1}\sum_{k=1}^K\I[z,z'\in\Rk](Kw_k)^2 &= (K-1)^{-1}\sum_{k=1}^K(Kw_k)^2-\frac{K}{K-1}+1+o_p(1)\\
& = K^{-1}\sum_{k=1}^K(Kw_k)^2+o_p(1).
\end{align*}

 \end{proof}

\section{Additional simulation results}\label{supp_mat:add sim}
In this section, we present additional simulation results.

\subsection{Results for the comparison between the Horvitz-Thompson and H\'ajek estimators}\label{supp_mat:len_ht_haj}
\begin{figure}[ht]
\centering 
\begin{subfigure}{\textwidth}
    \centering
    \includegraphics[width=\textwidth]{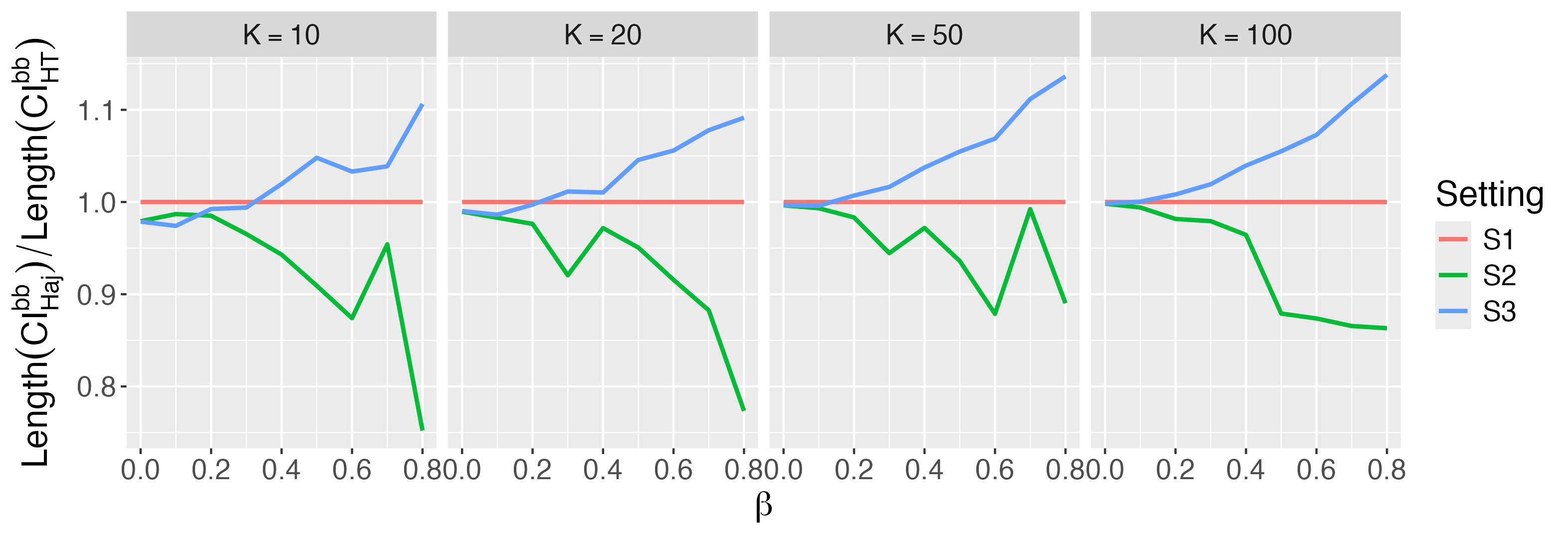}
  \end{subfigure}

   \begin{subfigure}{\textwidth}
    \centering
    \includegraphics[width=\textwidth]{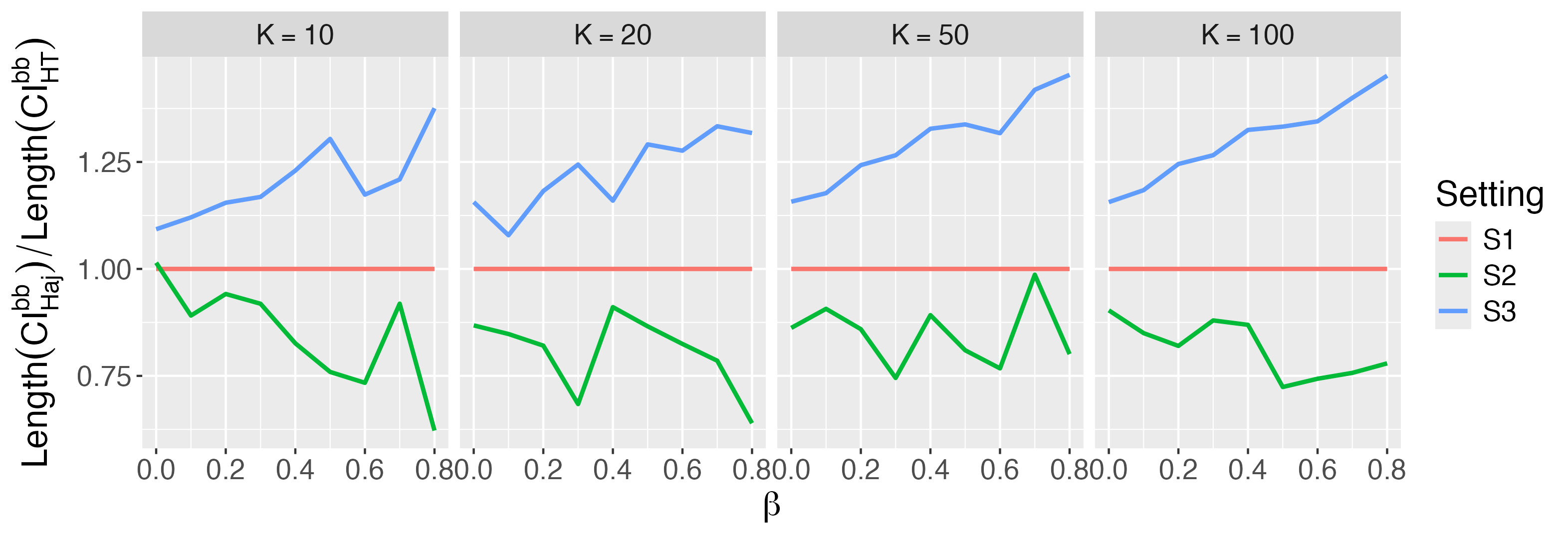}
  \end{subfigure}

 \begin{subfigure}{\textwidth}
    \centering
    \includegraphics[width=\textwidth]{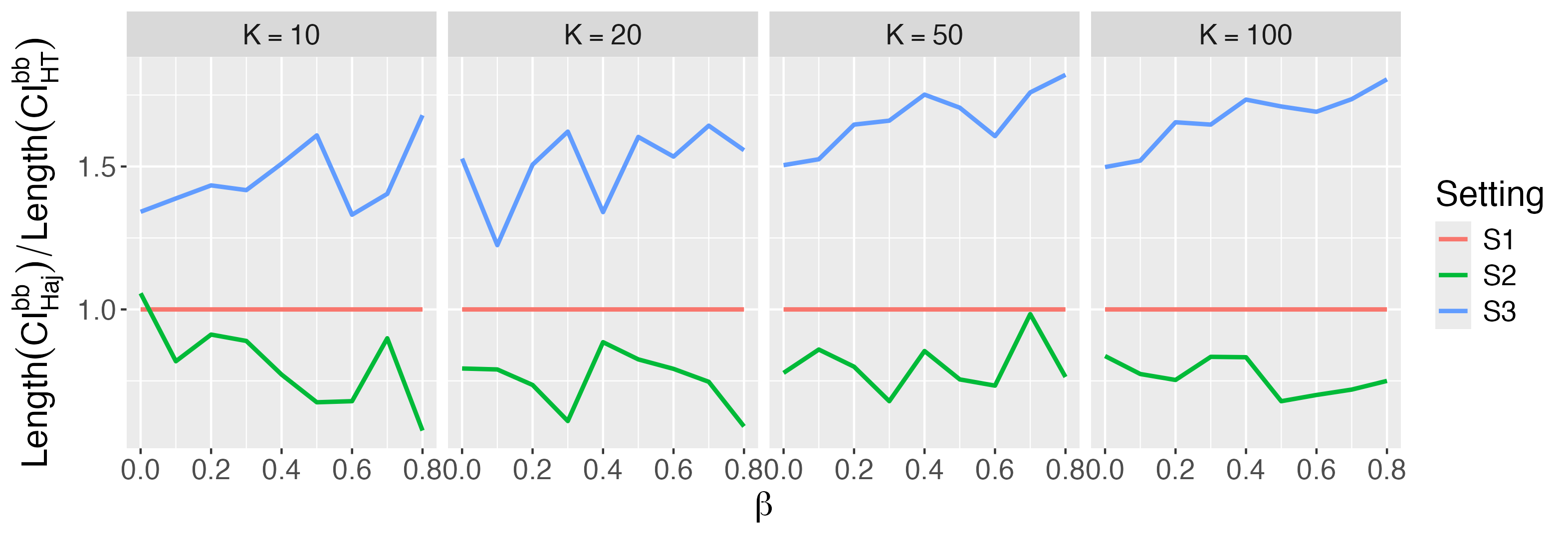}
  \end{subfigure}
\caption{Simulation results with $\rho = 0$. From top to bottom, the panels correspond to $\gamma = 0$, $0.5$, and $1$, respectively. Each line shows the empirical ratio of the average lengths of the confidence intervals $\text{CI}^{bb}_{*}$ in \eqref{eq:CI star bb} for $* = \HT$ and $\haj$, computed from a single simulation with between-block biases.}
\label{fig: len_haj_ht_bb_rho0} 
\end{figure}

\begin{figure}[ht]
\centering 
\begin{subfigure}{\textwidth}
    \centering
    \includegraphics[width=\textwidth]{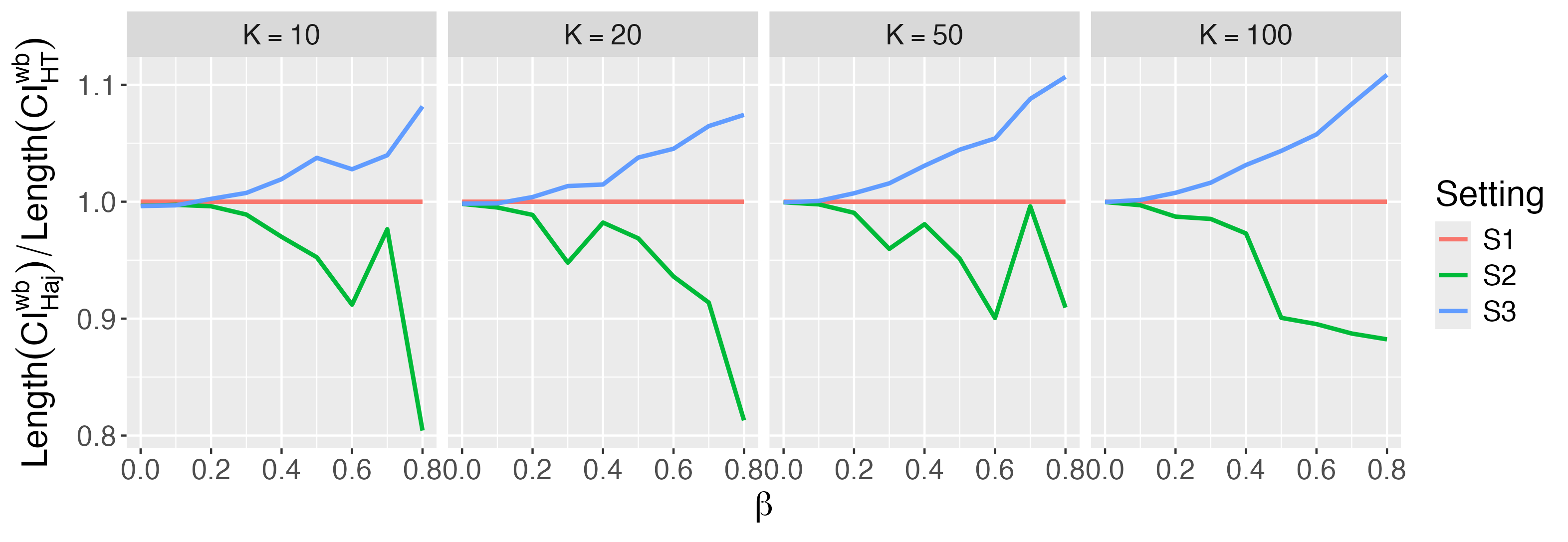}
  \end{subfigure}

   \begin{subfigure}{\textwidth}
    \centering
    \includegraphics[width=\textwidth]{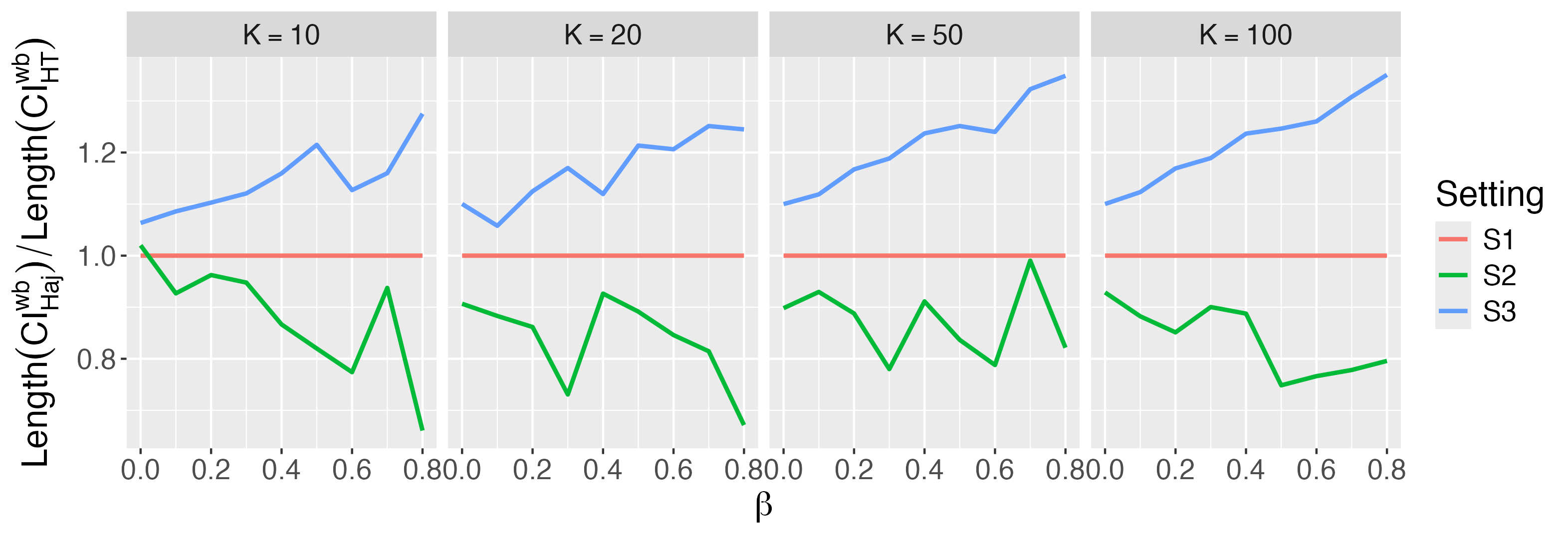}
  \end{subfigure}

 \begin{subfigure}{\textwidth}
    \centering
    \includegraphics[width=\textwidth]{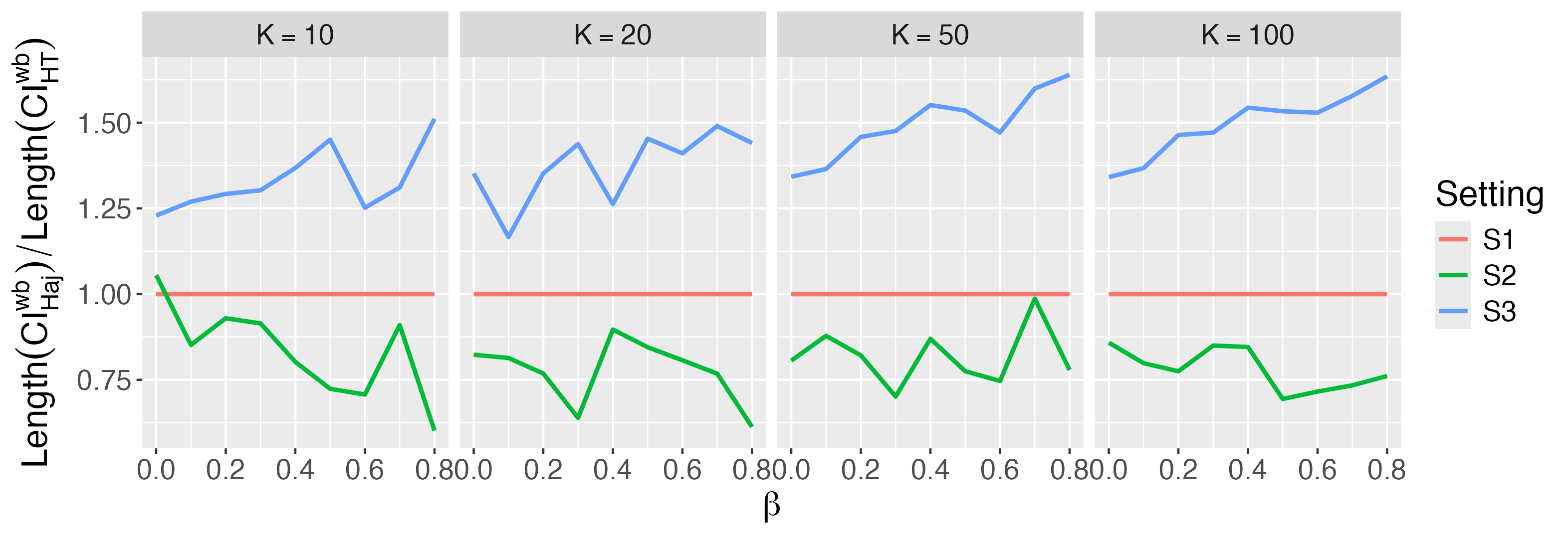}
  \end{subfigure}
\caption{Simulation results with $\rho = 0$. From top to bottom, the panels correspond to $\gamma = 0$, $0.5$, and $1$, respectively. Each line shows the empirical ratio of the average lengths of the confidence intervals $\text{CI}^{wb}_{*}$ in \eqref{eq:CI star wb} for $* = \HT$ and $\haj$, computed from a single simulation with between-block biases.}
\label{fig: len_haj_ht_wb_rho0} 
\end{figure}

We compare the lengths of confidence intervals from the HT and H\'ajek estimators.
While Corollary~\ref{cor:var comp} characterizes only the diagonal entries of the covariance matrices—rather than the full variances of the estimators—we find that the simulation results align well with the theoretical patterns described there.
Under \textbf{S1}, the two estimators are equivalent, and therefore yield identical confidence intervals.
However, under \textbf{S2}, the H\'ajek estimator is more efficient, whereas under \textbf{S3}, the HT estimator outperforms the H\'ajek estimator.
Under \textbf{S2}, the block sizes, $n_k$, are uncorrelated with the block index $k$, and thus the block means and treatment effects, making $\ybark(z)$ more stable across blocks compared to the weighted term $w_k \ybark(z)$ for $z \in \{1, \dots, T\}$.
In contrast, under \textbf{S3}, $n_k$ increases with $k$, so the weighted terms $w_k \ybark(z)$ are nearly constant across $k$, resulting in greater asymptotic efficiency of the HT estimator.
Because the results for different $\rho$ values given $\gamma$ are very similar, we report only the results for $\rho = 0$ in Figures~\ref{fig: len_haj_ht_bb_rho0} and \ref{fig: len_haj_ht_wb_rho0}.

\graphicspath{{./Images/}}

\graphicspath{{./Images/}}

\subsection{Results for the comparison between the variance estimators}\label{supp_mat:var_est_sims}

\begin{figure}[ht]
\centering 
\begin{subfigure}{\textwidth}
    \centering
    \includegraphics[width=\textwidth]{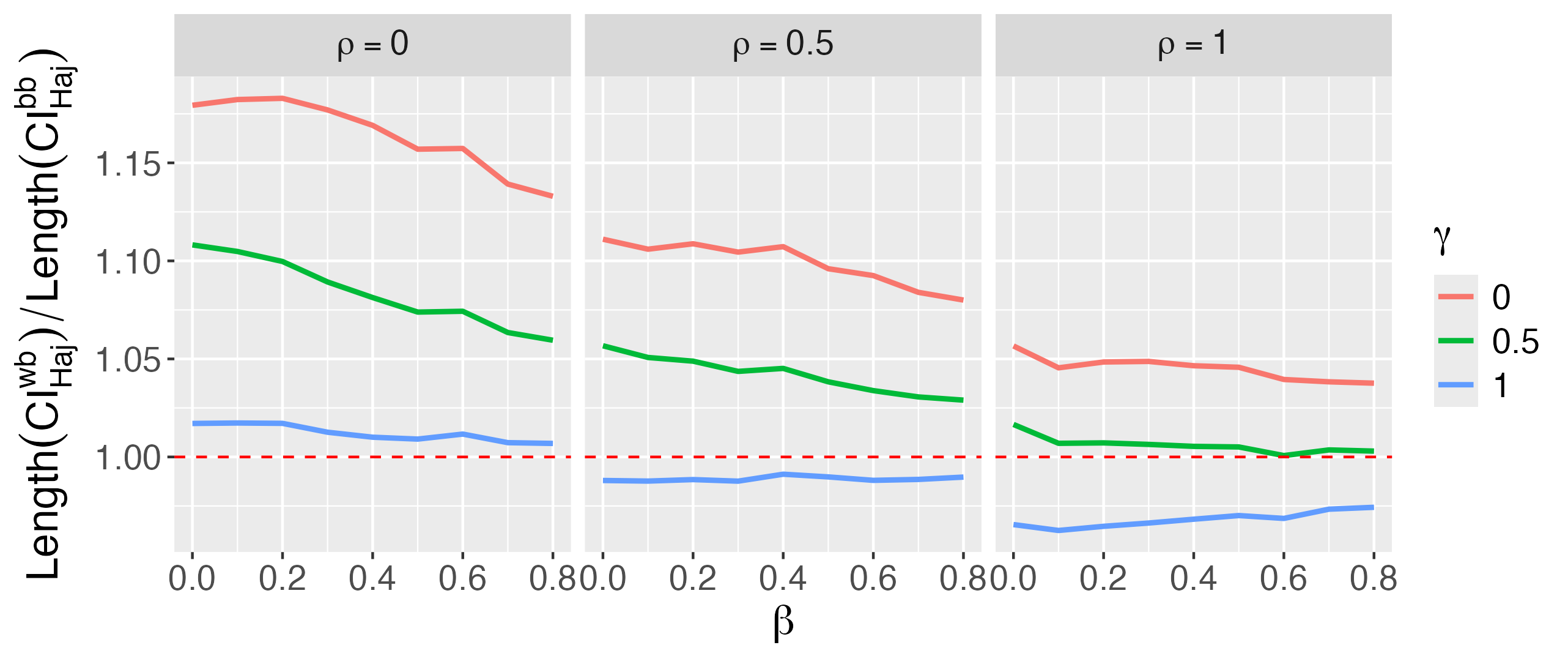}
  \end{subfigure}

   \begin{subfigure}{\textwidth}
    \centering
    \includegraphics[width=\textwidth]{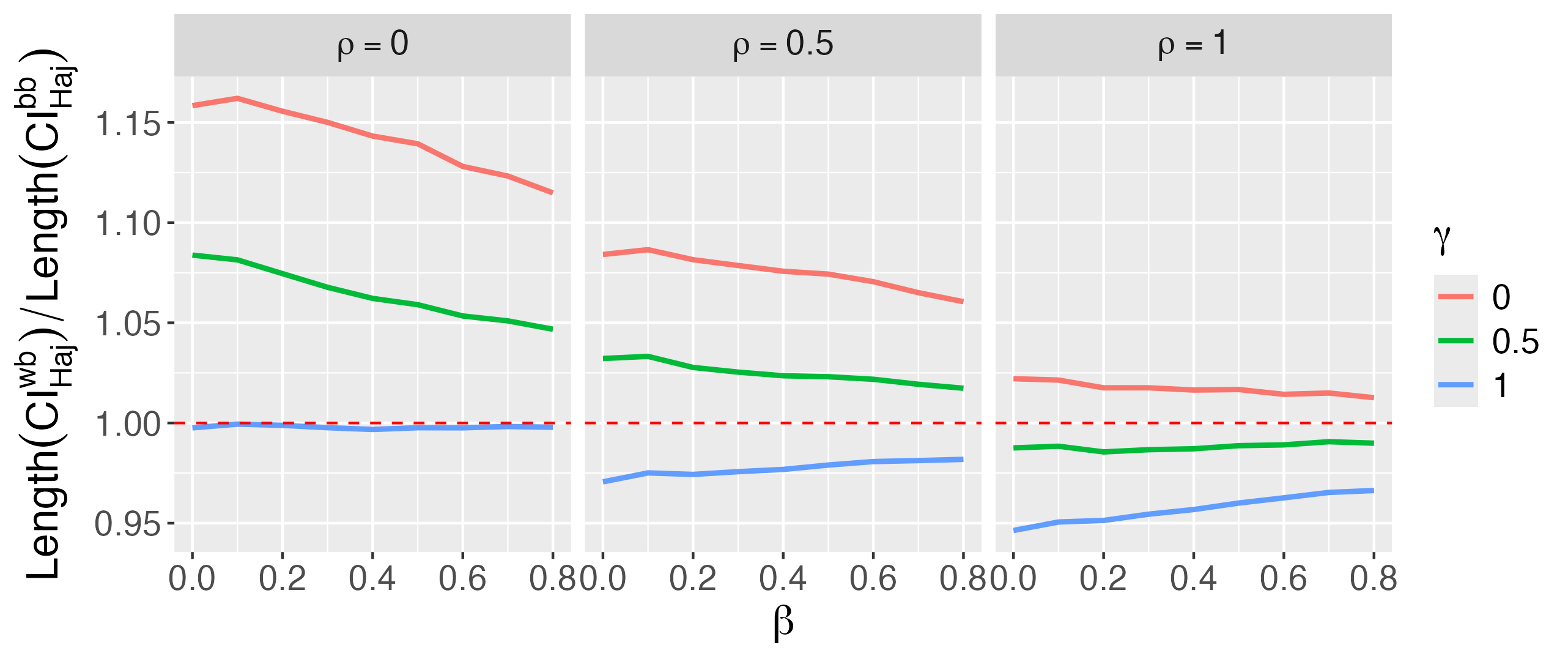}
  \end{subfigure}

 \begin{subfigure}{\textwidth}
    \centering
    \includegraphics[width=\textwidth]{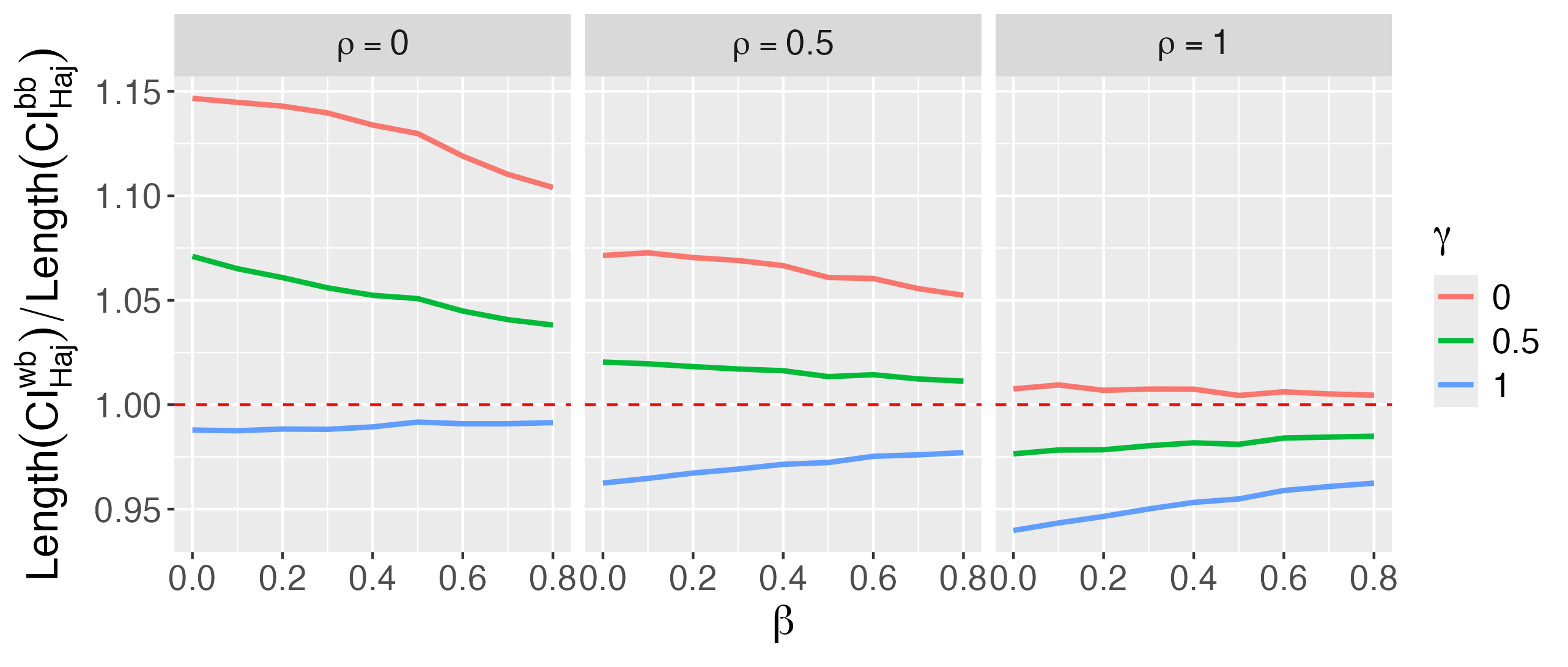}
  \end{subfigure}
\caption{Simulations under \textbf{S1}. From top to bottom, the panels correspond to $K = 10$, $20$, and $50$, respectively.
Each line corresponds to the empirical ratio of average length of confidence intervals  $\text{CI}^{wb}_{\haj} $ and  $\text{CI}^{bb}_{\haj} $ in one simulation.
The dashed horizontal red lines are at 1.}
\label{fig: len_wb_bb_haj_S1} 
\end{figure}

\begin{figure}[ht]
\centering 
\begin{subfigure}{\textwidth}
    \centering
    \includegraphics[width=\textwidth]{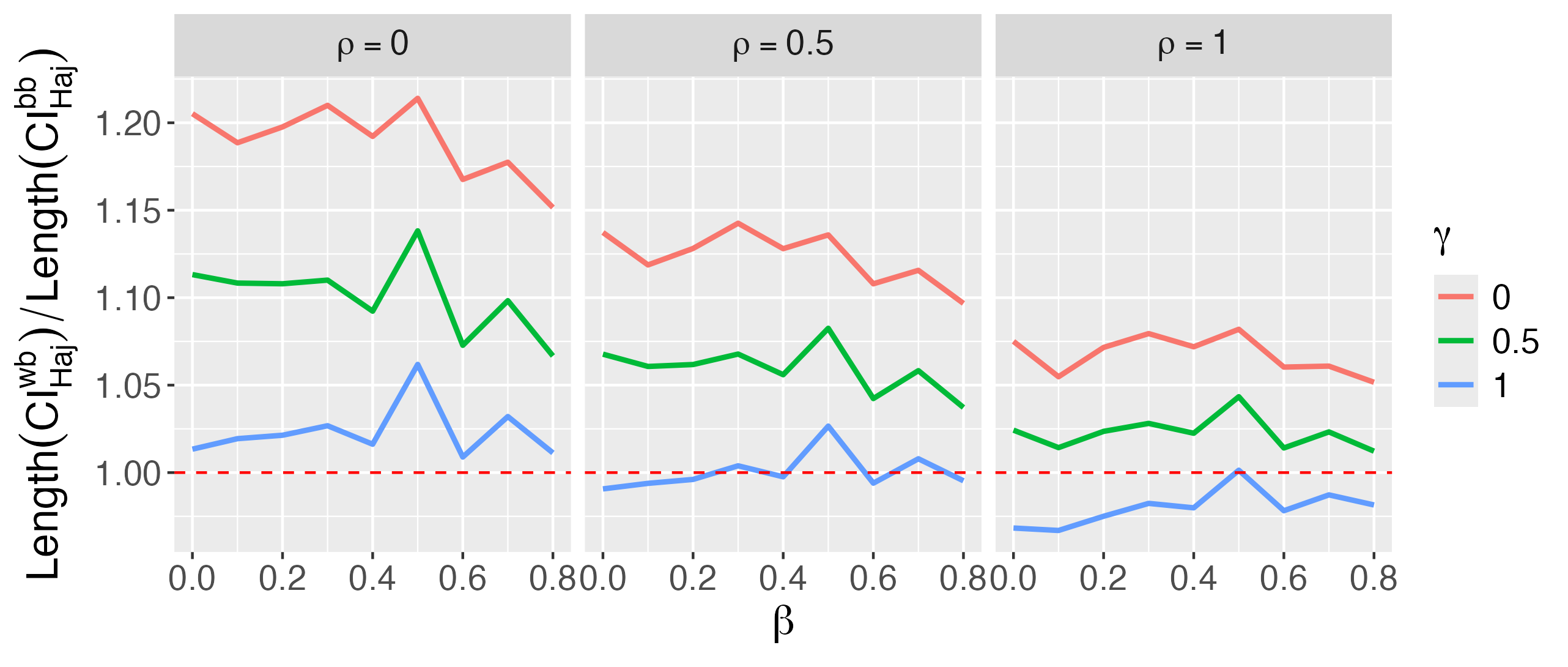}
  \end{subfigure}

   \begin{subfigure}{\textwidth}
    \centering
    \includegraphics[width=\textwidth]{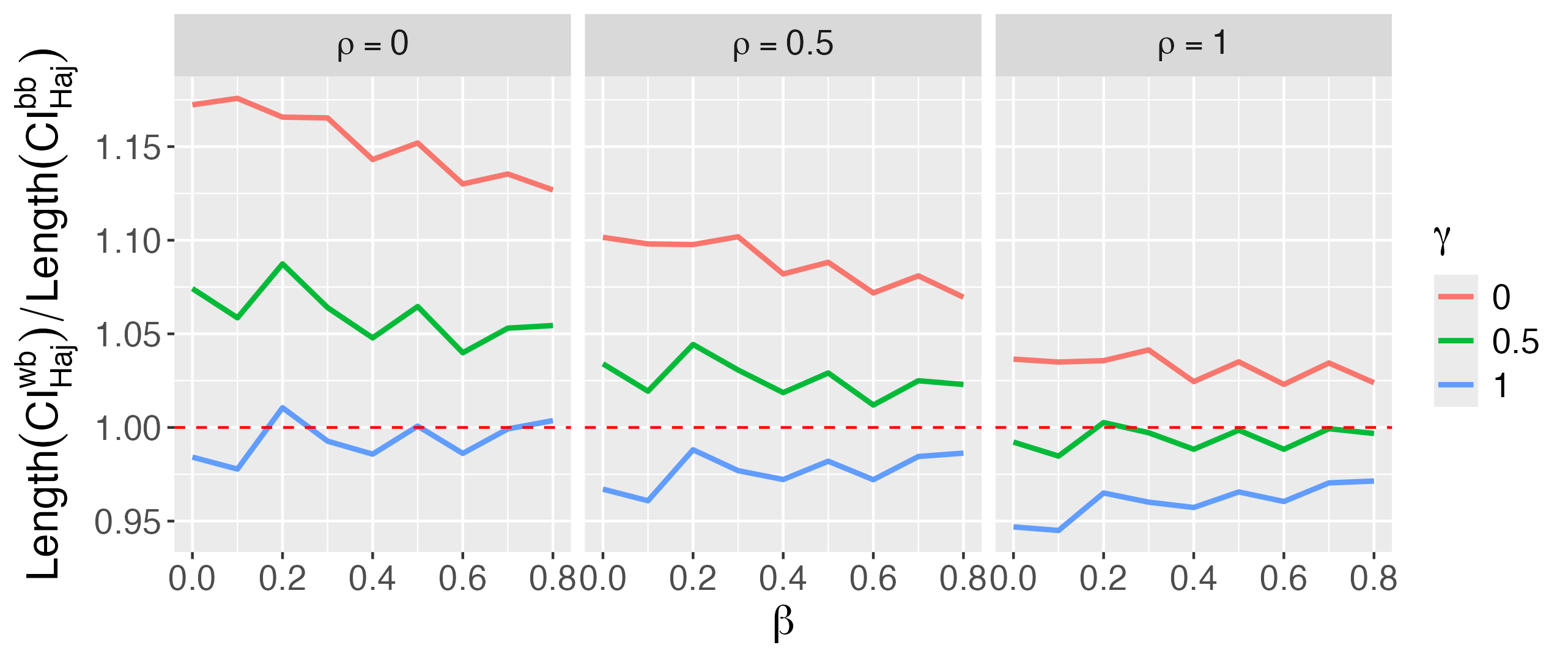}
  \end{subfigure}

 \begin{subfigure}{\textwidth}
    \centering
    \includegraphics[width=\textwidth]{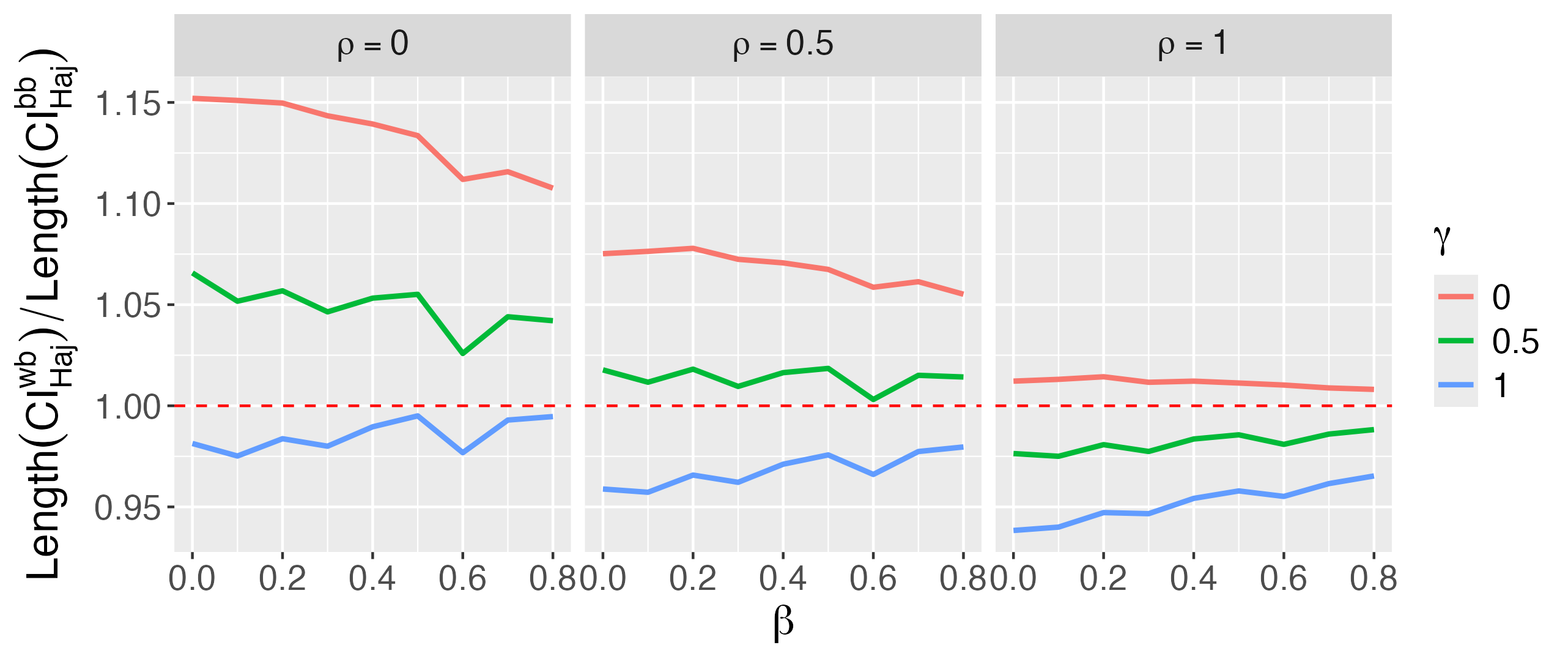}
  \end{subfigure}
\caption{Simulations under \textbf{S2}. From top to bottom, the panels correspond to $K = 10$, $20$, and $50$, respectively.
Each line corresponds to the empirical ratio of average length of confidence intervals  $\text{CI}^{wb}_{\haj} $ and  $\text{CI}^{bb}_{\haj} $ in one simulation.
The dashed horizontal red lines are at 1.}
\label{fig: len_wb_bb_haj_S2} 
\end{figure}

\begin{figure}[ht]
\centering 
\begin{subfigure}{\textwidth}
    \centering
    \includegraphics[width=\textwidth]{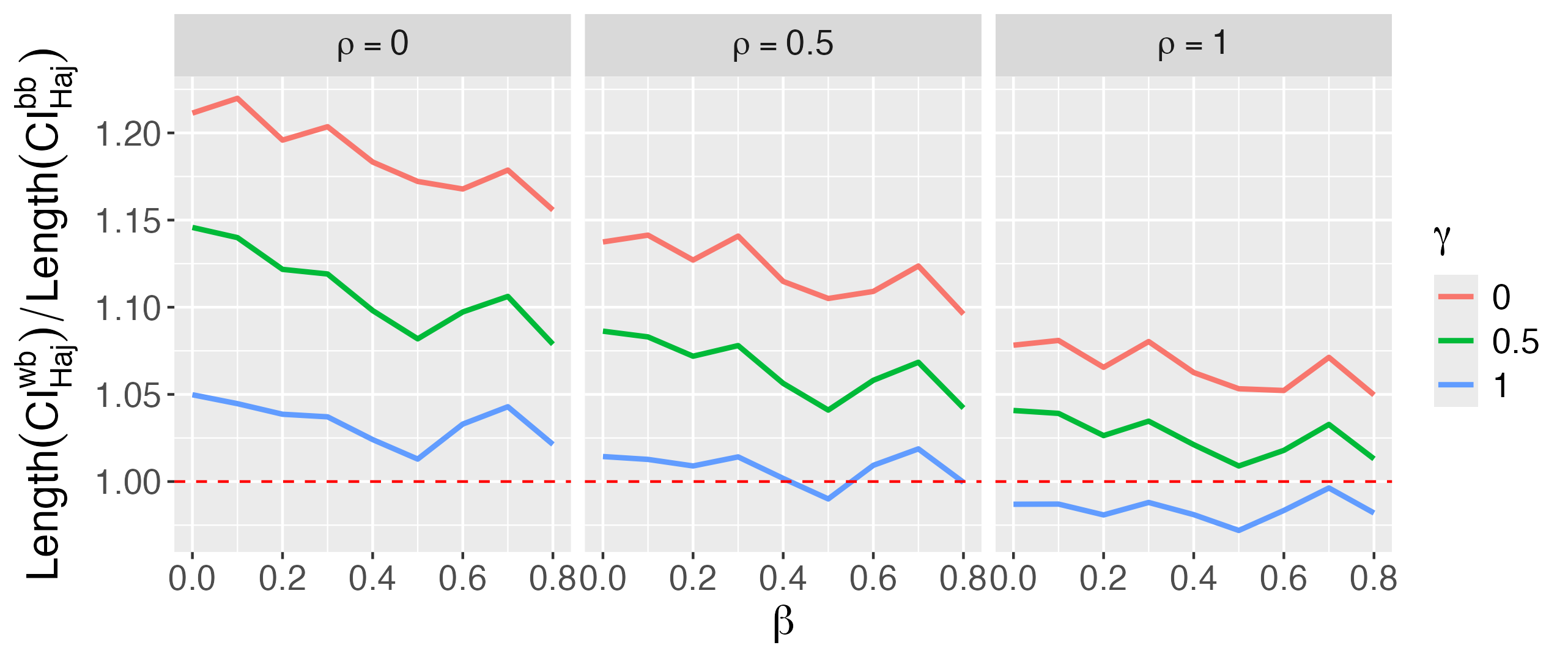}
  \end{subfigure}

   \begin{subfigure}{\textwidth}
    \centering
    \includegraphics[width=\textwidth]{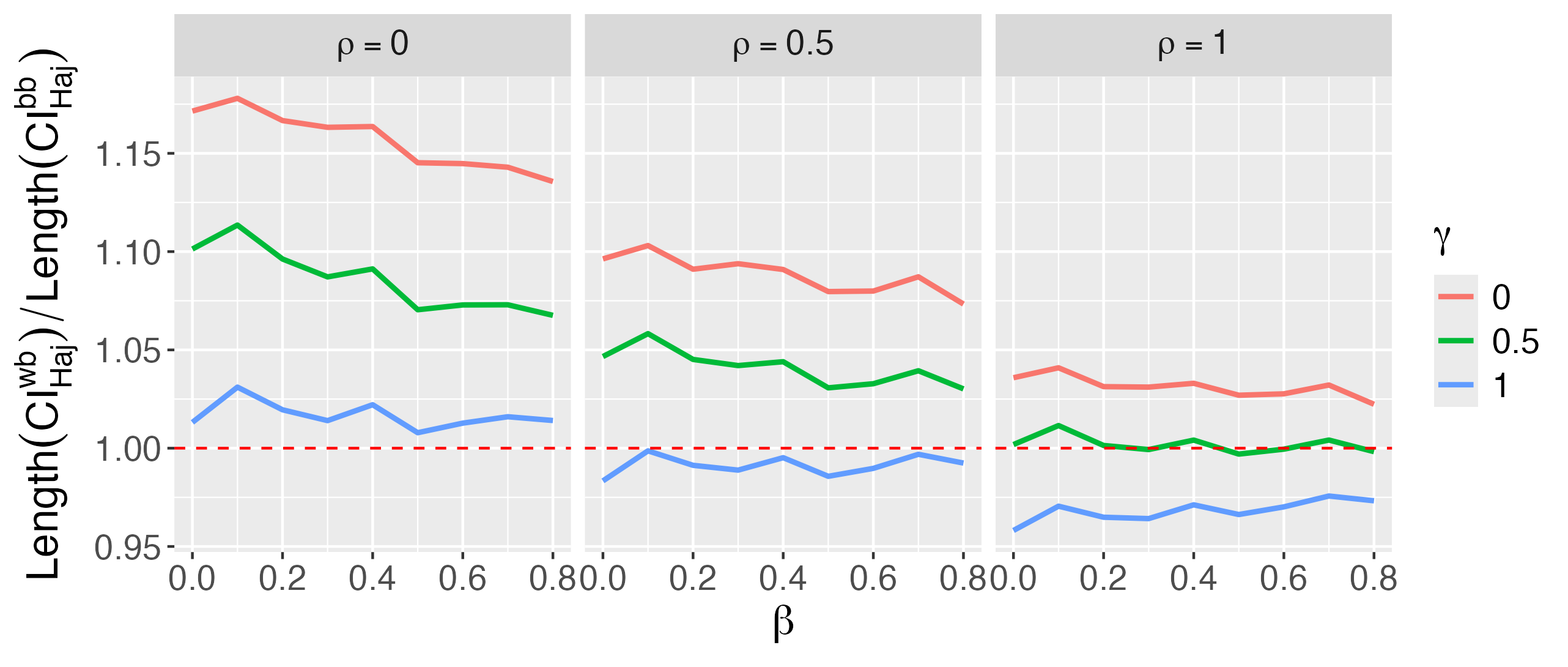}
  \end{subfigure}

 \begin{subfigure}{\textwidth}
    \centering
    \includegraphics[width=\textwidth]{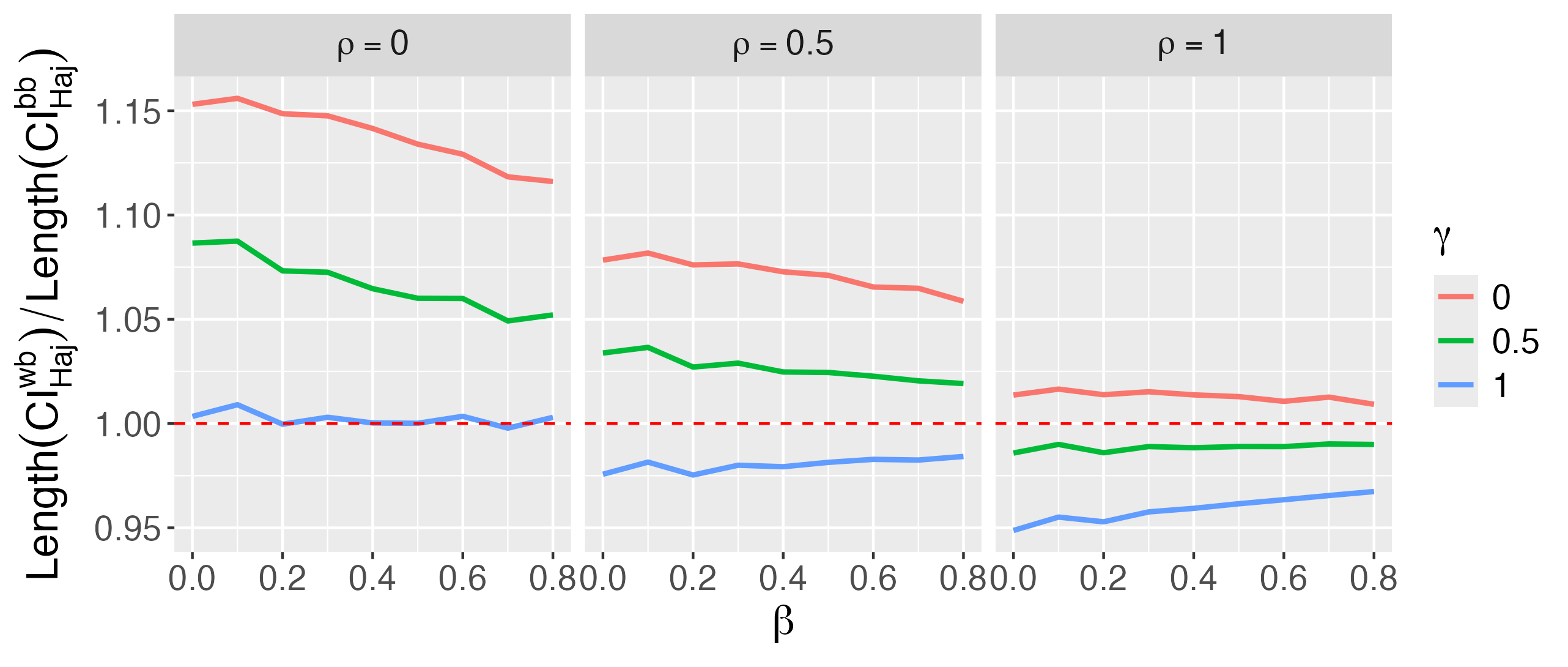}
  \end{subfigure}
\caption{Simulations under \textbf{S3}. From top to bottom, the panels correspond to $K = 10$, $20$, and $50$, respectively.
Each line corresponds to the empirical ratio of average length of confidence intervals  $\text{CI}^{wb}_{\haj} $ and  $\text{CI}^{bb}_{\haj} $ in one simulation.
The dashed horizontal red lines are at 1.}
\label{fig: len_wb_bb_haj_S3} 
\end{figure}

\begin{figure}[ht]
\centering 
\begin{subfigure}{\textwidth}
    \centering
    \includegraphics[width=\textwidth]{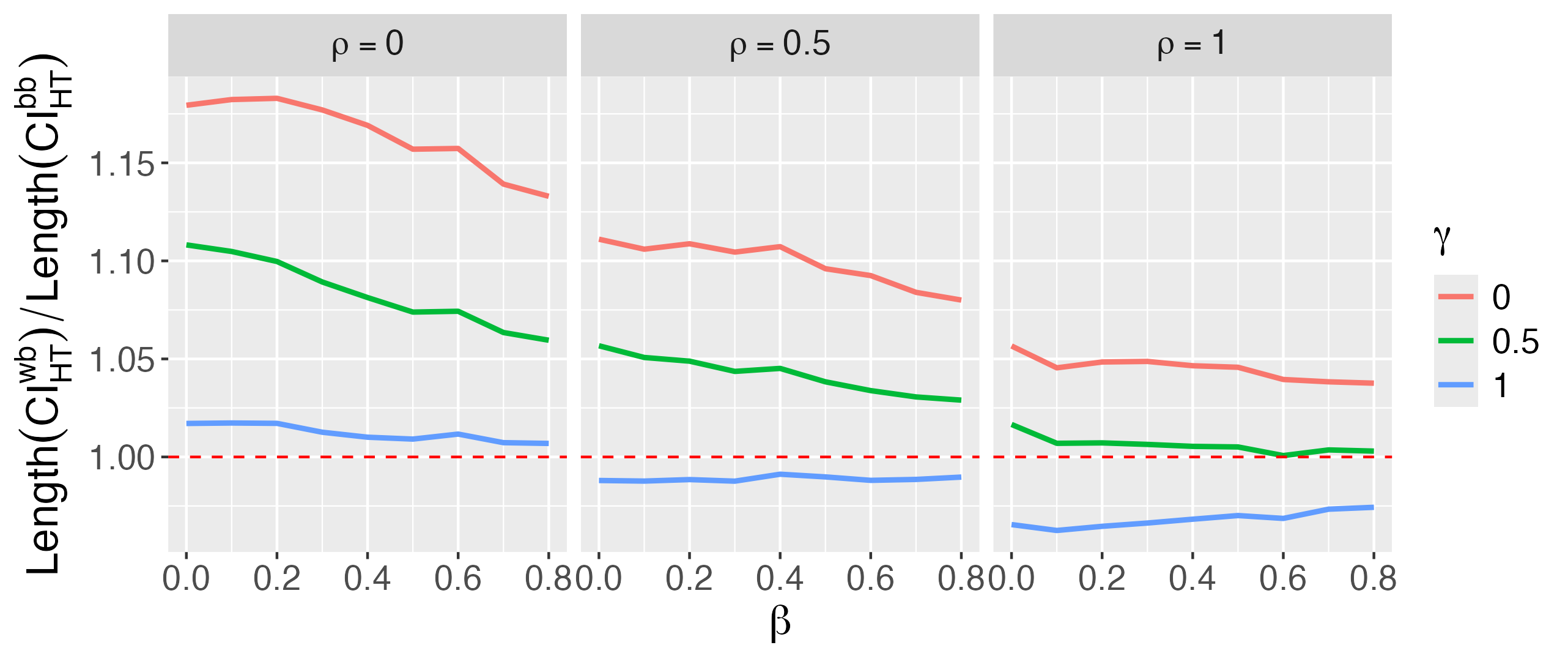}
  \end{subfigure}

   \begin{subfigure}{\textwidth}
    \centering
    \includegraphics[width=\textwidth]{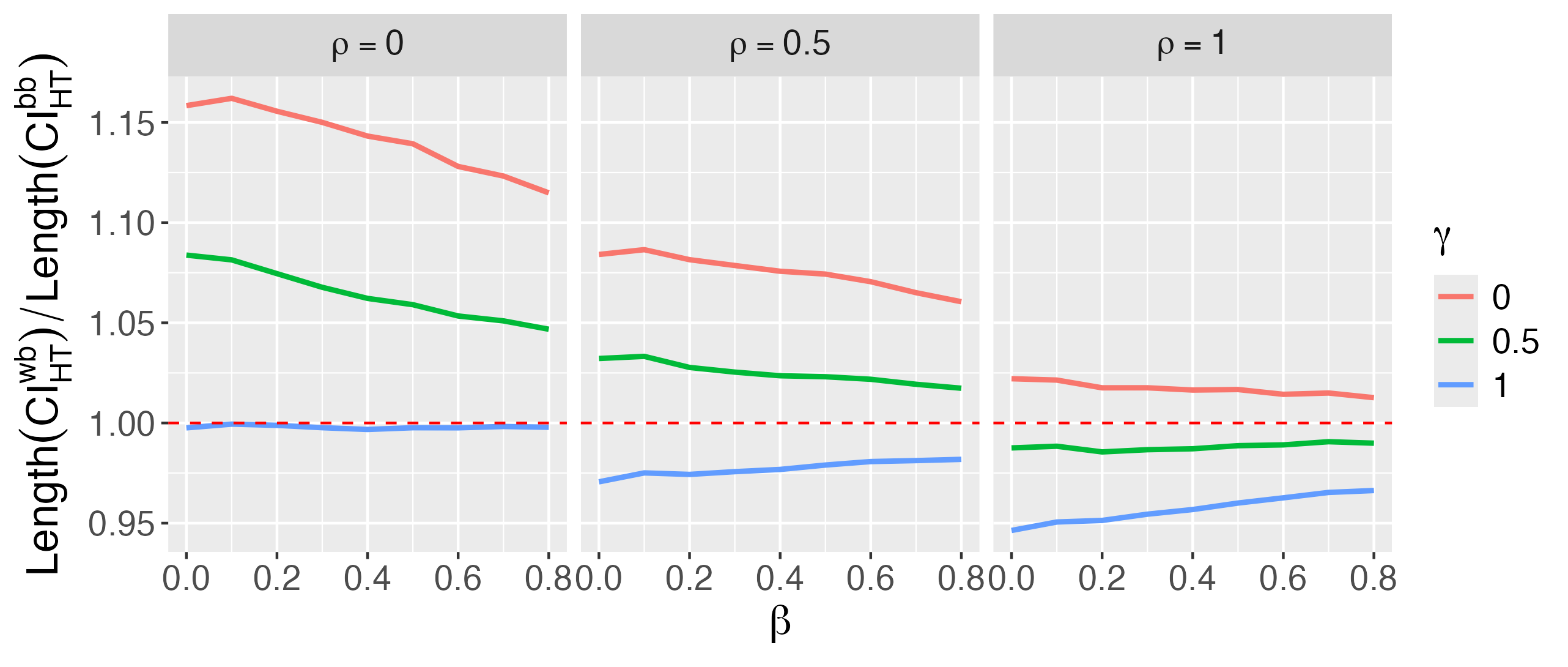}
  \end{subfigure}

 \begin{subfigure}{\textwidth}
    \centering
    \includegraphics[width=\textwidth]{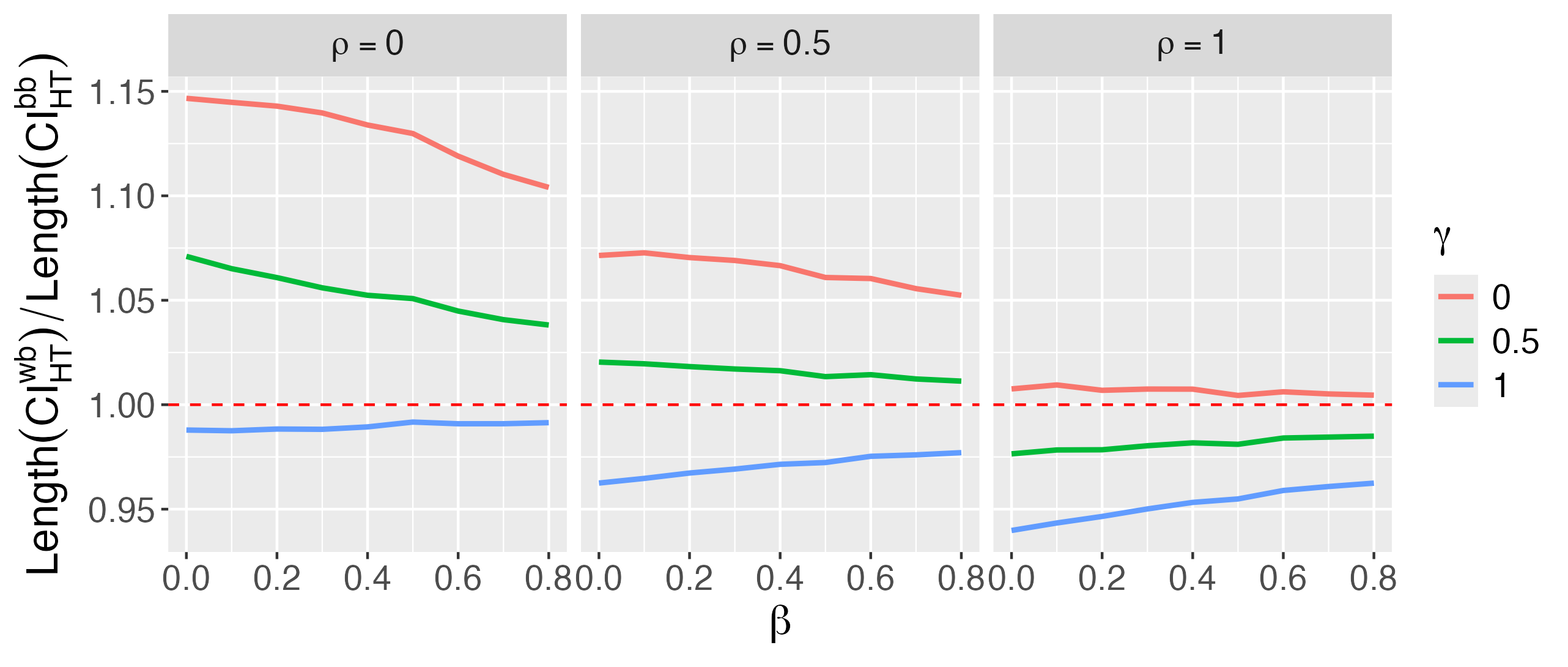}
  \end{subfigure}
\caption{Simulations under \textbf{S1}. From top to bottom, the panels correspond to $K = 10$, $20$, and $50$, respectively.
Each line corresponds to the empirical ratio of average length of confidence intervals  $\text{CI}^{wb}_{\HT} $ and  $\text{CI}^{bb}_{\HT} $ in one simulation.
The dashed horizontal red lines are at 1.}
\label{fig: len_wb_bb_HT_S1} 
\end{figure}

\begin{figure}[ht]
\centering 
\begin{subfigure}{\textwidth}
    \centering
    \includegraphics[width=\textwidth]{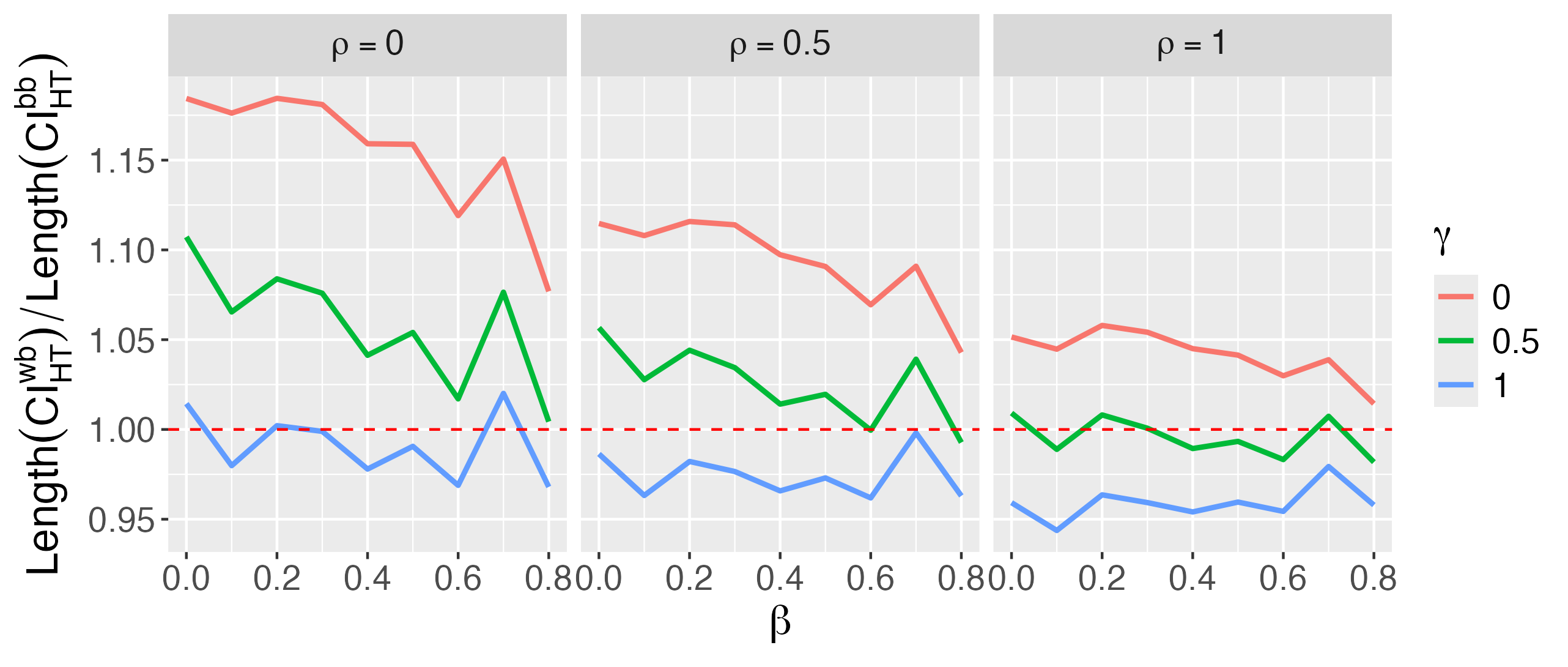}
  \end{subfigure}

   \begin{subfigure}{\textwidth}
    \centering
    \includegraphics[width=\textwidth]{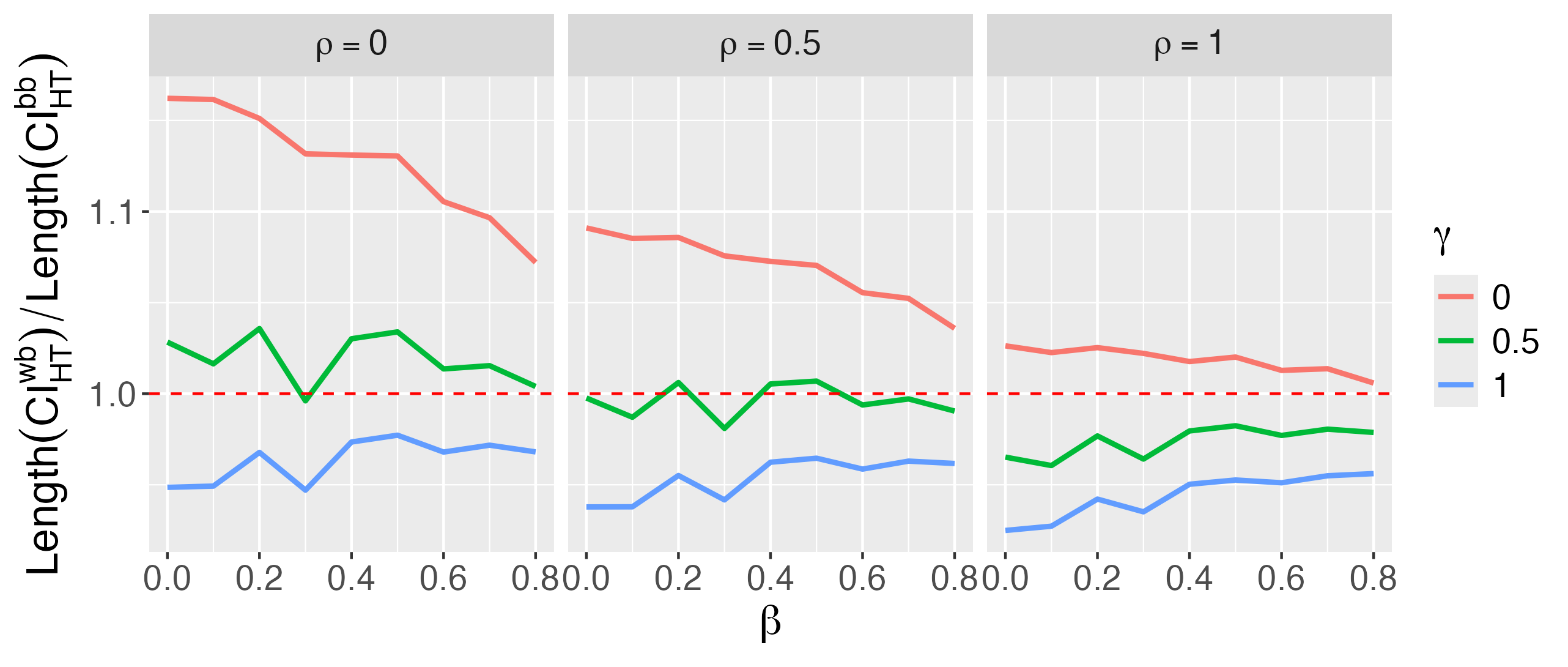}
  \end{subfigure}

 \begin{subfigure}{\textwidth}
    \centering
    \includegraphics[width=\textwidth]{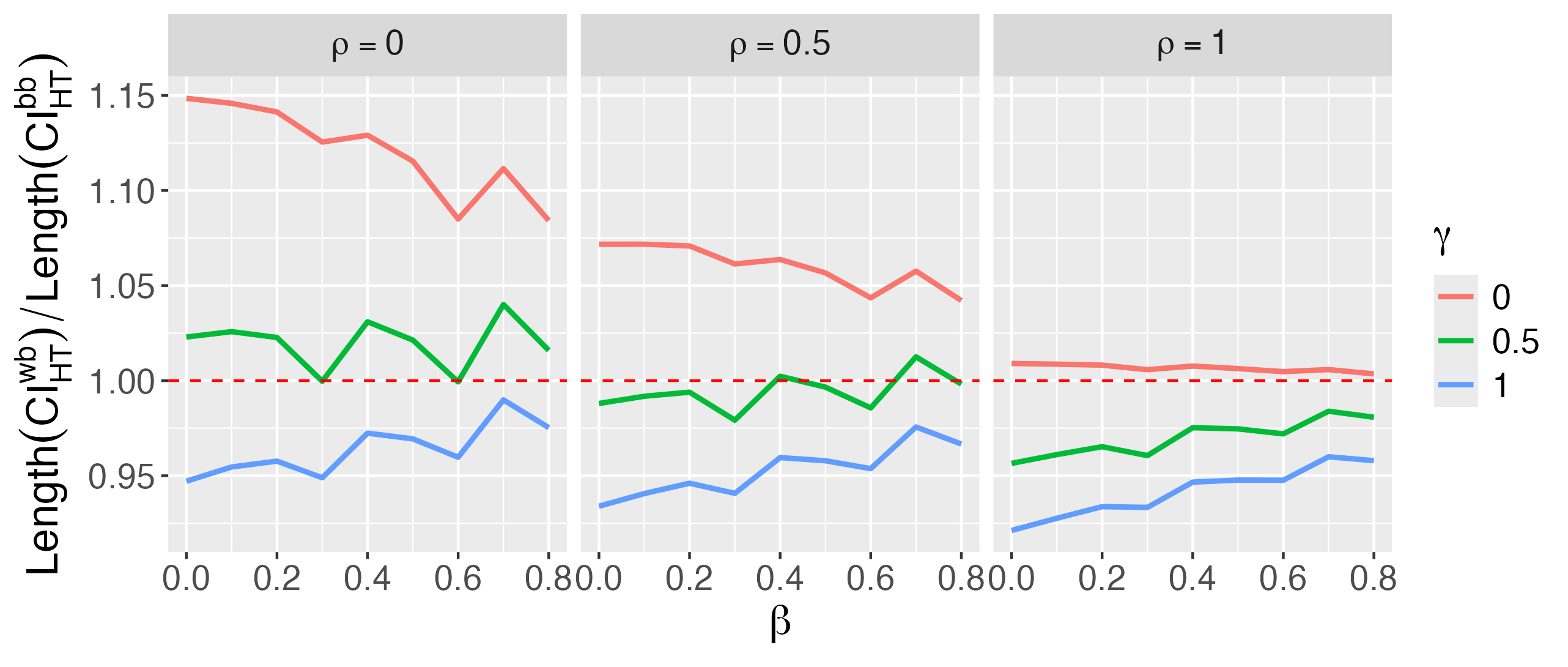}
  \end{subfigure}
\caption{Simulations under \textbf{S2}. From top to bottom, the panels correspond to $K = 10$, $20$, and $50$, respectively.
Each line corresponds to the empirical ratio of average length of confidence intervals  $\text{CI}^{wb}_{\HT} $ and  $\text{CI}^{bb}_{\HT} $ in one simulation.
The dashed horizontal red lines are at 1.}
\label{fig: len_wb_bb_HT_S2} 
\end{figure}

\begin{figure}[ht]
\centering 
\begin{subfigure}{\textwidth}
    \centering
    \includegraphics[width=\textwidth]{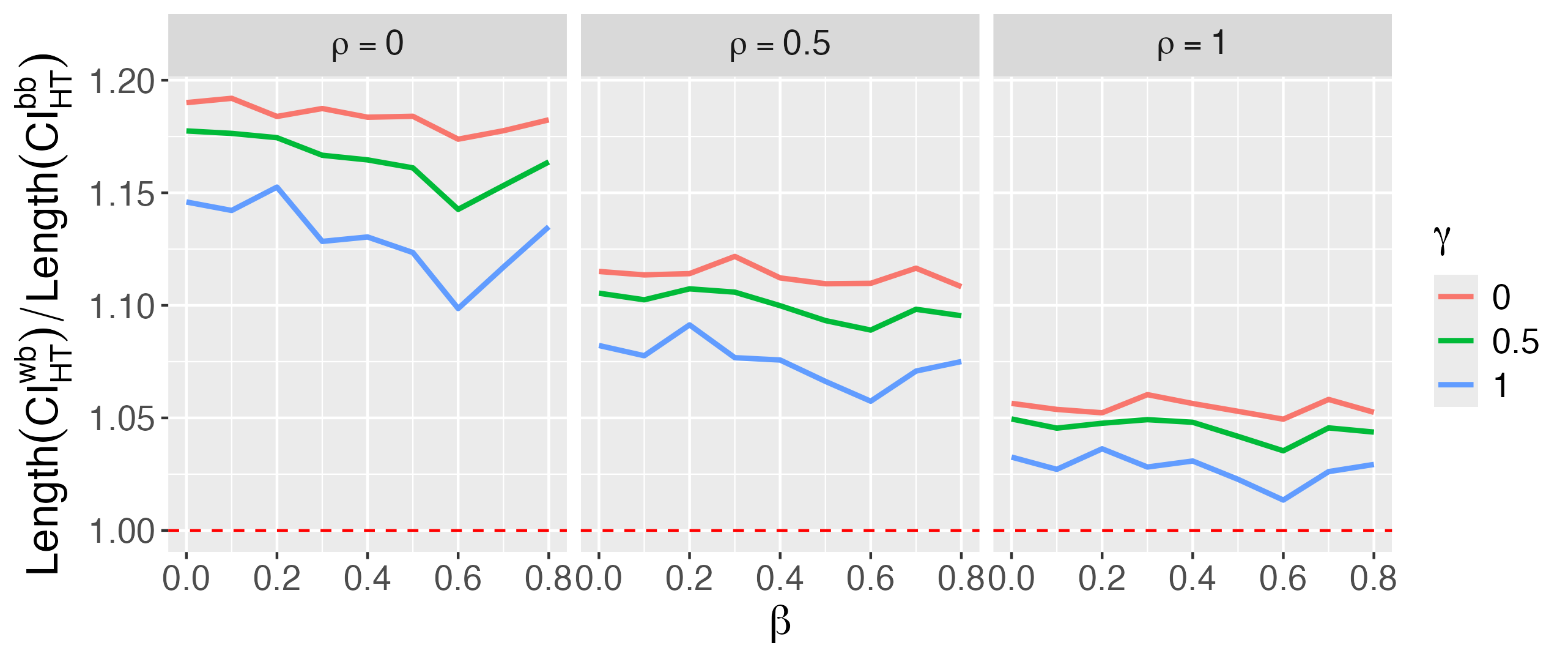}
  \end{subfigure}

   \begin{subfigure}{\textwidth}
    \centering
    \includegraphics[width=\textwidth]{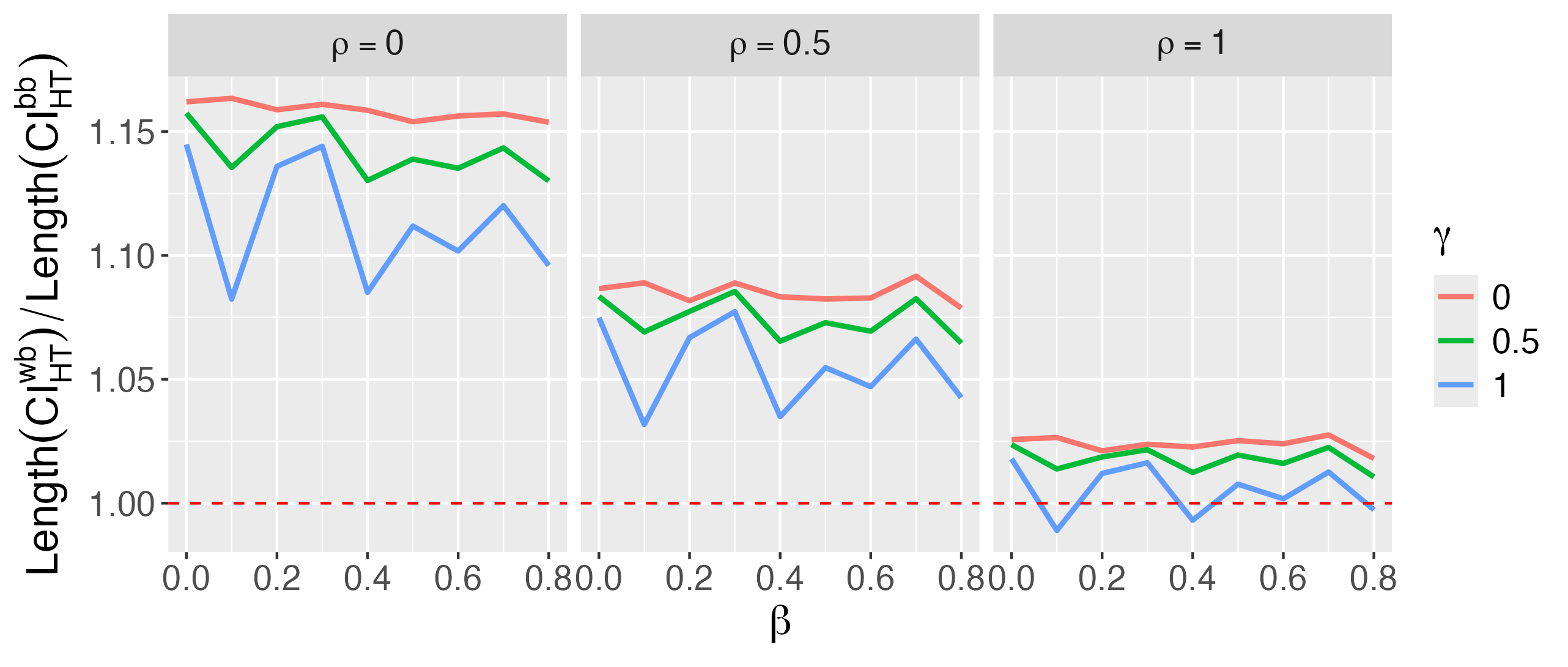}
  \end{subfigure}

 \begin{subfigure}{\textwidth}
    \centering
    \includegraphics[width=\textwidth]{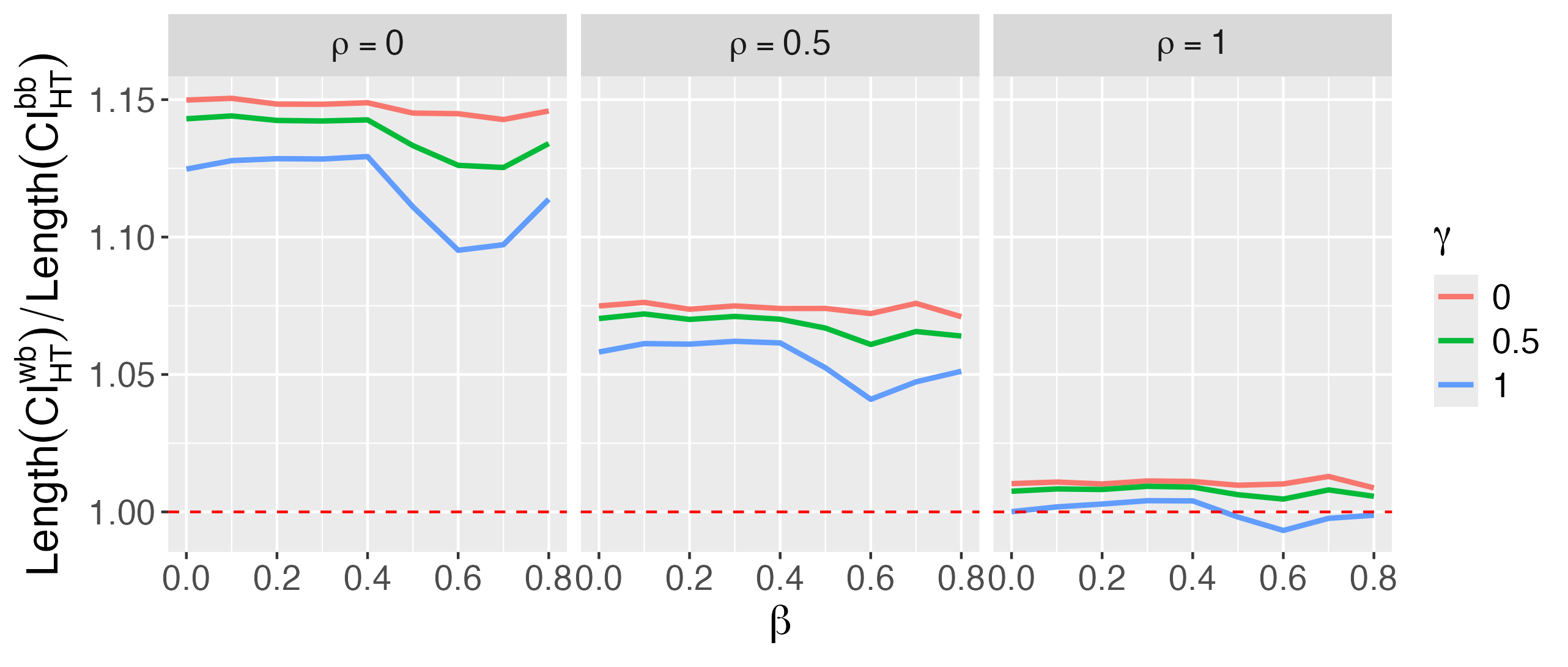}
  \end{subfigure}
\caption{Simulations under \textbf{S3}. From top to bottom, the panels correspond to $K = 10$, $20$, and $50$, respectively.
Each line corresponds to the empirical ratio of average length of confidence intervals  $\text{CI}^{wb}_{\HT} $ and  $\text{CI}^{bb}_{\HT} $ in one simulation.
The dashed horizontal red lines are at 1.}
\label{fig: len_wb_bb_HT_S3} 
\end{figure}

Next, we investigate how the choice of variance estimator affects the efficiency of confidence intervals.
When $\rho = 0$, the length of ${\rm CI}_{\haj}^{bb}$ is smaller than that of ${\rm CI}_{\haj}^{wb}$ across all values of $\gamma$, likely due to the substantial within-block bias in variance estimation under $\rho = 0$.
In contrast, when $\rho = 1$, the length of ${\rm CI}_{\haj}^{wb}$ is smaller than that of ${\rm CI}_{\haj}^{bb}$ when $\gamma = 0$, and the two are similar when $\gamma = 0.5$, reflecting the fact that the within-block bias vanishes under $\rho = 1$.
This emphasizes that the choice of variance estimator should be guided by the specific context and block structure.
Since the results with $K=50$ and $K=100$ are similar, we omit the results with $K=100$ and presents the results in Figures~\ref{fig: len_wb_bb_haj_S1}, \ref{fig: len_wb_bb_haj_S2}, \ref{fig: len_wb_bb_haj_S3}, \ref{fig: len_wb_bb_HT_S1}, \ref{fig: len_wb_bb_HT_S2}, and \ref{fig: len_wb_bb_HT_S3}.

\graphicspath{{./Images/}}

\graphicspath{{./Images/}}

\graphicspath{{./Images/}}

\graphicspath{{./Images/}}

\graphicspath{{./Images/}}

\graphicspath{{./Images/}}

\subsection{Additional coverage results}\label{supp_mat:add_cov}
\begin{figure}[ht]
\centering 
\begin{subfigure}{\textwidth}
    \centering
    \includegraphics[width=\textwidth]{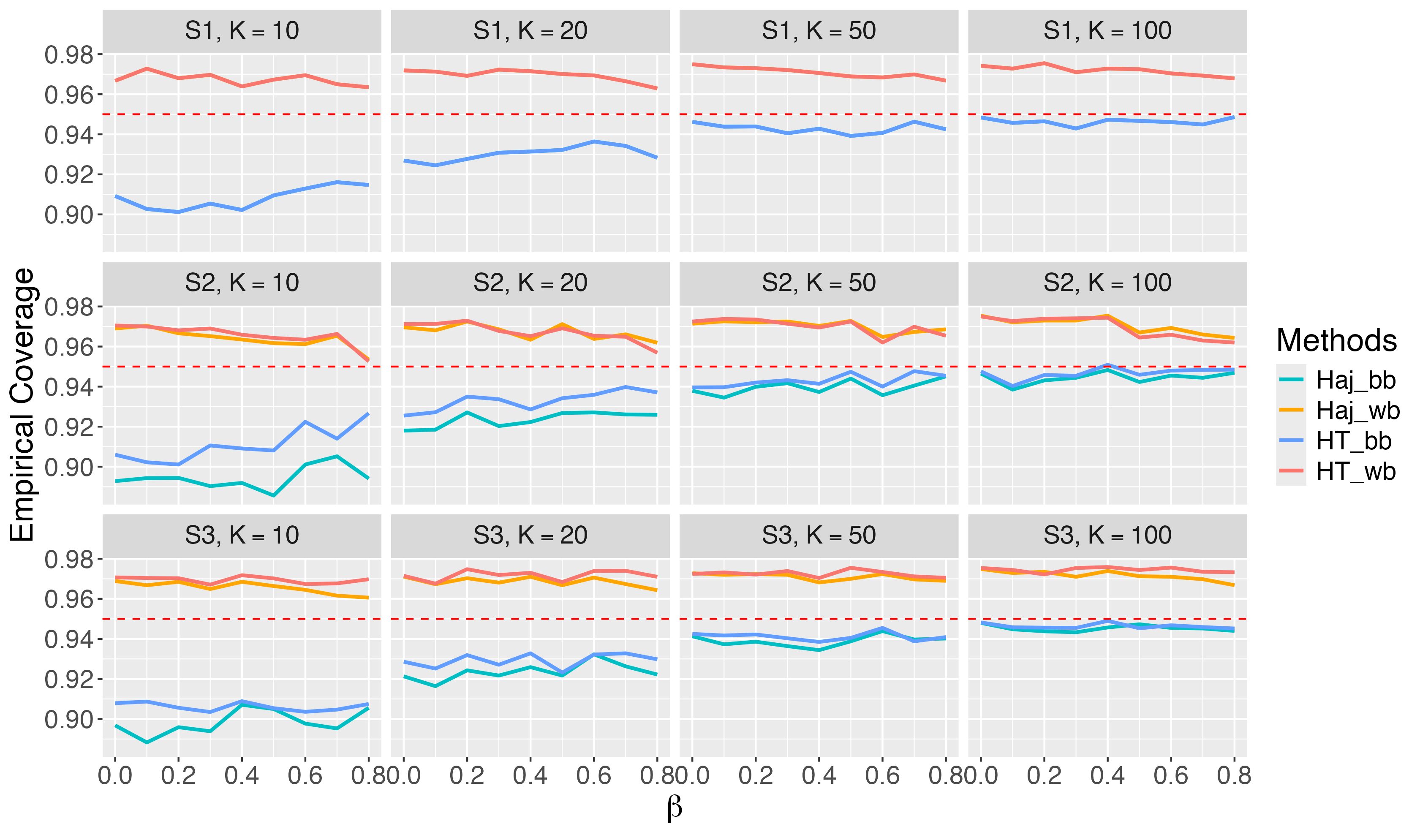}
  \end{subfigure}

 \begin{subfigure}{\textwidth}
    \centering
    \includegraphics[width=\textwidth]{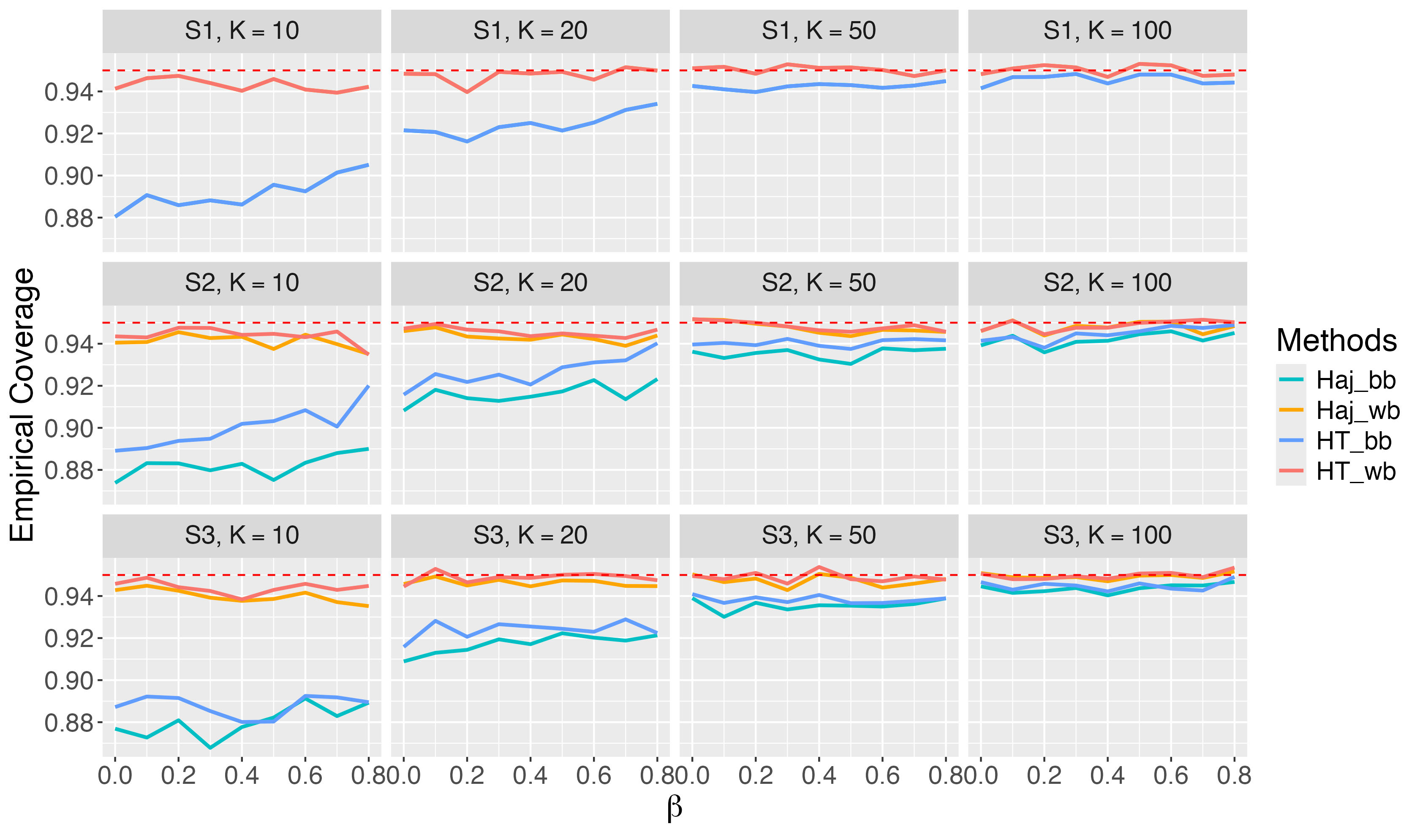}
  \end{subfigure}
\caption{Simulations with $\gamma=0$. From top to bottom, the panels correspond to $\rho = 0$, $0.5$, and $1$, respectively.
Each line corresponds to the empirical coverage of confidence interval in one simulation.
\texttt{HT} and \texttt{Haj} denote the Horvitz-Thompson and H\'ajek estimators respectively, and \texttt{bb} and \texttt{wb} denote corresponding variance estimators with between-block and within-block biases respectively.
The dashed horizontal red lines are at 0.95.}
\label{fig: cov_gamma0} 
\end{figure}

\begin{figure}[ht]
\centering 
\begin{subfigure}{\textwidth}
    \centering
    \includegraphics[width=\textwidth]{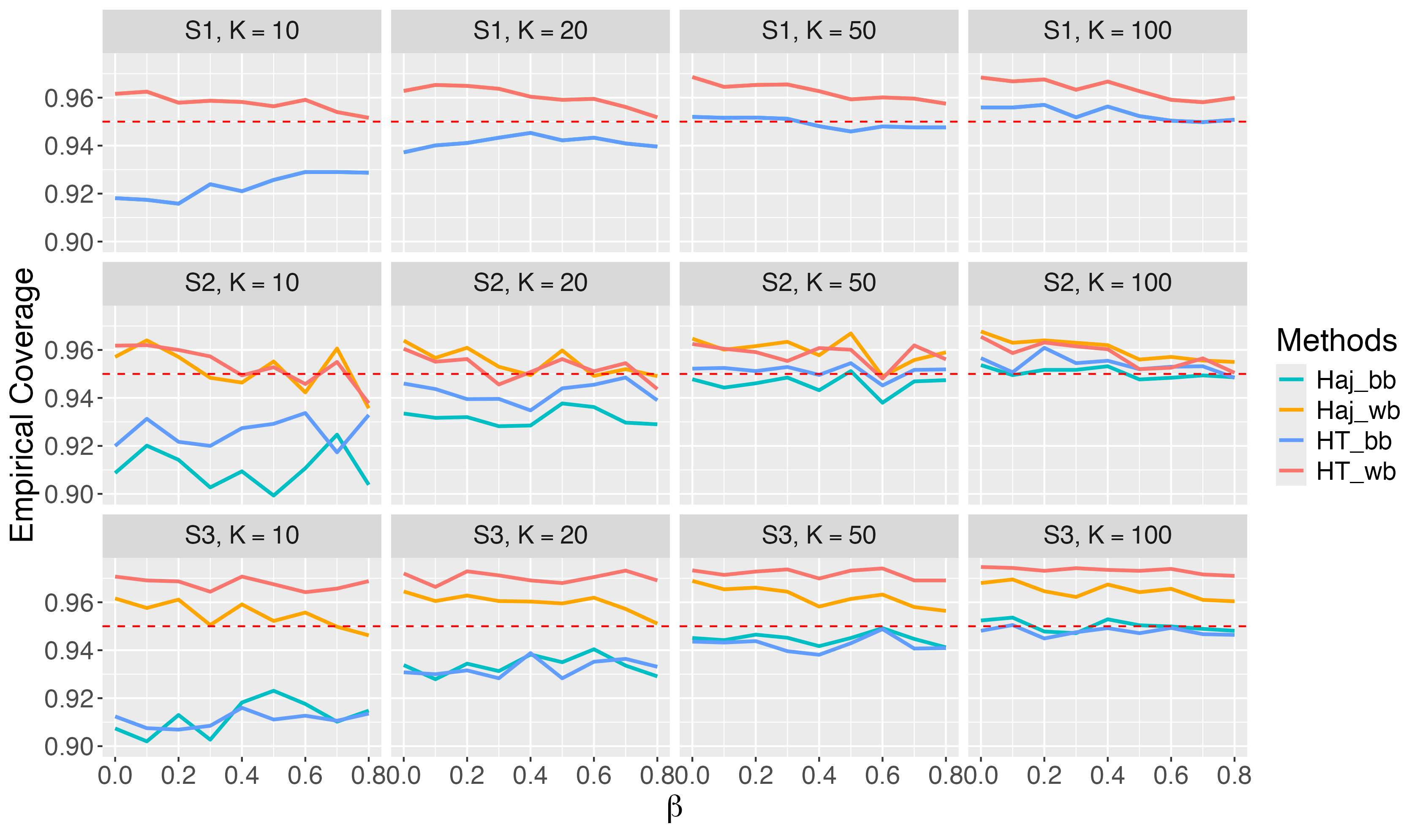}
  \end{subfigure}

 \begin{subfigure}{\textwidth}
    \centering
    \includegraphics[width=\textwidth]{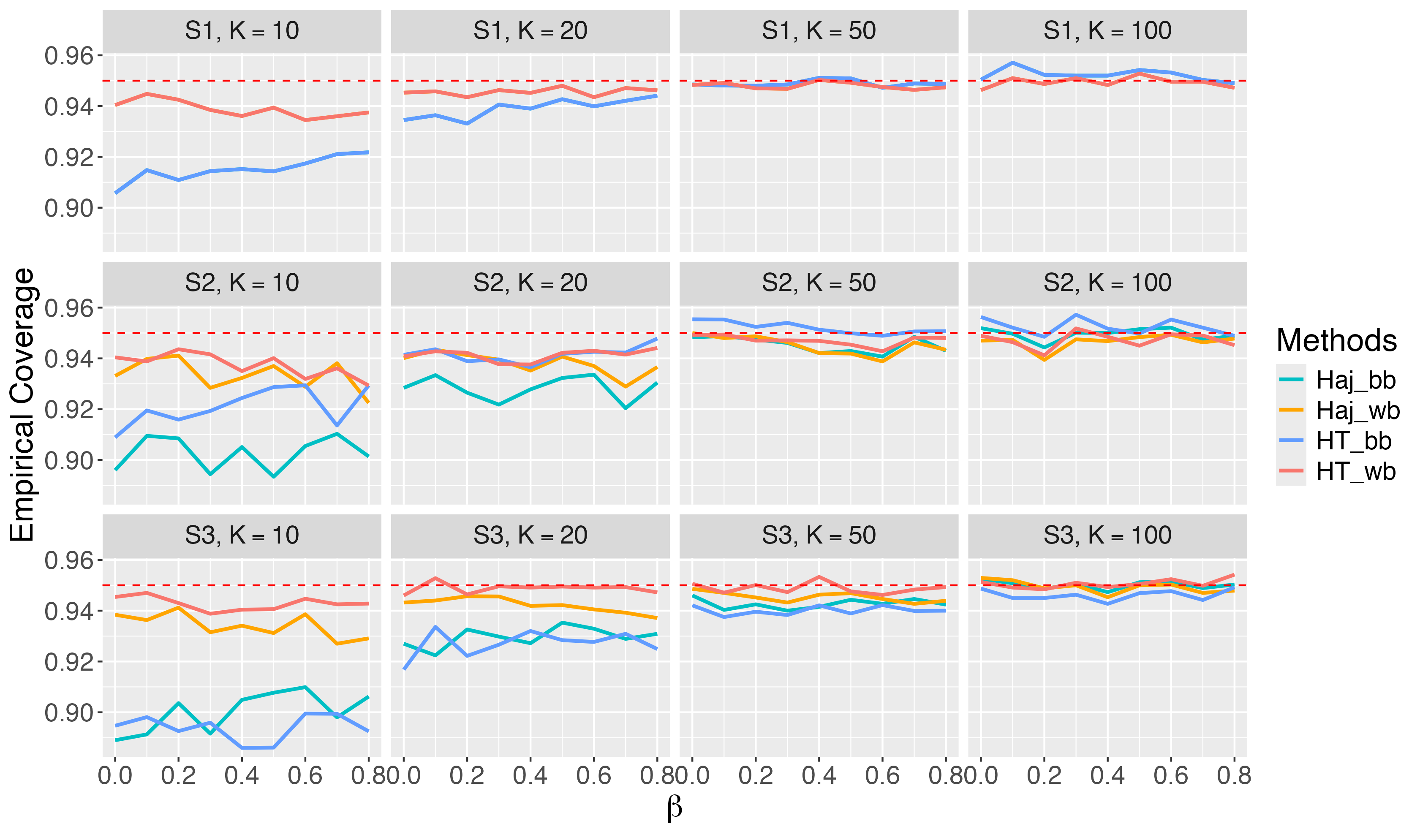}
  \end{subfigure}
\caption{Simulations with $\gamma=0.5$. From top to bottom, the panels correspond to $\rho = 0$, $0.5$, and $1$, respectively.
Each line corresponds to the empirical coverage of confidence interval in one simulation.
\texttt{HT} and \texttt{Haj} denote the Horvitz-Thompson and H\'ajek estimators respectively, and \texttt{bb} and \texttt{wb} denote corresponding variance estimators with between-block and within-block biases respectively.
The dashed horizontal red lines are at 0.95.}
\label{fig: cov_gamma0.5} 
\end{figure}

\begin{figure}[ht]
\centering 
\begin{subfigure}{\textwidth}
    \centering
    \includegraphics[width=\textwidth]{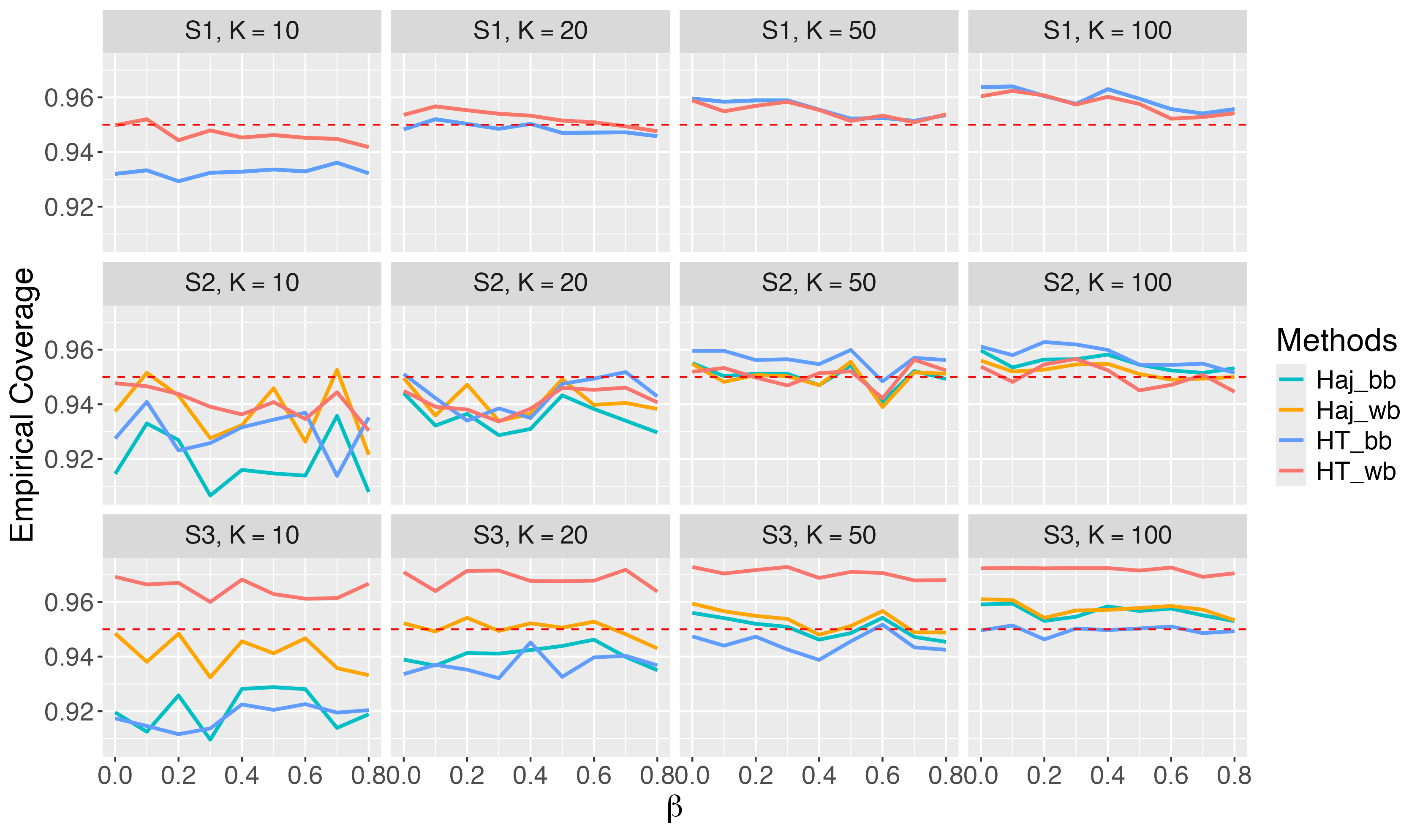}
  \end{subfigure}

   \begin{subfigure}{\textwidth}
    \centering
    \includegraphics[width=\textwidth]{Coverage-gamma1-rho1.png}
  \end{subfigure}

\caption{Simulations with $\gamma=1$. From top to bottom, the panels correspond to $\rho = 0$ and $0.5$, respectively.
Each line corresponds to the empirical coverage of confidence interval in one simulation.
\texttt{HT} and \texttt{Haj} denote the Horvitz-Thompson and H\'ajek estimators respectively, and \texttt{bb} and \texttt{wb} denote corresponding variance estimators with between-block and within-block biases respectively.
The dashed horizontal red lines are at 0.95.}
\label{fig: cov_gamma1} 
\end{figure}

In this section, we provide the additional results for the coverage of CIs ${\rm CI}_{*}^{bb}$, defined in \eqref{eq:CI star bb} using $\Shatbbstar$, and ${\rm CI}_{*}^{wb}$, defined in \eqref{eq:CI star wb} using $\Shatwbstar$, with the HT and H\'ajek estimators. For simplicity, we only presents the results with $\rho=0$ and $\rho=1$ in Figures~\ref{fig: cov_gamma0}, \ref{fig: cov_gamma0.5}, and \ref{fig: cov_gamma1}.

\graphicspath{{./Images/}}

\graphicspath{{./Images/}}

\graphicspath{{./Images/}}

\end{appendices}

\end{document}